\documentclass[twocolumn,secnumarabic,amssymb, nobibnotes, aps, prd]{revtex4-2}

\usepackage{graphicx}%
\usepackage{multirow}%
\usepackage{amsmath,amssymb,amsfonts}%
\usepackage{amsthm}%
\usepackage{mathrsfs}%
\usepackage[title]{appendix}%
\usepackage{xcolor}%
\usepackage{textcomp}%
\usepackage{booktabs}%
\usepackage{algorithm}%
\usepackage{algorithmicx}%
\usepackage{algpseudocode}%
\usepackage{listings}%
\usepackage{tikz}
\usepackage{adjustbox}
\usepackage{subfig}
\usepackage{pgfplots}
\usepgfplotslibrary{groupplots,dateplot}
\usetikzlibrary{shapes,positioning,fit,calc}
\usepackage{circuitikz}
\usetikzlibrary{decorations}
\usetikzlibrary{decorations.pathmorphing}
\usetikzlibrary{decorations.pathreplacing}
\usetikzlibrary{decorations.shapes}
\usetikzlibrary{decorations.text}
\usetikzlibrary{decorations.markings}
\usetikzlibrary{decorations.fractals}
\usetikzlibrary{decorations.footprints}
\usetikzlibrary{babel}
\usetikzlibrary{math}
\tikzset{>=latex}
\usetikzlibrary{arrows,automata}

\pgfplotsset{compat=newest}

\DeclareMathOperator{\real}{Re}

\begin{document}

\title{Stochastic quantum models for the dynamics of power grids }%

%
\begin{abstract}
While electric power grids play a key role in the decarbonization of society, it remains unclear how recent trends, such as the strong integration of renewable energies, can affect their stability. Power oscillation modes, which are key to the stability of the grid, are traditionally  studied numerically with the conventional view-point  of two regimes of extended  (inter-area) or localized  (intra-area) modes. In this article we introduce an analogy based on  stochastic quantum models and demonstrate its applicability  to power systems. We show from simple models that at low frequency the mean free path induced by disorder is inversely cubic in the frequency. This stems from the Courant-Fisher-Weyl theorem, which predicts a strong protection of the lowest frequency modes from disorder. As a consequence  a power oscillation, induced by some local disruption of the grid,  can propagate in a ballistic, diffusive or localised regime. In contrast with the conventional view-point, the existence of these three regimes is confirmed in a realistic model of the European power grid.  
\end{abstract}

\author{Pierrick Guichard}%
\altaffiliation{Univ. Grenoble Alpes, CNRS, Institut NEEL, F-38042, Grenoble, France}
\email[ ]{pierrick.guichard@neel.cnrs.fr}
\author{Nicolas Retière}%
\altaffiliation{Univ. Grenoble Alpes, CNRS, Grenoble INP, G2Elab, F-38000, Grenoble, France}
\email[]{nicolas.retiere@univ-grenoble-alpes.fr}
\author{Didier Mayou}%
\altaffiliation{Univ. Grenoble Alpes, CNRS, Institut NEEL, F-38042, Grenoble, France}
\email[]{didier.mayou@neel.cnrs.fr}
\date{June 2024}%

\maketitle

\section{Introduction}

While the proportion of electricity in final consumption has increased up to around 20 \% in 150 years, it is projected to reach over 50 \% by 2050. This second "electrical revolution" is expected to present considerable challenges for power systems stability. In particular, the massive rise of renewable energy production \cite{directorate-general_for_energy_european_commission_penetration_2020} introduces a high spatio-temporal variability, reduces system inertia and degrades stability \cite{international_energy_agency_integrating_2019}. Therefore new methods for monitoring power systems require advanced models and simulation tools for a better understanding of the dynamic of power grids \cite{entso-e_stability_2022}. Eigenmodes analysis are typically performed by power engineers for assessing dynamics of power systems, on a time scale of the order of a second and two regimes are conventionally identified. The inter-area regime is driven by long-range modes and gives power oscillations between distant parts of the grid. The intra-area regime is about oscillations inside limited areas.

In this paper, we present a new theoretical  approach inspired by quantum analogies. We examine how a local disruption of the grid induces a power oscillation which propagates in the power system, at the regional and continental scales. A study of simplified models based on the Lieb and the Honeycomb-Kagome lattices for modeling the geometry of the grid is presented first. The complex distribution of coupling and inertia of the nodes which exists at a regional or continental  scale, is treated as a disorder. Therefore the power oscillations propagate in a disordered medium and are characterized  by their mean free path and their localization length. The strong dependence on frequency of these two lengths  is established using a Shiba mean field theory and the scaling theory of Anderson localization. As a function of frequency  we establish the existence of three propagation regimes, ballistic, diffusive and localized, which is confirmed on a realistic model of the European grid.

\subsection{Swing equations and quantum analogy}

The power transmission grid is modeled by a planar graph. The voltage at every node $n$ is a unit complex number, described uniquely by its phase $\theta_n$ and all the phases are stored in a vector $\theta$. At every node $n$, the electric power $P_n$ is either generated or consumed by rotating machines of inertia $M_n$ and damping $D_n$. Applying the laws of dynamics to their rotors, subject to mechanical and electrical forces yields the swing equations \cite{dorfler2013synchronization} \cite{tyuryukanov} \cite{mukherjee} \cite{wang_2} modeling the physics of the power grid on time scales of the order of $1~\text{s}$

\begin{equation}
    \mathbb{M}\ddot{\theta} + \mathbb{D}\dot{\theta} = P - \mathbb{L}\theta,
    \label{LF_dyn}
\end{equation}
where  $\mathbb{M} := \text{Diag}(M_1, \dots, M_{N_\mathrm{nodes}})$ and  $\mathbb{D} := \text{Diag}(D_1, \dots, D_{N_\mathrm{nodes}})$.
Equation \eqref{LF_dyn} involves the square Laplacian matrix $\mathbb{L}$ \cite{andersson} \cite{kundur2017power} of dimension $N_\mathrm{nodes}\times N_\mathrm{nodes}$. Its non zero coefficients are $\mathbb{L}_{n_1 n_2} = -B_{l}$ where $B_l$ is the electrical susceptance of the line $l$ connecting the nodes $n_1$ and $n_2$. The diagonal coefficients of $\mathbb{L}$ are $\mathbb{L}_{n_1n_1} = -\sum_{n_2\neq n_1} \mathbb{L}_{n_1n_2}$. For the main part of the paper we neglect the damping term and discuss its influence only for the European model. The above equations represent coupled classical oscillators, on a two-dimensional space, that are submitted to external forces $P$   but the displacement of each oscillator, given by $\theta_n$, is one-dimensional. Note that the equations are invariant by a global shift of all phases $\theta$ which implies the existence of Goldstone modes at low frequencies for large systems \cite{goldstone1962broken} \cite{esposito2020effective} \cite{tulipman2021strongly}. Finally we note that in their original form the dimension of the pseudo-force $P$ is a power. Usually $P$ is measured in $\text{MW}$ and the pseudo-mass in $\text{MW}\text{s}^2$.

The structure of the low frequency eigenmodes is important and we comment here on the simplest case of uniform mass distribution. The spectrum of $\mathbb{L}$ can be analysed as follows. According to the Courant-Fisher-Weyl theorem, which holds for hermitian matrices, the eigenvector of $\mathbb{L}$ corresponding to its k-th smallest eigenvalue by $X_k = \underset{X \in \mathbb{C}^n, \; X\perp X_1\dots X_{k-1}}{\text{argmin}}\frac{X^\dagger \mathbb{L}X}{X^\dagger X}$. Because $\mathbb{L}$ is a Laplacian matrix, we have $X^\dagger \mathbb{L}X = \sum_{i<j}(-\mathbb{L}_{ij}) (X_i - X_j)^2$, for any vector $X$. Consequently, the ground state $X_1$ whose eigenvalue is $0$ is a vector of constant coefficients and is totally independent from the repartition of susceptances $B_l$. Since the next eigenvectors $X_k$ are orthogonal to $X_1$, they also benefit from a protection from disorder, even if weaker. In particular, the sign of the components of $X_k$ presents less than $k$ variations over connected subsets of nodes called nodal domains. A disorder may only reshape the domains, but not change their maximal number. All these considerations suggest that eigenvectors of $\mathbb{L}$  with small eigenvalues are protected from local disorder and only  sensitive to large scale variations of susceptances $B_l$. This is in agreement with the existence of Goldstone modes for large grids in the limit of low frequencies.

\begin{figure*}[ht!]
    \centering
 \subfloat[The Lieb lattice]{\begin{adjustbox}{clip,trim=1cm 1cm 0.65cm 0.65cm, raise=1cm,max width=0.25\linewidth}
\begin{tikzpicture}
\tikzmath{\l=10;}
\foreach \X in {0,2,...,\l}
{\foreach \Y in {0,2,...,\l}
 {\draw (\X,\Y) -- ++(0,1);
\draw (\X,\Y+1) -- ++(0,1);
 \draw (\X,\Y) -- ++(1,0);
\draw (\X+1,\Y) -- ++(1,0);
\node[state, fill=yellow, minimum size=20pt] at (\X,\Y){};}}
\tikzmath{\l=10;}
\foreach \X in {0,2,...,\l}
{\foreach \Y in {0,2,...,\l}
 {
 \node[state, fill=orange, minimum size=12pt] at (\X,\Y+1){};
 \node[state, fill=orange, minimum size=12pt] at (\X+1,\Y){};
 }}
\end{tikzpicture}
\end{adjustbox}}
\quad\quad\quad\quad
\subfloat[The Honeycomb-Kagome lattice]{\begin{adjustbox}{clip,trim=0cm 0cm 0cm 0cm,max width=0.5\linewidth} 
\begin{tikzpicture}
\tikzmath{\l=3; \t=-1; \ux=0.5; \uy=-0.8660254;
\zx=1-\ux;
\zy=-\uy;
\Xx=2*(\ux -\t);
\Xy=2*\uy;
\Yx=2*(\zx -\t);
\Yy=2*(\zy);
}
\foreach \X in {0,1,...,\l}
{\foreach \Y in {0,1,...,\l}
 {

\draw (\X*\Xx +\Y*\Yx,\X*\Xy + \Y*\Yy) -- (\X*\Xx + \Y*\Yx +1 , \X*\Xy + \Y*\Yy);
\draw (\X*\Xx +\Y*\Yx+1,\X*\Xy + \Y*\Yy) -- (\X*\Xx + \Y*\Yx +2 , \X*\Xy + \Y*\Yy);

\draw (\X*\Xx +\Y*\Yx,\X*\Xy + \Y*\Yy) -- (\X*\Xx + \Y*\Yx -\ux , \X*\Xy + \Y*\Yy -\uy);
\draw (\X*\Xx +\Y*\Yx-\ux,\X*\Xy + \Y*\Yy-\uy) -- (\X*\Xx + \Y*\Yx -2*\ux , \X*\Xy + \Y*\Yy -2*\uy);

\draw (\X*\Xx +\Y*\Yx +2,\X*\Xy + \Y*\Yy) -- (\X*\Xx + \Y*\Yx+2 +\zx , \X*\Xy + \Y*\Yy +\zy);
\draw (\X*\Xx +\Y*\Yx +2+\zx,\X*\Xy + \Y*\Yy+\zy) -- (\X*\Xx + \Y*\Yx+2 +2*\zx , \X*\Xy + \Y*\Yy +2*\zy);

}
}

\foreach \X in {0,1,...,\l}
{\foreach \Y in {0,1,...,\l}
 {
 \node[state, fill=yellow, minimum size=20pt] at (\X*\Xx +\Y*\Yx,\X*\Xy + \Y*\Yy){};
\node[state, fill=yellow, minimum size=20pt] at (\X*\Xx + \Y*\Yx +2 , \X*\Xy + \Y*\Yy){};
\node[state, fill=orange, minimum size=12pt] at (\X*\Xx + \Y*\Yx +1 , \X*\Xy + \Y*\Yy){};
\node[state, fill=orange, minimum size=12pt] at (\X*\Xx + \Y*\Yx -\ux , \X*\Xy + \Y*\Yy-\uy){};
\node[state, fill=orange, minimum size=12pt] at (\X*\Xx + \Y*\Yx +2 +\zx , \X*\Xy + \Y*\Yy+\zy){};
 }}
\end{tikzpicture}
\end{adjustbox}}
\caption{\textbf{The toy models.} For both lattices, the node states (yellow) and the line states (orange) alternate. A line state is always connected to two node states.}
\label{fig:lattices_lieb_hk}
\end{figure*}
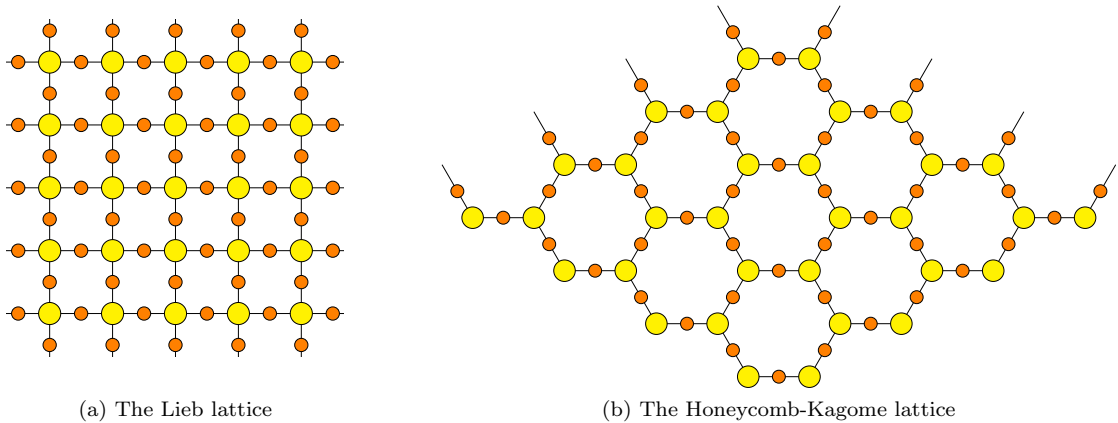

Figure \ref{fig:lattices_lieb_hk} shows two bipartite lattices, used in this work as toy models for the quantum analogy (see appendix \ref{appendix_hamiltonian}): the Lieb lattice (left panel) and the Honeycomb-Kagome (HK) lattice (right panel) \cite{cui2020realization} \cite{barreteau2017bird}. Note that these lattices are those of many real materials \cite{ding2013covalent} \cite{feng2012covalent}. The quantum analogy, introduced in \cite{guichard}, extends previous work, as the dual theory of transmission line outages \cite{ronellenfitsch2017dual}.
On a bipartite lattice, composed of $N_\mathrm{nodes}$ node states and $N_\mathrm{lines}$ line states,  the instantaneous electrical state of the grid is represented by a wavefunction $\psi:= \left[\begin{array}{c}
         i\mathbb{M}^{\frac{1}{2}}\dot{\theta} \\
         \hline
          \mathbb{H}_\mathrm{LN}\theta
    \end{array}\right]$  of size $N_\mathrm{nodes} + N_\mathrm{lines}$,
where $\mathbb{H}_\mathrm{LN}$ is defined from the parameters of the grid (see also appendix \ref{appendix_hamiltonian})). The instantaneous set of injected powers at the nodes is represented by the vector $\mathbb{P}:=\left[\begin{array}{c}\mathbb{M}^{-\frac{1}{2}}P\\
         \hline 0
    \end{array}\right]$ of the same size $N_\mathrm{nodes} + N_\mathrm{lines}$.
The  evolution of $\psi$ is driven by

\begin{equation}
    i\frac{\partial\psi}{\partial t} = \mathbb{H}\psi - \mathbb{P}.
    \label{schrodinger}
\end{equation}

This is a Schrödinger-like equation with $\hbar=1$ and we note that the matrix elements of $\mathbb{H}$ have the dimension of an angular frequency. Yet the eigenvalues of $\mathbb{H}$ are named energies in the rest of the text.

The tight-binding like Hamiltonian $\mathbb{H}$, defined on a lattice that mimics a grid, is built in the following way. One associates one node state $n$ for each node of the grid and one adds a line state $l$ for each line of the grid that connects two nodes. There is no matrix element of the Hamiltonian between two node states  or between two line states. Only line and node states can be coupled which means that the Hamiltonian $\mathbb{H}$ is bipartite. The coupling between one node state $n$ and a line state $l$ is given by

\begin{equation}
    \mathbb{H}_{ln} = \alpha_n \beta^0_{nl}\alpha_l,
    \label{H_alpha}
\end{equation}
where $\alpha_n := \sqrt{\frac{M}{M_n}}$, $\alpha_l = \sqrt{\frac{B_l}{B}}$,  $\beta^0_{nl} = \pm \sqrt{\frac{B}{M}}$, and $M$ and $B$ are reference values for inertia and susceptance. The two coefficients $\beta^0_{nl}$ related to the same line $l$ have an opposite sign chosen arbitrarily in accordance with the orientation of the line. All the possible configurations of parameters of a given lattice can be explored by varying $\alpha_n$ and $\alpha_l$ which are dimensionless positive parameters.
For the sake of simplicity of the presentation we consider two types of disorder, which affect only either the coefficients $\alpha_n$ on the nodes (type ($\mathcal{N}$)) or the coefficients $\alpha_l$ on the line (type ($\mathcal{L}$)). 
For the type ($\mathcal{N}$) we set the values of the node parameters to $\alpha_1 = \sqrt{\frac{M}{M_1}}\gg 1 $ and $\alpha_2 = \sqrt{\frac{M}{M_2}}=1$. This corresponds to a small mass $M_1$ on sites $1$ of concentration $c_1=c$ and to unchanged mass on sites $2$ of concentration $c_2=1-c$. Indeed the mass of nodes that correspond to consumers, or renewable energy generators, have usually a much lower mass than the sites corresponding to other generators. For the disorder of type ($\mathcal{L}$) there is a removal of a given concentration $c_1=c$ of lines in the network, corresponding to $\alpha_1=0$, while the other lines are unchanged with $\alpha_2 = 1$.

The eigenvectors of $\mathbb{H}$ with eigenvalues close to zero are expected to be protected from local disorder.
Indeed, since the product of the two non-diagonal blocks of $\mathbb{H}$ is equal to $\mathbb{L}_\mathrm{M} :=\mathbb{M}^{-\frac{1}{2}}\mathbb{L}\mathbb{M}^{-\frac{1}{2}}$ (see equation \eqref{laplacian_mass}), the eigenvalues of $\mathbb{H}$ are square roots of the eigenvalues of $\mathbb{L}_\mathrm{M}$. The $N_\mathrm{Nodes}$ first components of the eigenvectors of $\mathbb{H}$ are eigenvectors of $\mathbb{L}_\mathrm{M}$. Again in the simple case of a uniform mass this means that the spectral properties of $\mathbb{H}$ are inherited from the Laplacian structure of $\mathbb{L}$. Therefore, the low energy modes for large lattices are Goldstone modes. This is  confirmed, in the next section, by the expression of the mean free path $\ell$ as a function of frequency, even in the case of non uniform inertia distribution.

\section{Spectral properties of the toy models}


Our approach is based on the toy models presented in figure \ref{fig:lattices_lieb_hk}. Perfect crystalline lattices, like the Lieb and Honeycomb-Kagome lattice, are indeed a convenient starting point for our work, since the related Hamiltonian is invariant by translation.
Because of this, Bloch's theorem \cite{bloch1929quantenmechanik} and the corresponding Bloch waves can be used to diagonalize the Hamiltonian, leading to analytical expressions of the eigenvalues, eigenvectors and densities of states of this operator.

\subsection{The Lieb lattice}

\begin{figure*}[ht!]
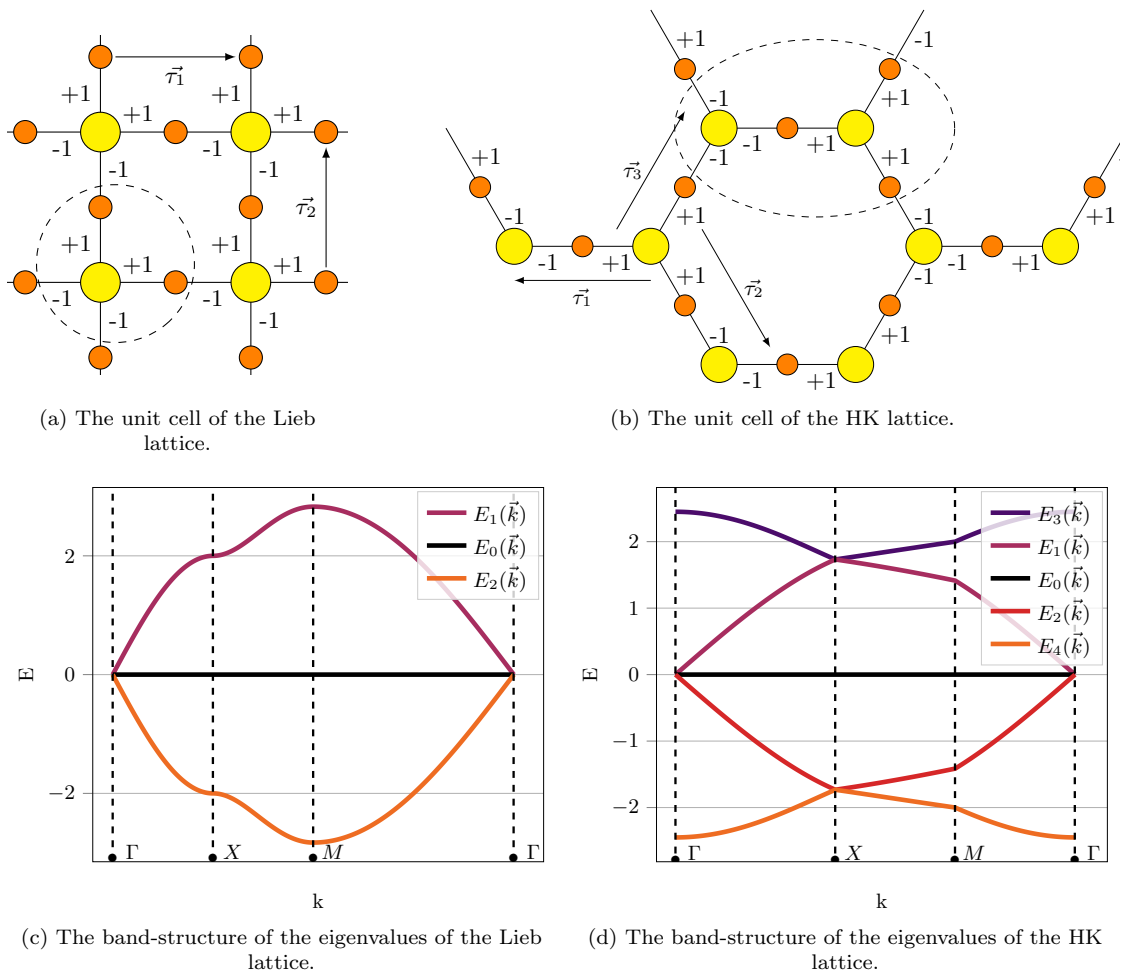

    \centering
     \subfloat[The unit cell of the Lieb lattice.]{\begin{adjustbox}{clip,trim=1.4cm 1.2cm 0.72cm 0.72cm, raise=0.3cm,max width=0.3\linewidth} 
    \begin{tikzpicture}
    \tikzmath{\l=4;}
    \foreach \X in {0,2,...,\l}
    {\foreach \Y in {0,2,...,\l}
     {\draw (\X,\Y) -- ++(0,1) node [midway, left] {{\small +1}};
    \draw (\X,\Y+1) -- ++(0,1) node [midway, right] {{\small -1}};
     \draw (\X,\Y) -- ++(1,0) node [midway, above] {{\small +1}};
    \draw (\X+1,\Y) -- ++(1,0)node [midway, below] {{\small -1}};
    \node[state, fill=yellow, minimum size=15pt] at (\X,\Y){};}}
    \tikzmath{\l=4;}
    \foreach \X in {0,2,...,\l}
    {\foreach \Y in {0,2,...,\l}
     {
     \node[state, fill=orange, minimum size=7pt] at (\X,\Y+1){};
     \node[state, fill=orange, minimum size=7pt] at (\X+1,\Y){};
     }}
      \node[ellipse, draw = black, minimum width = 2.1cm, minimum height = 2.1cm, rotate=-50, dashed] (e) at (2.2, 2.25){};
      \draw [->] (2.2, 5) -- (3.8, 5) node [midway, below] {{\small $\vec{\tau_1}$}};
      \draw [->] (5, 2.2) -- (5, 3.8) node [midway, left] {{\small $\vec{\tau_2}$}};
    \end{tikzpicture}
    \end{adjustbox}}
    \quad\quad\quad\quad
     \subfloat[The unit cell of the HK lattice.]{\begin{adjustbox}{clip,trim=0cm 0.cm 0.4cm 0cm,raise=0.cm, max width=0.5\linewidth} 
    \begin{tikzpicture}
    \tikzmath{\l=1; \t=-1; \ux=0.5; \uy=-0.8660254;
    \zx=1-\ux;
    \zy=-\uy;
    \Xx=2*(\ux -\t);
    \Xy=2*\uy;
    \Yx=2*(\zx -\t);
    \Yy=2*(\zy);
    }
    \foreach \X in {0,1,...,\l}
    {\foreach \Y in {0,1,...,\l}
     {
    
    \draw (\X*\Xx +\Y*\Yx,\X*\Xy + \Y*\Yy) -- (\X*\Xx + \Y*\Yx +1 , \X*\Xy + \Y*\Yy) node [midway, below] {{\normalsize -1}};
    \draw (\X*\Xx +\Y*\Yx+1,\X*\Xy + \Y*\Yy) -- (\X*\Xx + \Y*\Yx +2 , \X*\Xy + \Y*\Yy)node [midway, below] {{\normalsize +1}};
    
    \draw (\X*\Xx +\Y*\Yx,\X*\Xy + \Y*\Yy) -- (\X*\Xx + \Y*\Yx -\ux , \X*\Xy + \Y*\Yy -\uy) node [midway, right] {{\normalsize -1}};
    \draw (\X*\Xx +\Y*\Yx-\ux,\X*\Xy + \Y*\Yy-\uy) -- (\X*\Xx + \Y*\Yx -2*\ux , \X*\Xy + \Y*\Yy -2*\uy) node [midway, right] {{\normalsize +1}};

    \draw (\X*\Xx +\Y*\Yx +2,\X*\Xy + \Y*\Yy) -- (\X*\Xx + \Y*\Yx+2 +\zx , \X*\Xy + \Y*\Yy +\zy) node [midway, right] {{\normalsize +1}};
    \draw (\X*\Xx +\Y*\Yx +2+\zx,\X*\Xy + \Y*\Yy+\zy) -- (\X*\Xx + \Y*\Yx+2 +2*\zx , \X*\Xy + \Y*\Yy +2*\zy) node [midway, right] {{\normalsize -1}};
    
    }
    }
    
    \foreach \X in {0,1,...,\l}
    {\foreach \Y in {0,1,...,\l}
     {
     \node[state, fill=yellow, minimum size=15pt] at (\X*\Xx +\Y*\Yx,\X*\Xy + \Y*\Yy){};
    \node[state, fill=yellow, minimum size=15pt] at (\X*\Xx + \Y*\Yx +2 , \X*\Xy + \Y*\Yy){};
    \node[state, fill=orange, minimum size=7pt] at (\X*\Xx + \Y*\Yx +1 , \X*\Xy + \Y*\Yy){};
    \node[state, fill=orange, minimum size=7pt] at (\X*\Xx + \Y*\Yx -\ux , \X*\Xy + \Y*\Yy-\uy){};
    \node[state, fill=orange, minimum size=7pt] at (\X*\Xx + \Y*\Yx +2 +\zx , \X*\Xy + \Y*\Yy+\zy){};
     }}
      \node[ellipse, draw = black, minimum width = 4.1cm, minimum height = 2.6cm, rotate=-0, dashed] (e) at (4.4, 1.73){};
      
      \draw [->] (2, -0.5) -- (0, -0.5) node [midway, below] {{\small $\vec{\tau_1}$}};
      \draw [->] (2.5+0.25, 0+0.25) -- (2.5+0.25+2*\ux, 2*\uy+0.25) node [midway, right] {{\small $\vec{\tau_2}$}};
      \draw [->] (1.5, 0+0.25) -- (1.5+2*\ux, -2*\uy+0.25) node [midway, left] {{\small $\vec{\tau_3}$}};
    \end{tikzpicture}
    \end{adjustbox}}
    \\
      \subfloat[The band-structure of the eigenvalues of the Lieb lattice.]{\begin{adjustbox}{clip,trim=0cm 0.cm 0.cm 0.cm,max width=0.4\linewidth}
    \input{lieb_dispersion}
    \end{adjustbox}}
    \quad
      \subfloat[The band-structure of the eigenvalues of the HK lattice.]{\begin{adjustbox}{clip,trim=0cm 0.cm 0.cm 0.cm,max width=0.4\linewidth}
    \input{hk_dispersion}
    \end{adjustbox}}

         \caption{\textbf{Spectral properties of the toy models.}
         The upper panels shows the unit cell of the lattices and the signs of the hopping coefficients.
          The bottom panels give the band structure of the lattice, showing its eigenvalues.
          $X$, $M$ and $\Gamma$ are here the usual high symmetry points.}
    \label{fig:dos_cdt}
    \end{figure*}

The Lieb lattice, represented in figure \ref{fig:lattices_lieb_hk}, is derived from the square lattice by addition of the line states.
The unit cell of this lattice is composed of two line states and one node states, giving a mesh density coefficient $\gamma := \frac{N_\mathrm{nodes}}{N_\mathrm{lines}} =  \frac{1}{2}$.
In particular, there are three states in the unit cell.
Therefore, in a basis of Bloch waves, the Hamiltonian is block-diagonal with blocks of size $3\times3$ and is a function of the wave vector
 $\vec{k} = \begin{bmatrix} k_1 & k_2\end{bmatrix}^\dagger := \begin{bmatrix} \vec{k}\cdot \vec{\tau_1} & \vec{k}\cdot \vec{\tau_2}\end{bmatrix}^\dagger$
 ($\tau_1$ and $\tau_2$ are represented in figure \ref{fig:dos_cdt}).
 
In the basis of Bloch waves, the Hamiltonian is

\begin{equation}
    \mathbb{H} = 2 \sqrt{\frac{B}{M}}\begin{bmatrix}
        0 & \sin(\frac{k_1}{2}) & \sin(\frac{k_2}{2})\\
        \sin(\frac{k_1}{2}) & 0 & 0 \\
        \sin(\frac{k_2}{2}) & 0 & 0
    \end{bmatrix}.
    \label{H_block_lieb}
\end{equation}

Diagonalising the block presented in equation \eqref{H_block_lieb} yields the eigenvalues and eigenvectors of $\mathbb{H}$.
For the Lieb lattice, the eigenvalues are approximated, near the zero energy,
 by 

\begin{align}
    &E_0(\vec{k}) = 0, \\
    &E_1(\vec{k}) = -E_2(\vec{k}) = \sqrt{\frac{B}{M}} E(\vec{k}),
\end{align}
where $E(\vec{k}) :=\sqrt{k_1^2 + k_2^2}$ is the first eigenvalue normalised with respect to the unit of energy $\sqrt{\frac{B}{M}}$. The vector

\begin{equation}
    X(\vec{k}) \simeq \frac{1}{\sqrt{2N_\mathrm{nodes}} }\begin{bmatrix}
    1\\
    \frac{ k_1}{E(\vec{k})}\\
    \frac{ k_2}{E(\vec{k})}
    \end{bmatrix},
    \label{eigenvector_approx_lieb}
\end{equation}
is an eigenvector of $\mathbb{H}$ associated with the eigenvalue $\sqrt{\frac{B}{M}} E(\vec{k})$.

\subsection{The Honeycomb-Kagome lattice}

Similarly to the Lieb lattice, it is also possible to use the Bloch theorem for computing the eigenvalues and eigenvectors of the Honeycomb-Kagome lattice presented in figure  \ref{fig:lattices_lieb_hk},
The unit cell of the Honeycomb-Kagome lattice is composed of two node states and three line states. So in this case, there are five states in the unit cell.
Therefore, in the basis of Bloch waves, the Hamiltonian is block diagonal with blocks of size $5\times 5$.
For this lattice, the wave vector $\vec{k}$ is written in $\mathbb{R}^3$ in the following manner

\begin{equation}\vec{k} := \begin{bmatrix}
    k_1 \\ k_2 \\ k_3
\end{bmatrix} := \begin{bmatrix}
    \vec{k} \cdot \vec{\tau_1}\\
    \vec{k} \cdot \vec{\tau_2}\\
    \vec{k} \cdot \vec{\tau_3}
\end{bmatrix},
\end{equation}
where $\tau_1$, $\tau_2$ and $\tau_3$ are represented on figure \ref{fig:dos_cdt}.

\begin{figure*}[ht!]
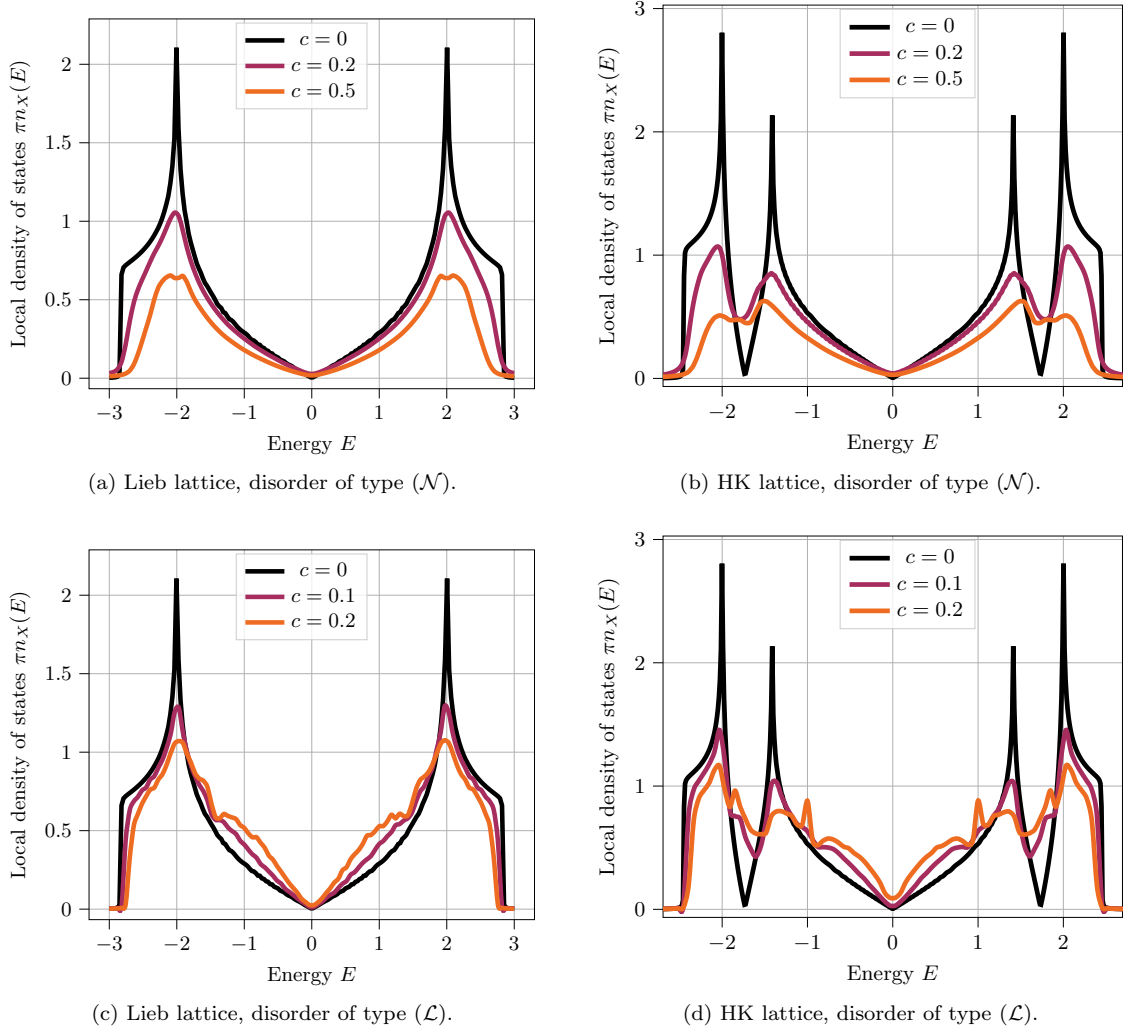

    \centering
     \subfloat[Lieb lattice, disorder of type ($\mathcal{N}$).]{\begin{adjustbox}{clip,trim=0cm 0.cm 0.cm 0.cm,max width=0.4\linewidth}
    \input{lieb_dos_cdt_nodes}
    \end{adjustbox}}\quad\quad
      \subfloat[HK lattice, disorder of type ($\mathcal{N}$).]{\begin{adjustbox}{clip,trim=0cm 0.cm 0.cm 0.cm,max width=0.4\linewidth}
    \input{hk_dos_cdt_nodes}
    \end{adjustbox}}\\
   \subfloat[Lieb lattice, disorder of type ($\mathcal{L}$).]{\begin{adjustbox}{clip,trim=0cm 0.cm 0.cm 0.cm,max width=0.4\linewidth}
    \input{lieb_dos_cdt_rupture_nodes}
    \end{adjustbox}}
        \quad\quad
      \subfloat[HK lattice, disorder of type ($\mathcal{L}$).]{\begin{adjustbox}{clip,trim=0cm 0.cm 0.cm 0.cm,max width=0.4\linewidth}
    \input{hk_dos_cdt_nodes_rupture}
    \end{adjustbox}}\\
         \caption{\textbf{Densities of states of the toy models.}
          The panels shows the local density of states $\pi n_X(E)$ on a node state $X$ for a disorder of type ($\mathcal{N}$) and ($\mathcal{L}$) for various values of disorder $c$.}
    \label{fig:dos_cdt_hk}
    \end{figure*}

In the basis of Bloch waves, the Hamiltonian is

\begin{equation}
    \mathbb{H} = \sqrt{\frac{B}{M}}\begin{bmatrix}
    0 & 0 & -e^{-i\frac{k_1}{2}} & -e^{-i\frac{k_2}{2}} & -e^{-i\frac{k_3}{2}} \\
    0 & 0 & e^{i\frac{k_1}{2}} & e^{i\frac{k_2}{2}} & e^{i\frac{k_3}{2}} \\
    -e^{i\frac{k_1}{2}} & e^{-i\frac{k_1}{2}} & 0 & 0 & 0 \\
    -e^{i\frac{k_2}{2}} & e^{-i\frac{k_2}{2}}  & 0 & 0 & 0\\
    -e^{i\frac{k_3}{2}} & e^{-i\frac{k_3}{2}} & 0 & 0 & 0\\
    \end{bmatrix}.
\end{equation}

Its eigenvalues can be written in function of the coefficient $\mu(\vec{k}) := e^{-ik_1} + e^{-ik_2} + e^{-ik_3} \simeq 3 - \frac{1}{2}(k_1^2 + k_2^2 + k_3^2)$
 by
 \begin{align}
     &E_0(\vec{k}) = 0, \\
     &E_1(\vec{k}) = -E_2(\vec{k}) =  \sqrt{\frac{B}{M}}E(\vec{k}),\\
     &E_3(\vec{k}) = -E_4(\vec{k}) =  \sqrt{\frac{B}{M}}\sqrt{3 + \vert \mu(k)\vert},
 \end{align}
 where $E(\vec{k}) := \sqrt{3 - \vert \mu(k)\vert}$ is again the first eigenvalue, normalised with respect to the unit of energy $\sqrt{\frac{B}{M}}$. It can be approximated by $E(\vec{k})  \simeq \sqrt{\frac{1}{2}(k_1^2 + k_2^2 + k_3^2)} = \frac{\sqrt{3}}{2} \vert \vert \vec{k} \vert \vert$,
 hence the vector

\begin{align}
      X(\vec{k}) := \frac{1}{\sqrt{2N_\mathrm{nodes}}}\begin{bmatrix}
      i\\
      i\\
      \frac{k_1}{E(\vec{k})}\\
      \frac{k_2}{E(\vec{k})}\\
      \frac{k_3}{E(\vec{k})}
      \end{bmatrix},
    \label{h_k_eigenvalue}
\end{align}
is an eigenvector of $\mathbb{H}$ associated with the eigenvalue $\sqrt{\frac{B}{M}}E(\vec{k})$.

The Lieb and Honeycomb-Kagome lattice have in common the presence of conical bands near the zero energy.
This regime of frequencies corresponds to the states that are expected to be particularly well-protected from disorder, as 
indicated by the Courant-Fisher-Weyl principle, hence it is of particular interest here.
In the same range of frequencies, the density of states is linear in $E$, and is given for both lattices by
\begin{equation}
    n(E) \simeq \frac{2 \eta N_\mathrm{nodes}}{\pi} \frac{M}{B}\vert E\vert,
    \label{dos_hk_approx}
\end{equation}
where $\eta = \frac{1}{4}$ for the Lieb lattice and $\eta=\frac{\sqrt{3}}{4}$ for the Honeycomb-Kagome lattice.
The linearity of the density of states near the zero energy is characteristic of a power system.
The densities of states, computed numerically, are shown in the panels of figure \ref{fig:dos_cdt_hk}.
 They are linear in the limit of low energy, even in configurations including disorder.

\section{The Shiba mean field theory}

Shiba's theory \cite{shiba1971reformulation} \cite{julien2001real} introduces an effective medium replicating the effects of stochastic disorder on the propagation of signals at a complex angular frequency $z$.
It applies for Hamiltonians of the same form than given in equation \eqref{H_alpha}.
In a nutshell, this theory is constructed by expanding the Green function of the disordered medium using a series of operators, called locators.
The average of each locator is then computed self-consistently, for taking into account any possible configuration of disorder.
 As a result, it gives an effective Hamiltonian with coefficients that are invariant by translation, mimicking the effects of disorder on the propagation of signals.
 This operator has a complex diagonal coefficient (called a self-energy) $\sigma(z)$ on all the sites affected by the disorder,
 while all the hopping terms are renormalised by $\alpha := \sqrt{c_1 \alpha_1^2 + c_2 \alpha_2^2}$.
 This means that for the disorder of type ($\mathcal{N}$) (respectively, type ($\mathcal{L}$)),
 $\sigma(z)$ is located on the node (respectively, on the line) states and the hopping terms $\mathbb{H}_{ln}$ from equation \eqref{H_alpha} are equal to $\alpha \sqrt{\frac{B}{M}}$.

\subsection{The mean field}

The self-energy $\sigma(z)$ is related self-consistently to the parameters of the problem by

\begin{equation}
    \frac{c_1 \alpha_1^2}{z - \alpha_1^2\Delta(z)} + \frac{c_2 \alpha_2^2}{z  - \alpha_2^2\Delta(z)} = \frac{\alpha^2}{z - \sigma(z) - \alpha^2\Delta(z)}
    \label{shiba},
\end{equation}
where $\Delta(z)$ is the effective interactor, representing the global effect of the medium on a given state. The calculations of $\sigma(z)$ can be started from an equivalent factorized form of Shiba's equation \eqref{shiba}, namely

  \begin{equation}
      \sigma(z) =- \frac{V^2 z\left(\frac{\alpha}{\alpha_1 \alpha_2}\right)^2}{1 - \frac{z}{\Delta(z)}\left(\frac{\alpha }{\alpha_1 \alpha_2}\right)^2},
      \label{sigma_S_AS}
  \end{equation}
where $V :=  \sqrt{c_1 c_2}\frac{\alpha_1^2 - \alpha_2^2}{\alpha}$.
As pointed out by appendix \ref{appendix_hamiltonian}, the density of states of the physical lattice can be deduced from the average of the Green function on the sites (1) and (2)
\begin{equation}
    <g>(z) := \frac{c_1}{z - \alpha_1^2 \Delta(z)} + \frac{c_2}{z - \alpha_2^2 \Delta(z)},
    \label{g_shiba_physical}
\end{equation}
But, importantly, $<g>(z)$ is not the Green function of the effective medium, and is not used for the calculations of $\sigma(z)$.
On the other hand, the diagonal coefficient $\Tilde{g}(z)$ of the Green function of the effective medium $\Tilde{G}(z)$,
 on a site subject to disorder (a node site for a disorder of type ($\mathcal{N}$), a line site for a disorder of type ($\mathcal{L}$)),
is defined by
  

  \begin{equation}
      \Tilde{g}(z) := \frac{1}{z - \sigma(z) -  \alpha^2 \Delta(z)}.
      \label{g_eff}
  \end{equation}

By definition, $\Tilde{g}(z)$ is a diagonal coefficient of $\mathcal{P}\Tilde{G}(z) \mathcal{P}$, where $\mathcal{P}$ is 
the projector on all the nodes (respectively, lines) for a disorder of type ($\mathcal{N}$) (respectively, type ($\mathcal{L}$)).
Applying formula \eqref{PGP}, we obtain

  \begin{equation}
      \Tilde{g}(z) = \frac{1}{z - \sigma(z) -  \alpha^2 \Delta(z)} = \frac{z}{\alpha^2} g_\mathrm{lap}(Z),
      \label{g_lap}
  \end{equation}
  where $Z :=\frac{z(z - \sigma(z))}{\alpha^2}$ and $g_\mathrm{lap}$ is the Green function of the Laplacian matrix connecting the sites together (nodes or lines, depending on the type of disorder).
  For a disorder of type ($\mathcal{N}$) (respectively, type ($\mathcal{L}$)), the Laplacian matrix is $\mathbb{L}_\mathrm{M}$ (respectively, $\Tilde{\mathbb{L}}_\mathrm{M}$), defined in equation \eqref{laplacian_mass}
  (respectively, equation \eqref{laplacian_mass_2}). This means that depending on the type of disorder, $g_\mathrm{lap}(Z) = X^\dagger \frac{1}{z - \mathbb{L}_\mathrm{M}}X$ or $g_\mathrm{lap}(Z) = X^\dagger \frac{1}{z - \Tilde{\mathbb{L}}_\mathrm{M}}X$, where $X$ is a unit vector located on a single state (node or line).

  At low frequency, equation \eqref{g_lap} gives an accurate estimation of $\Tilde{g}(z)$, as shown in the next sections.
  Therefore, the values of $\sigma(z)$ and $\Delta(z)$ can be retrieved from the system of two equations \eqref{sigma_S_AS} and \eqref{g_lap}.

  \subsection{Analytic computation of the speed and scattering time at low frequency}
  \label{appendix_low_freq}

  In equation \eqref{H_alpha}, the sign of $\beta^0_{nl}$ is given by the arbitrary orientation of the lines of the network.
  Reversing the orientation changes the signs of all the coefficients of the Hamiltonian but does not change the system.
  Therefore, all the Green function-like quantities of equation \eqref{sigma_S_AS} and \eqref{g_eff} are odd functions.
  As a consequence, $\sigma(z)$ and $\Delta(z)$ are also odd, meaning that

  \begin{align}
    \sigma(-z) &= -\sigma(z), \\
    \Delta(-z) &= -\Delta(z).
  \end{align}
At $z=0$, $\sigma(z)$ is also expected to be continuous, which implies that $\sigma(0) = 0$.
This is consistent with the calculations of the next paragraphs. 
Indeed, for the disorders of type ($\mathcal{N}$) and ($\mathcal{L}$), we show later that only odd powers of $z$ are involved in the approximation of $\sigma(z)$ at low frequency,
which is
     \begin{equation}
         \sigma(z) = \sigma_rz + i\sigma_iz^3,
     \end{equation}
where $\sigma_r$ and $\sigma_i$ are real constants.

The propagation of a signal in the effective medium is characterised by the eigenvalues $z$ of the effective Hamiltonian.
Because $\sigma(z)$ takes complex values, $z$ is also complex and its imaginary (respectively, real) part characterises the scattering time $\tau$
  (respectively, the velocity $v$) of the disordered system.
  Because in the effective medium, the hopping coefficients are renormalized by $\alpha$ and the self energies on the perturbed sites are equal to $\sigma(z)$, the eigenvalues $z$ of the effective Hamiltonian 
  are solutions of
  \begin{equation}
    z(z - \sigma(z)) = \alpha^2 \frac{B}{M}E^2,
    \label{autoco}
\end{equation}
where $E$ is an eigenvalue of the Hamiltonian with $\sqrt{\frac{B}{M}}=1$.
  After linearisation for low values of $\frac{\alpha E}{(1-\sigma_r)}$, we get

  \begin{equation}
      z \simeq \sqrt{\frac{B}{M}}\frac{\alpha  E}{\sqrt{1-\sigma_r}} + i \left(\frac{B}{M}\right)^{\frac{3}{2}}\frac{\sigma_i}{2}  \frac{(\alpha E)^3}{(1-\sigma_r)^{\frac{5}{2}}}.
  \end{equation}
   
  From this formula we can deduce the group velocity $v$ of a signal at low energy
   
  \begin{equation}
      v = \sqrt{\frac{B}{M}}\frac{\alpha}{\sqrt{1-\sigma_r}} av_0,
      \label{v_shiba}
  \end{equation}
  where $v_0$ is the velocity of the signal in the lattice with $a=1$ and $\sqrt{\frac{B}{M}}=1$.
  The scattering time $\tau$ is deduced from the imaginary part of $z$ by
   
  \begin{equation}
      \frac{1}{\tau} = -\left(\frac{B}{M}\right)^{\frac{3}{2}}\sigma_i \frac{(\alpha E)^3}{(1-\sigma_r)^{\frac{5}{2}}}.
      \label{tau_shiba}
  \end{equation}

The mean free path can be deduced from the velocity $v$ and the scattering time $\tau$ by $\ell = v\tau$.

\subsection{Analytic expressions of the self-energy at low frequency, for a disorder of type ($\mathcal{N}$)}
\label{appendix_low_freq_1}

For a disorder of type ($\mathcal{N}$), the value of $\sigma(z)$ can be computed,
 starting from the spectral properties of the laplacian matrix $\mathbb{L}_\mathrm{M}$.
 We suppose also that $\sigma(z)$ tends to zero when $z$ tends to zero.
This is a supposition that matches the physical reality, as the signals are less and less sensitive to disorder when the frequency decreases.
 Because of that, $Z =\frac{z(z - \sigma(z))}{\alpha^2}$ also tends to zero when $z$ tends to zero.
In this regime the real part of $g_\mathrm{lap}(Z)$ is known to behave like a logarithm, while the imaginary part tends to a constant \cite{kogan2020green}.
Therefore, we can write

\begin{equation}
    \Tilde{g}(z) \simeq \frac{z}{\alpha^2} (iC + D\ln(Z)),
\end{equation}
where $C:=-\eta \frac{M}{B}$ and $D$ is a real constant. In particular, $\Tilde{g}(z)$ tends to zero when $z$ tends to zero.
 Because $\sigma(z)$ also tends to zero, we conclude that $\Delta(z)$ tends to infinity when $z$ goes to zero and that for small values of $z$

\begin{equation}
    \frac{z}{\Delta(z)} \simeq -z^2(iC + D\ln(Z)).
    \label{Delta_approx}
\end{equation}

 Now, the value of $\Delta(z)$ is inserted in formula \eqref{sigma_S_AS}. It gives

\begin{align}
    \sigma(z) &\simeq -V^2 z \left(\frac{\alpha}{\alpha_1\alpha_2}\right)^2 \left[1 +  \left(\frac{\alpha}{\alpha_1 \alpha_2}\right)^2 \frac{z}{\Delta(z)}\right],\\
    & \simeq -V^2 z \left(\frac{\alpha}{\alpha_1\alpha_2}\right)^2\left[1 -  i\left(\frac{\alpha}{\alpha_1 \alpha_2}\right)^2 C z^2\right].
\end{align}

In this calculation, we have neglected the contribution related to $D\ln(Z)$ because $Dz^2\ln(Z)$ is negligible in front of $1$, for small values of $z$.
The values of $\sigma_r$ and $\sigma_i$ can be identified from this expression. It gives

\begin{equation}
    \sigma_r = -V^2 \left(\frac{\alpha}{\alpha_1\alpha_2}\right)^2 = -c_1 c_2\left( \frac{\alpha_1}{\alpha_2} - \frac{\alpha_2}{\alpha_1}\right)^2,
    \label{A_inertia_def}
\end{equation}
and 

\begin{equation}
    \sigma_i = -\sigma_r \left(\frac{\alpha}{\alpha_1 \alpha_2}\right)^2 C  =  c_1 c_2 \left( \frac{1}{(\alpha_2)^2} - \frac{1}{(\alpha_1)^2}\right)^2 \alpha^2 C.
    \label{B_inertia_def}
\end{equation}

These results hold for any parameters $\alpha_1 = \sqrt{\frac{M}{M_1}}$ and $\alpha_2 = \sqrt{\frac{M}{M_2}}$,
 even if in the next sections we have a specific focus on the case $\alpha_1 \gg 1$ and $\alpha_2 = 1$.


\subsection{Analytic expressions of the self-energy at low frequency, for a disorder of type ($\mathcal{L}$)}
\label{appendix_low_freq_2}
For a disorder of type ($\mathcal{L}$), the value of $\sigma(z)$ can be computed analogously as for the disorder of type ($\mathcal{N}$), by replacing $\mathbb{L}_\mathrm{M}$ by $\Tilde{\mathbb{L}}_\mathrm{M}$.
Because $\Tilde{\mathbb{L}}_\mathrm{M}$ has a kernel of dimension $N_\mathrm{lines} -N_\mathrm{nodes}-1$, at low frequency the real part of the Green function of $\Tilde{\mathbb{L}}_\mathrm{M}$ is approximately equal to
\begin{equation}
    \real(g_\mathrm{lap}(z)) = \frac{1- \gamma}{z}.
\end{equation}
Meanwhile, the imaginary part of $g_\mathrm{lap}(z)$ tends to $\Tilde{C} := -\gamma \eta \frac{M}{B}$. 
Keeping the main contributions of the real and imaginary part, an approximation of $\Tilde{g}(z)$ at low frequency can be deduced from equation \eqref{g_lap} 

\begin{equation}
    \Tilde{g}(z) = i\Tilde{C}\frac{z}{\alpha^2} + \frac{1-\gamma}{z - \sigma(z)}.
    \label{g_rupture}
\end{equation}

We suppose now, for simplifying the calculations, that $\alpha_1=0$, which corresponds to the removal of lines of the grid.
Using equation \eqref{g_eff}, equation \eqref{sigma_S_AS} can be reformulated as

\begin{equation}
    \sigma(z) = c(z - \frac{1}{\Tilde{g}(z)}),
    \label{shiba_lines_moyennee}
\end{equation}

which gives

\begin{equation}
    \sigma_r = \frac{-c \gamma}{1- \gamma-c},
\end{equation}

and
\begin{equation}
    \sigma_i =  \frac{c(1-c)}{(1-\gamma -c)^2} \frac{1}{(\alpha_2)^2}\Tilde{C}.
    \label{sigma_i_outage}
\end{equation}

All these calculations hold for $\alpha_1=0$ and $\alpha_2 = 1$. For a more general line disorder,
 it is still possible to give approximate analytic expressions for $\sigma_r$ and $\sigma_i$.
 Replacing the value of $\Delta(z)$ from equation \eqref{g_rupture} in equation \eqref{sigma_S_AS}, and identifying the real part gives $\sigma_r$ as the negative solution of the second order equation

\begin{equation}
     \left(\frac{\alpha_1 \alpha_2}{\alpha}\right)^2 \sigma_r^2 + \left[ \frac{\alpha^2(\gamma-1)}{\gamma} -  \left(\frac{\alpha_1 \alpha_2}{\alpha}\right)^2 + V^2\right]\sigma_r - V^2 = 0,
\end{equation}

while the imaginary part yields the following approximation of $\sigma_i$

\begin{equation}
    \sigma_i = \frac{\Tilde{C} V^2}{\gamma^2 \left[ \frac{ \alpha^2(\gamma-1)}{\gamma(1-\sigma_r) } -  \left( \frac{\alpha_1 \alpha_2}{\alpha}\right)^2\right]^2}.
    \label{sigma_i_gal}
\end{equation}

\section{Theoretical results for the toy models}

The formulas derived with the mean field are now specifically applied to our toy models.
We analyse in particular the numerical approximation of the density of states given by equation \eqref{g_shiba_physical},
 as well as the analytic expressions of the velocity, the mean free path and the length of localization induced by the disorder.

\subsection{The spectral density of states}

At low energy, the densities of states shown in the panels of figure \ref{fig:dos_cdt_hk} are linear, even in configurations including disorder.
 Their slopes are inversely proportional to $v^2$, where $v$ is the group velocity, and are decreased or increased, depending if the disorder is of type ($\mathcal{N}$) or ($\mathcal{L}$). 

From the mean field of Shiba's theory and equation  \ref{v_shiba} one calculates exactly the velocity at zero energy. For a disorder of type ($\mathcal{N}$) and ($\mathcal{L}$), the velocity is respectively

\begin{equation}
    v = \begin{cases}
       &\sqrt{\frac{B}{<M>}} a v_0 \quad (\mathcal{N}),\\
      &\sqrt{\frac{<B>}{M}} a v_0 \quad (\mathcal{L}),
    \end{cases}
\end{equation}

 where $<M> :=M(1-c)$ is the average mass and $<B>:=B\left(1-\frac{c}{1-\gamma}\right)$ is the average susceptance. Here $a$ is the distance between nearest neighbor nodes,  $v_0$ is the velocity of the reference periodic lattice with $B=M=a=1$, and $\gamma = \frac{N_\mathrm{nodes}}{N_\mathrm{lines}}$. For the Lieb lattice, $\gamma = \frac{1}{2}$ and $v_0 = 1$, while $\gamma = \frac{2}{3}$ and $v_0 = \frac{\sqrt{3}}{2}$ for the Honeycomb-Kagome lattice. The velocity is increased when the global inertia of the grid is reduced and decreased when lines are cut in the grid. Note that $<B>$ cancels when the number of line states equals the number of node states. This is the minimum  concentration of line states that allows to have a percolating circuit. 

We note that the total number of states in the energy range of figure \ref{fig:dos_cdt_hk} decreases for the disorder of type ($\mathcal{N}$) when the concentration $c$ of small mass increases. This is because a node state with a small mass couples strongly to  a combination of line states. This gives rise to two eigenstates with energies of high  modulus,  which are beyond the energy range shown here.

\subsection{The mean free path}

In physics, when a wave propagates in a medium, its direction may be impacted by the presence of obstacles or heterogeneities.
The mean free path is the average distance between two consecutive shocks,
 in other words it is the distance the wave can travel ballistically  before interacting with the disorder, which is called scattering.
 Analogously, for power systems, the mean free path is the distance reached by a signal before being diffused by the heterogeneities of the grid. 
For taking into account the effect of scattering in our model, the third-order approximation $\sigma(z) = \sigma_r z + i\sigma_i z^3$, valid at low frequency, is used.
According to equation \eqref{tau_shiba}, the scattering time $\tau$ at low frequency is given by $\frac{1}{\tau} = -\left(\frac{B}{M}\right)^{\frac{3}{2}}\sigma_i \frac{(\alpha E)^3}{(1-\sigma_r)^{\frac{5}{2}}}$.
 From $v$ and $\tau$ can eventually be computed the mean free path $\ell = v\tau$, for the two types of disorder




\begin{equation}
     \frac{\ell}{a}= \begin{cases}
        &\frac{(1-c)v_0}{c \eta} E^{-3}\quad (\mathcal{N}), \\ & \frac{\left[ 1-\gamma\right]^2 v_0}
    {c\gamma \eta} 
    E^{-3}\quad (\mathcal{L}).
     \end{cases} 
    \label{mfp_2}
\end{equation}


Formula \eqref{mfp_2}, which holds at low frequency, is consistent with the intuition given by the Courant-Fisher-Weyl theorem. Indeed the mean free path is proportional to the inverse of the cube of $E$ and the width due to disorder, which is also is proportional to the inverse of the cube of $E$, becomes negligible compared to the energy. In the Shiba mean field theory this dependence as the inverse cube of $E$, in the low energy limit is valid for any disorder either on node states, or on line states or both. Perturbation theory, presented in appendix \ref{appendix_fermi}, confirms this result. In addition it indicates that when the Laplacian structure is lost (that is when the Hamiltonian is not of the separable form presented in equation \eqref{H_alpha}), the mean free-path is much smaller and is inversely proportional to $E$.

\subsection{The localization length}

For the local uncorrelated disorder considered here, and in the presence of time reversal symmetry, we expect that the scaling theory of Anderson's localization \cite{anderson1958absence} applies. In a two-dimensional system all states are localized on a characteristic length $\xi$. The ratio of $\xi$ to the mean free path $\ell$ at the energy $E$ is well given by the approximate formula

\begin{equation}
    \frac{\xi}{\ell} \simeq \exp( \frac{\pi^2 n(E)\ell^2}{2\tau}).
    \label{localisation}
\end{equation}
where $n(E)$ is the density of states per unit area and $\tau$ the elastic scattering time \cite{lee1985disordered}. Because $\frac{\ell^2}{\tau}\propto E^{-3}$ and $n(E)\propto E$, a divergence of $\frac{\xi}{\ell}$ is expected in the limit of low frequency.


\section{Numerical results of the regimes of propagation}

We consider now the propagation of a power oscillation represented by the wave-function  $\psi(t)$, and generated by a power source $\mathbb{P}_n(t) = \cos(Et)$ located at a node $n$. The average distance $R(t)$ travelled by the power oscillation is 
\begin{equation}
R(t)= \left[\sum_{k} \left(\frac{\vert \psi_k\vert}{\vert\vert \psi \vert\vert}\right)^2 d^2(n, k) \right]^{\frac{1}{2}},
\end{equation}
where $d(n, k)$ is the geographical distance between the node $n$ and the node or line $k$. At sufficiently large time $Et\gg 1$, the propagation reflects the diffusion properties of states of energy $E$.



\subsection{Results for the toy model}




\begin{figure*}[ht!]
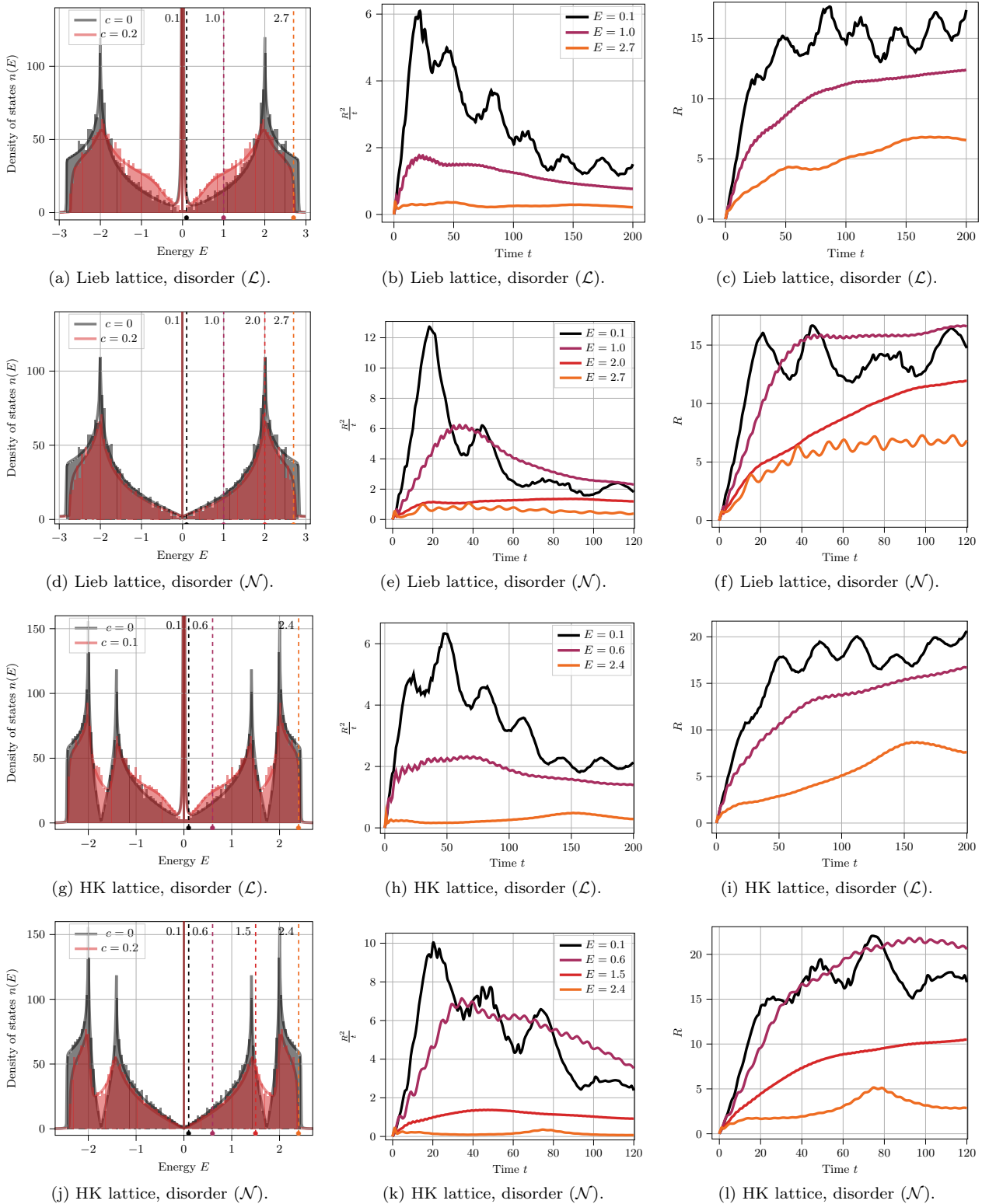

    \centering
     \subfloat[Lieb lattice, disorder ($\mathcal{L}$).]{\begin{adjustbox}{clip,trim=0cm 0.cm 0.cm 0.cm,max width=0.3\linewidth}
\input{finite_dos}
\end{adjustbox}}\quad
    \subfloat[Lieb lattice, disorder ($\mathcal{L}$).]{\resizebox{0.3\linewidth}{!}{\input{multiple_w_lieb_40_localisation_0.2}}}\quad
    \subfloat[Lieb lattice, disorder ($\mathcal{L}$).]{\resizebox{0.3\linewidth}{!}{\input{multiple_w_lieb_40_localisation_0.2_R}}}\\
    \subfloat[Lieb lattice, disorder ($\mathcal{N}$).]{\begin{adjustbox}{clip,trim=0cm 0.cm 0.cm 0.cm,max width=0.3\linewidth}
\input{lieb_dos_mass}
\end{adjustbox}}\quad
    \subfloat[Lieb lattice, disorder ($\mathcal{N}$).]{\resizebox{0.3\linewidth}{!}{\input{multiple_w_lieb_40_mass_noise_2}}}\quad
        \subfloat[Lieb lattice, disorder ($\mathcal{N}$).]{\resizebox{0.3\linewidth}{!}{\input{multiple_w_lieb_40_mass_noise_2_R}}}\\
        \subfloat[HK lattice, disorder ($\mathcal{L}$).]{\begin{adjustbox}{clip,trim=0cm 0.cm 0.cm 0.cm,max width=0.3\linewidth}
            \input{hk_finite_dos_2}
            \end{adjustbox}}\quad
                \subfloat[HK lattice, disorder ($\mathcal{L}$).]{\resizebox{0.3\linewidth}{!}{\input{multiple_w_hk_30_localisation}}}\quad
                \subfloat[HK lattice, disorder ($\mathcal{L}$).]{\resizebox{0.3\linewidth}{!}{\input{multiple_w_hk_30_localisation_R}}}
                 \\
                 \subfloat[HK lattice, disorder ($\mathcal{N}$).]{\begin{adjustbox}{clip,trim=0cm 0.cm 0.cm 0.cm,max width=0.3\linewidth}
            \input{hk_dos_mass}
            \end{adjustbox}}
            \quad
                \subfloat[HK lattice, disorder ($\mathcal{N}$).]{\resizebox{0.3\linewidth}{!}{\input{multiple_w_hk_40_mass_noise}}}\quad
                 \subfloat[HK lattice, disorder ($\mathcal{N}$).]{\resizebox{0.3\linewidth}{!}{\input{multiple_w_hk_40_mass_noise_R}}}
    \caption{\textbf{Propagation of signals in the toy models.} 
    The panels of the two upper layers correspond to the Lieb lattice of size $40\times 40$,
     while the panels of the third and fourth layers correspond to the Honeycomb-Kagome (HK) lattice of size $30\times 30$. 
    The left panels give the density of states of the perfect system (gray histogram) and of the disordered system (red histogram).
    On the left panels, the continuous lines are the densities of states given by Shiba's theory, which match accurately the histogram of the exact diagonalisation.
     The central (respectively, right) panel represents $\frac{R^2}{t}$ (respectively, $R$) in function of $t$ for various angular frequencies $E$.}
    \label{fig:lieb_lattice_40_0.2_cdt}
 \end{figure*}

 For a finite lattice (with a source which is not close to the boundaries), three regimes are expected, depending on its size $L$, on the mean free path $\ell(E)$ and on the localisation length $\xi(E)$. For $\xi > \ell >L$ the propagation of the signal is ballistic until it reaches the boundaries. For $\xi > L> \ell $ the signal has time to become diffusive before it reaches the boundaries. For $ L> \xi >\ell $ the signal localizes before reaching the limit of the grid. 
 
 Figure \ref{fig:lieb_lattice_40_0.2_cdt} gives results for a Lieb lattice of size $40\times 40$ and for the Honeycomb-Kagome lattice of size $30\times 30$, with disorders of type ($\mathcal{L}$) and ($\mathcal{N}$). For both disorders, the three regimes are observed. 
  For the Lieb lattice, and for the disorder of type ($\mathcal{L}$) with $E=0.1$, $R(t)$ is ballistic at short time until it reaches the boundaries. Beyond this time $R(t)$ tends to saturate with oscillations which reflect the oscillations of the source. For $E=1.0$ the propagation becomes diffusive before reaching the boundaries. Then at larger times $R(t)$ saturates also. Finally for $E=2.7$ the wave is localized inside the grid. For the disorder of type ($\mathcal{N}$), $E=0.1$ and $E=1.0$ are in the  ballistic regime, $E=2.0$ is diffusive and $E=2.7$ is localized. 
 For the Honeycome-Kagome lattice, the same regimes of propagation are also observed, at the frequencies $E=0.1$, $E=0.6$, $E=2.4$ for the disorder of type ($\mathcal{L}$) and $E=0.1$, $E=1.5$, $E=2.4$ for the disorder of type ($\mathcal{N}$).

The finite size implies not only that signals are reflected by the boundaries of the network but also that the spectrum of the Hamiltonian is discrete, leading to the estimation $\frac{\xi}{\ell} \simeq \exp( \frac{\pi^2 \ell^2}{2\Delta E S\tau})$, where $\Delta E$ is the average distance between two consecutive eigenvalues near $E$, and $S$ is the surface of the grid.  For the Lieb lattice at $E=2.7$ and the Honeycomb-Kagome lattice at $E=2.4$, we find $\frac{\xi}{\ell} \simeq 2.5$, which matches the numerical results for localized signals.

\subsection{Results for the Pantagruel model of the European grid}

 \begin{figure*}[ht!]
    \centering
\subfloat{\begin{adjustbox}{clip,trim=6cm 17.3cm 5.7cm 4.2cm,max width=0.3\linewidth}
         \includegraphics[width=1.5\linewidth]{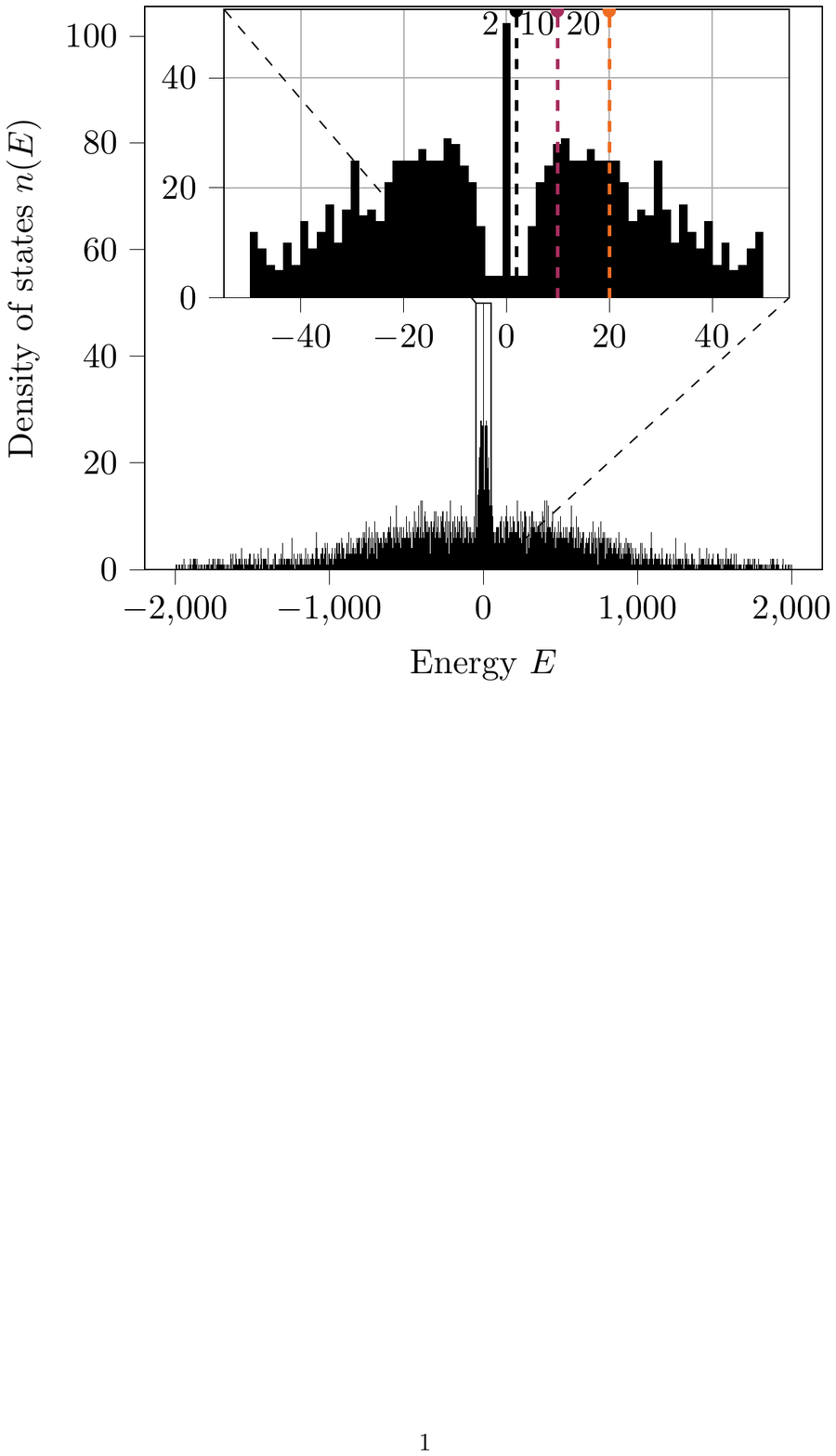}
     \end{adjustbox}}\quad
    \subfloat{\resizebox{0.3\linewidth}{!}{\input{w_2_10_20_fessenheim}}}\quad
    \subfloat{\resizebox{0.3\linewidth}{!}{\input{w_2_10_20_fessenheim_R}}}\\
     \subfloat[$E=2$, $t=1~\text{s}$.]{\begin{adjustbox}{clip,trim=1.4cm 3.8cm 1.2cm 1.3cm,max width=0.3\linewidth}
        \includegraphics[width=0.7\linewidth]{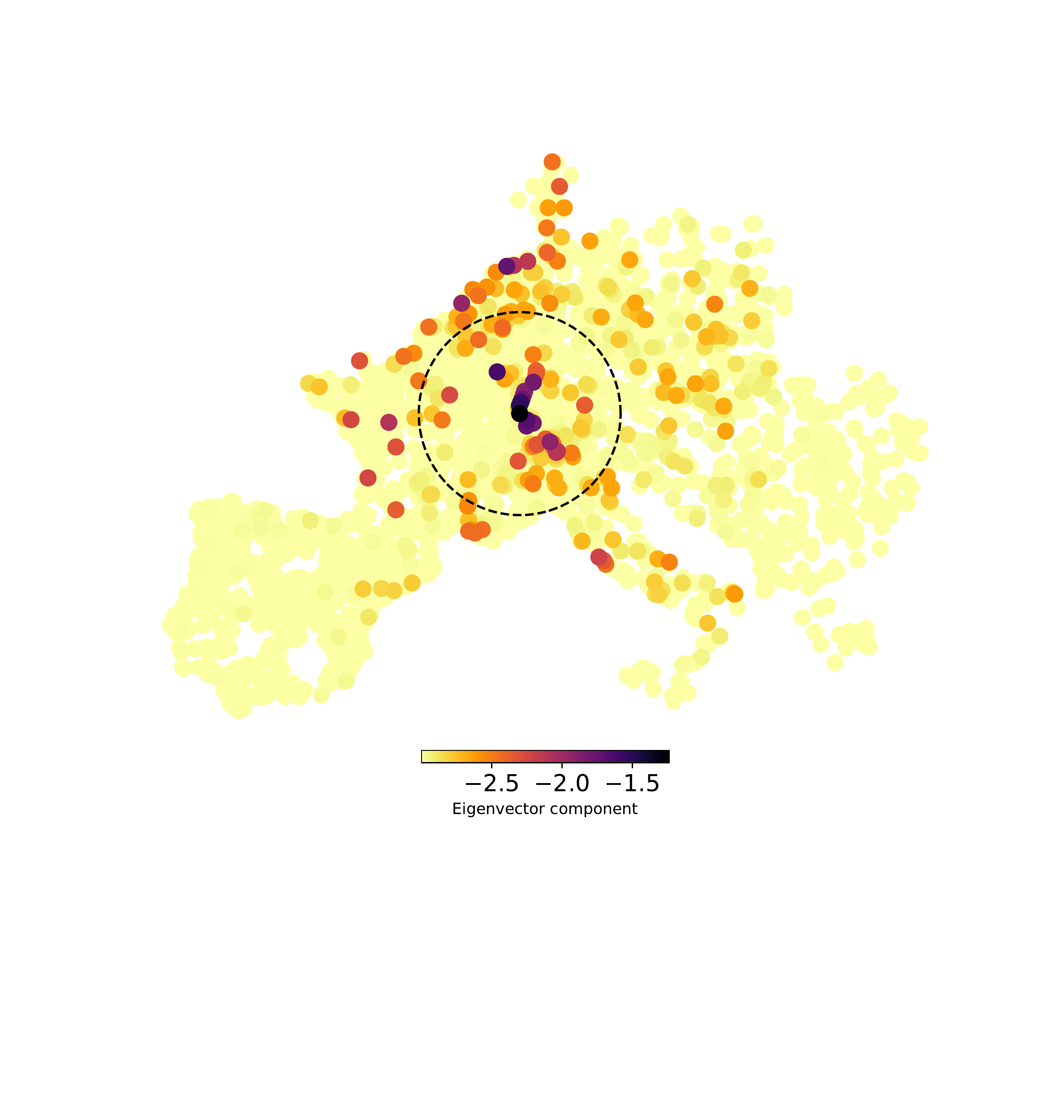}
    \end{adjustbox}}
    \subfloat[$E=10$, $t=1~\text{s}$.]{\begin{adjustbox}{clip,trim=1.4cm 3.8cm 1.2cm 1.3cm,max width=0.3\linewidth}
        \includegraphics[width=0.7\linewidth]{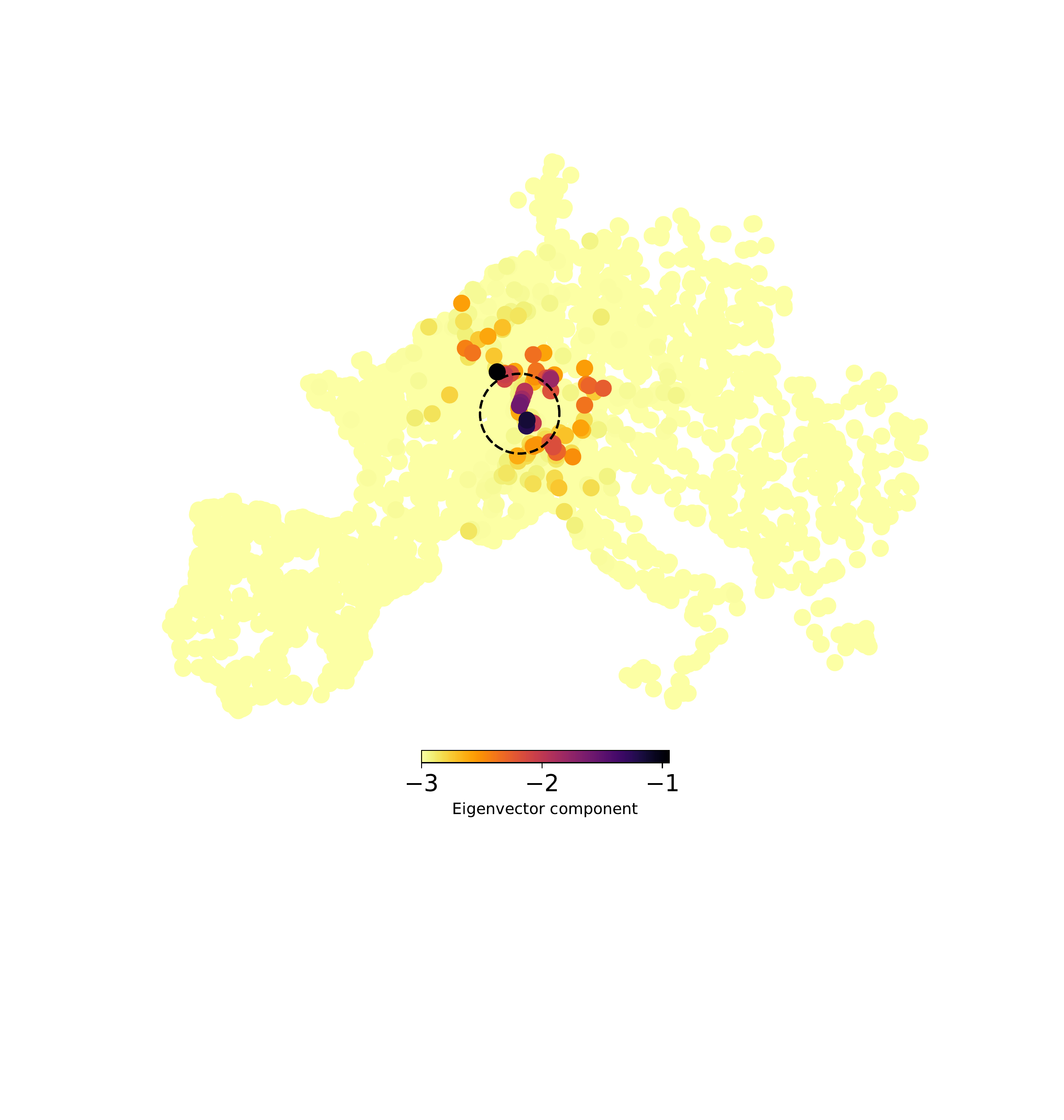}
    \end{adjustbox}}
    \subfloat[$E=20$, $t=1~\text{s}$.]{\begin{adjustbox}{clip,trim=1.4cm 3.8cm 1.2cm 1.3cm,max width=0.3\linewidth}
        \includegraphics[width=0.7\linewidth]{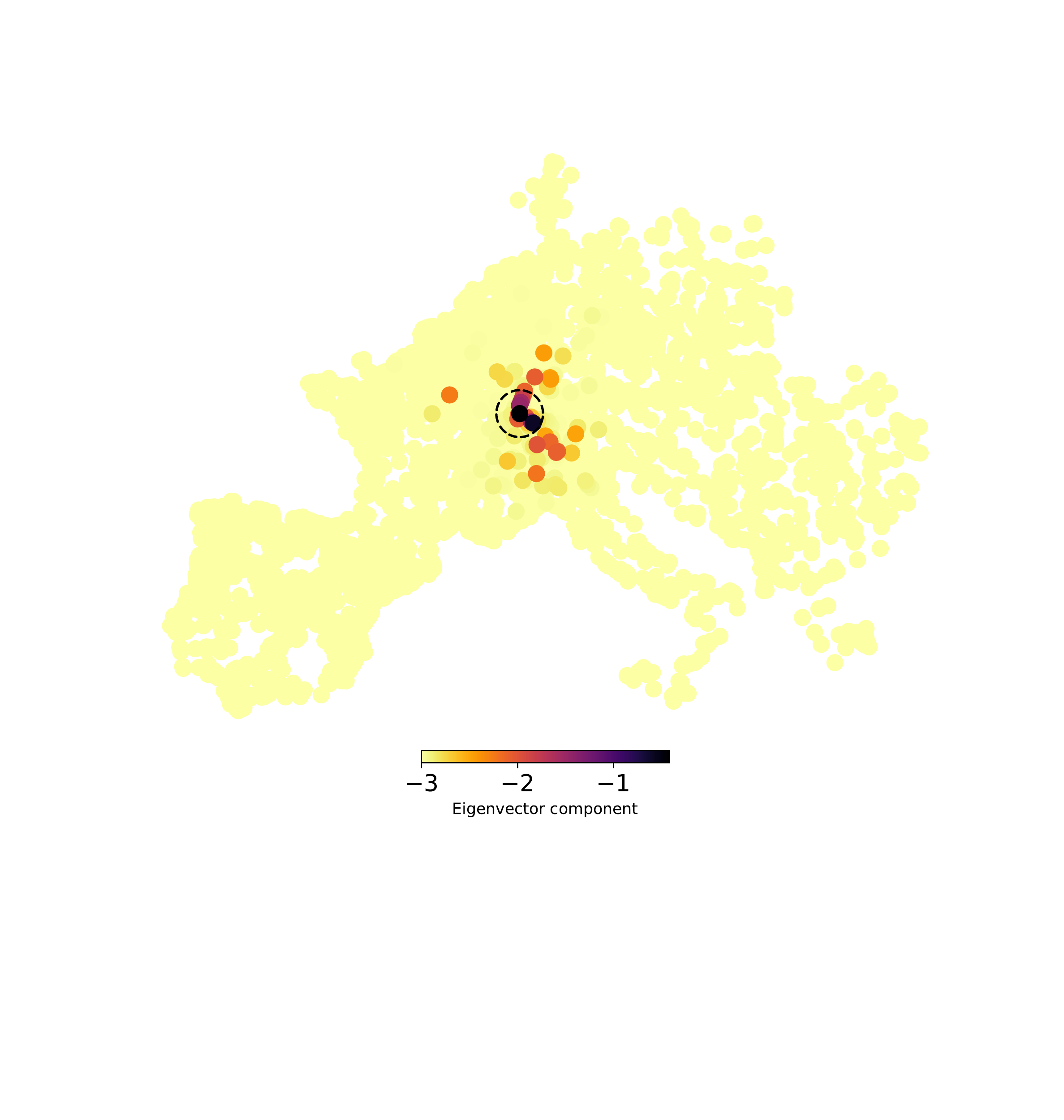}
    \end{adjustbox}}
    \\
    \subfloat[$E=2$, $t=5~\text{s}$.]{\begin{adjustbox}{clip,trim=1.4cm 3.8cm 1.2cm 1.3cm,max width=0.3\linewidth}
        \includegraphics[width=0.7\linewidth]{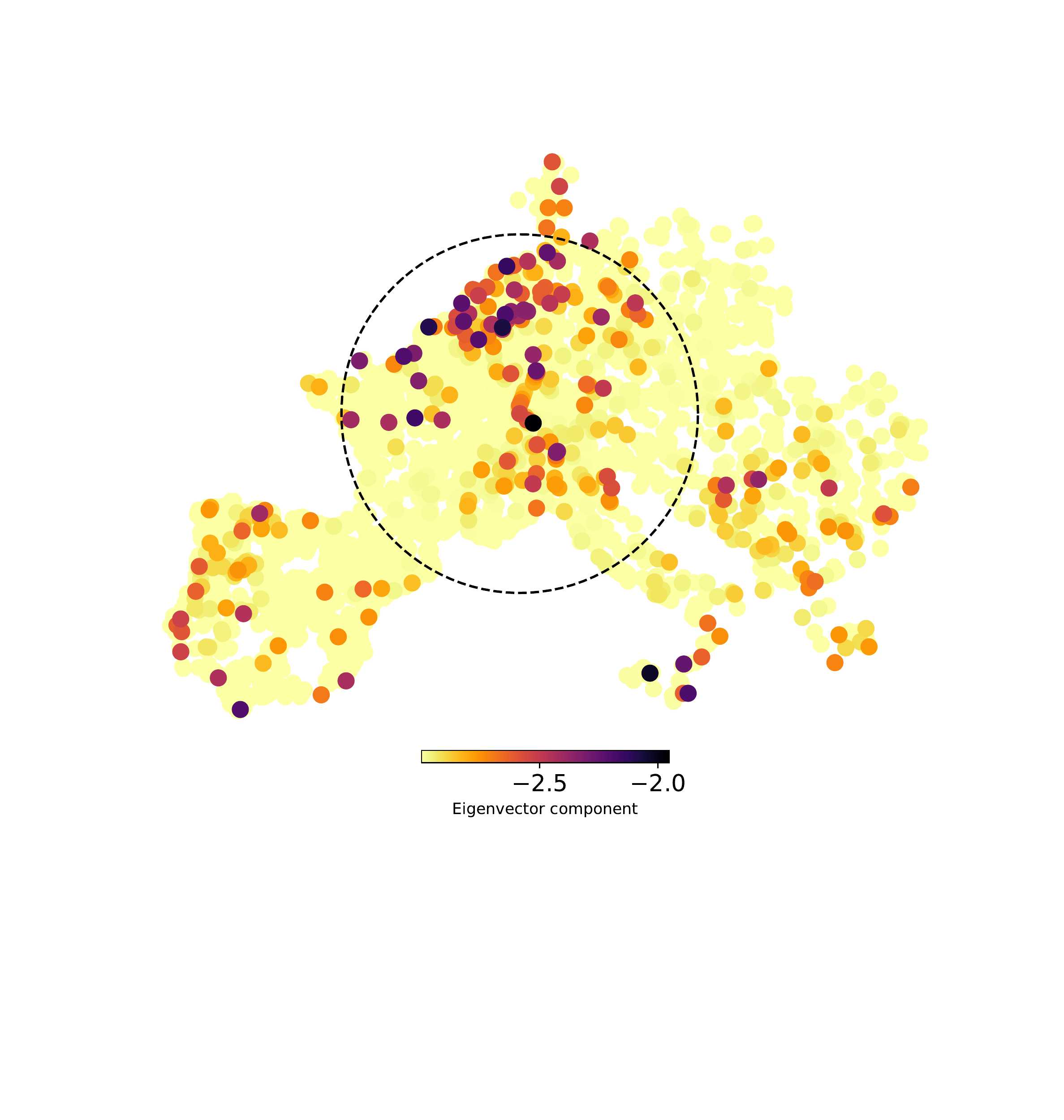}
    \end{adjustbox}}
    \subfloat[$E=10$, $t=5~\text{s}$.]{\begin{adjustbox}{clip,trim=1.4cm 3.8cm 1.2cm 1.3cm,max width=0.3\linewidth}
        \includegraphics[width=0.7\linewidth]{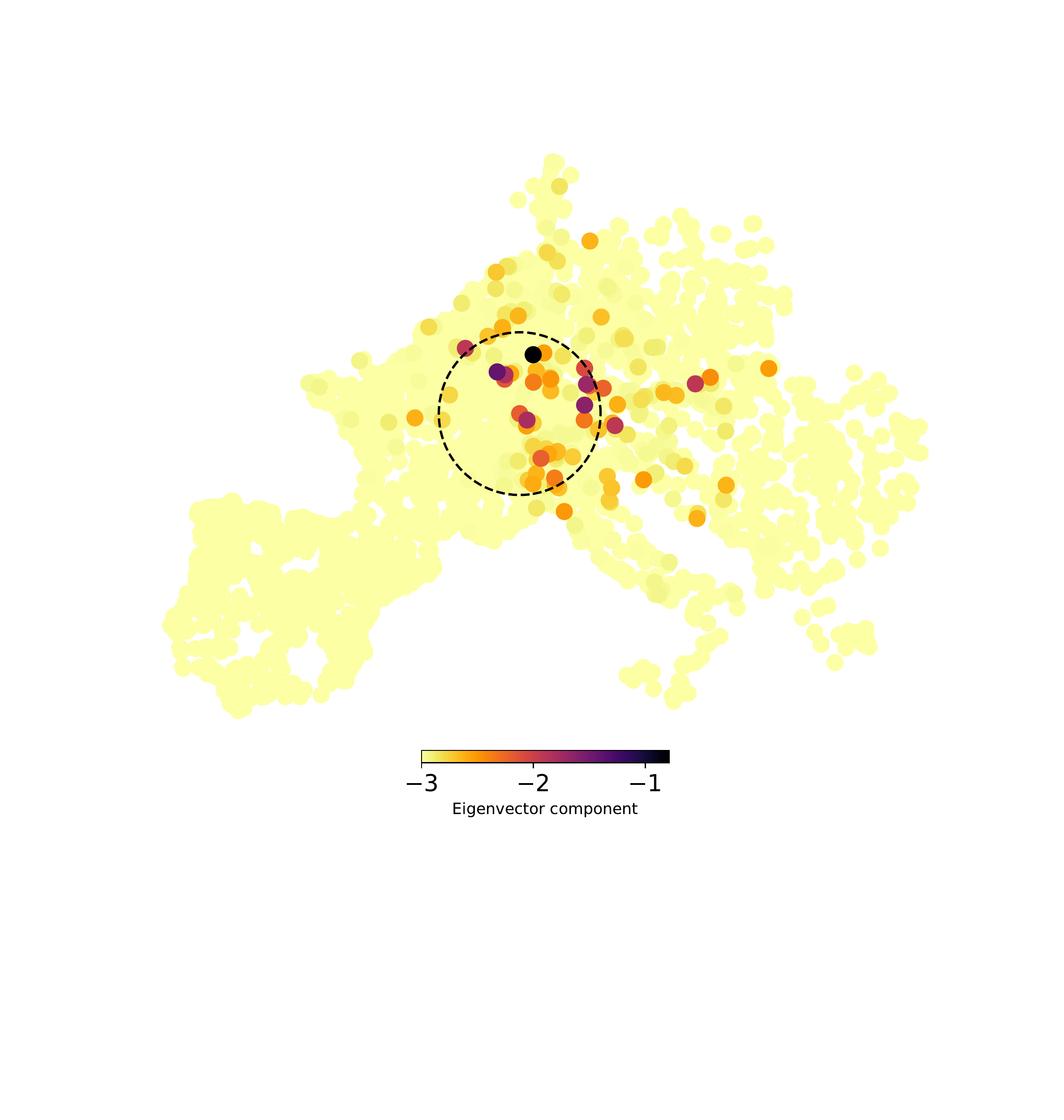}
    \end{adjustbox}}
    \subfloat[$E=20$, $t=5~\text{s}$.]{\begin{adjustbox}{clip,trim=1.4cm 3.8cm 1.2cm 1.3cm,max width=0.3\linewidth}
        \includegraphics[width=0.7\linewidth]{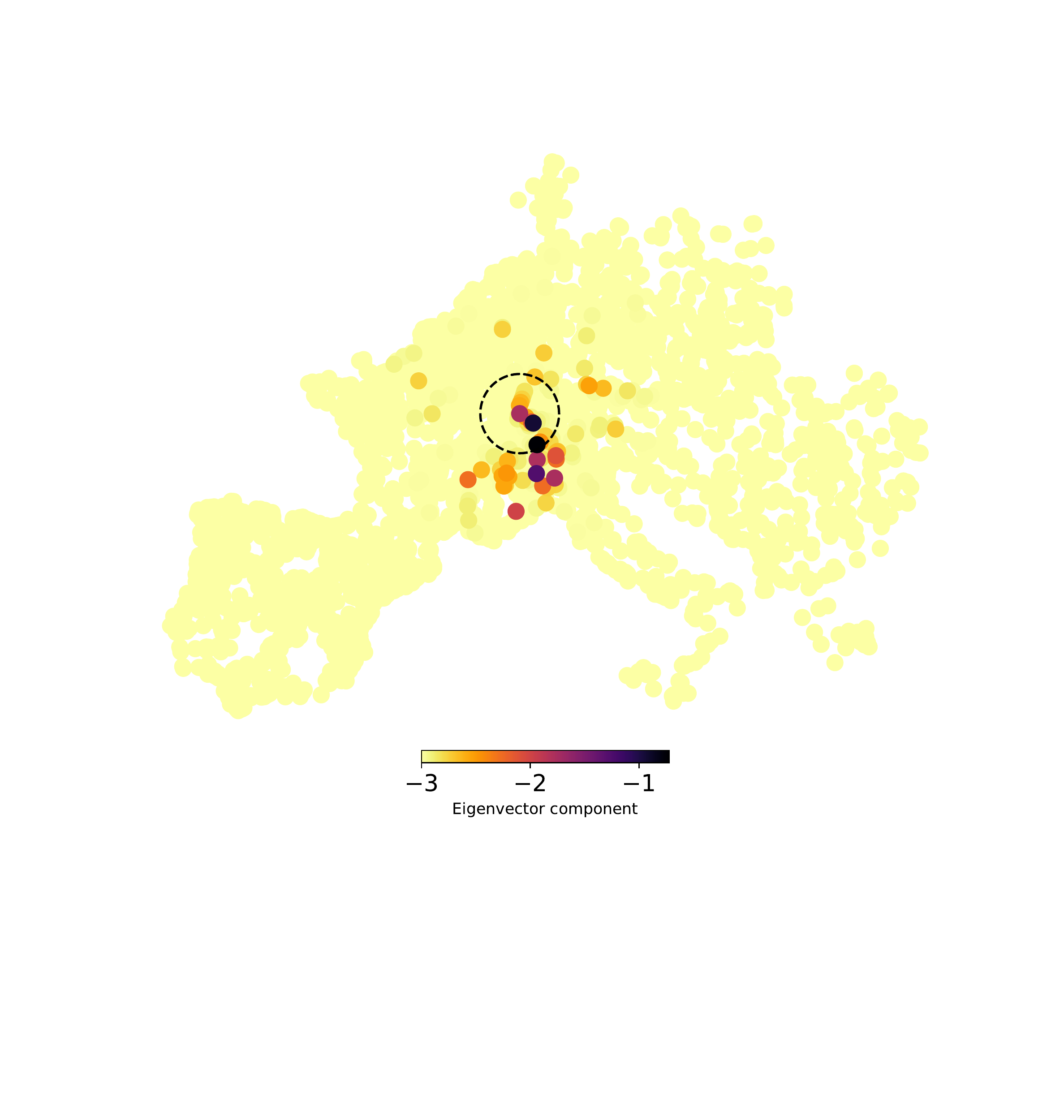}
    \end{adjustbox}}
        \caption{\textbf{Propagation of signals in the Pantagruel model of the European transmission grid, from the Fessenheim node.} The upper left panel gives the density of states of the model. The upper central (respectively, right) panel represents $\frac{R^2}{t}$ (respectively, $R$) in function of $t$ for various angular frequencies $E$.
    The central and bottom panels show as a colormap the value of $\log(\frac{\vert \psi_n\vert^2}{\vert\vert \psi\vert \vert^2})$ for every node $n$ of the network. The central panels represent the signals at $t=1~\text{s}$ for $E=2$, $E=10$ and $E=20$ (from left to right). The bottom panels represent the same quantities at $t=5~\text{s}$.}
    \label{fig:panta_fessenheim}
  \end{figure*}
  
 Figure \ref{fig:panta_fessenheim} extends the results of the previous sections to the Pantagruel model \cite{jaquod} \cite{jaquod2}, a dynamic description of the synchronous grid of continental Europe.
 It includes only the high voltage lines of the transmission grid, the distribution grid being aggregated to the parameters of the model.
 The network has $N_\mathrm{nodes} = 3809$ nodes (of which $20\%$ are generators), and $N_\mathrm{lines} = 7343$ lines, giving a mesh density coefficient $\gamma =\frac{N_\mathrm{nodes}}{ N_\mathrm{lines}} \simeq 0.52$, closer to the Lieb ($\gamma = \frac{1}{2}$) than to the Honeycomb-Kagome lattice ($\gamma = \frac{2}{3}$) . The geographical coordinates of the nodes are obtained via a stereographic projection onto the plane tangent to the earth, at a reference point which is close to the center of the grid. We have checked that the results presented here are insensitive to the precise position of the reference point.
This network is contained in a bounding box with a length of $3000$ km and a width of $2000$ km, while the average length of a line is around $30$ km.
The power grid is subject simultaneously to a disorder induced by its irregular topology, as well as a disorder induced by the mass and susceptance parameters. In particular,  the inertias of the generators are around $4.10^{-1}~\text{MWs}^2$. The inertias of the loads, that is the sites corresponding to consumers, are much smaller and are set to $10^{-3}~\text{MWs}^2$. Yet the results discussed here are essentially independent of the precise  value of the loads inertia provided that it is sufficiently small compared to the inertia of the generators. Finally the damping is neglected which is pertinent for the effects lasting around $5-10~\text{s}$ which are discussed here.


  The upper left panel of figure \ref{fig:panta_fessenheim} shows the spectral density of the Pantagruel grid and a zoom of the central part of it in the inset. The spectral density of the central part is qualitatively close to that of the Lieb lattice (see above), with a linear part at low frequencies due  to the Goldstone modes. The spectral density decreases beyond about $E=10$, which corresponds to a frequency $\frac{E}{2\pi} \simeq 1.6 ~\text{Hz}$. At much higher energies there is also a part of the spectrum  which corresponds to the effect of nodes with small inertia as discussed for the Lieb lattice case.
 
 The upper central (respectively, right) panel of figure \ref{fig:panta_fessenheim} represents the results of the propagation of a sinusoidal signal from a specific node of the grid. We choose the node of the grid associated to the french Fessenheim nuclear plant which is far from boundaries of the European grid. Similar results for other nodes (the Isar and Creys-Malville nodes) can be found in figures \ref{fig:panta_isar} and \ref{fig:panta_crey}, respectively. Ballistic, diffusive and localised regimes are observed at the frequencies $0.32~ \text{Hz}$ ($E=2$), $1.6~ \text{Hz}$ ($E=10$) and $3.2~\text{Hz}$ ($E=20$), which are located in the increasing linear part, maximum part, or decreasing part of the spectral density respectively. This is remarkably similar to the propagation regimes identified for the stochastic models on the Lieb and Honeycomb-Kagome lattices. The other panels indicate a precise repartition of the intensity of the wave-function $\psi$ that describes the power oscillation. The values are scattered in a way which is also reminiscent of the results for the purely stochastic models of Lieb and Honeycomb-Kagome lattices described in figure $\ref{fig:lieb_lattice_40_0.2}$.
 
 The analogy with a stochastic quantum model gives also an insight about the localization of modes. For example considering the energy $E=20$, we calculate the ratio between the localization length $\xi$ and the mean free path $\ell$ from the scaling theory of localization. From equation \eqref{localisation} and from the spectral density and the diffusivity $\frac{R^2}{t}$ we get $\frac{\xi}{\ell} \simeq 1.5$. As shown in figure \ref{fig:panta_fessenheim}, the mean free path, which is reached after a short ballistic period, is of the order of $100$ km, (i.e. a few inter-node distances) and therefore one expects a localization length of the order of $150$ km. The saturation of $R(t)\simeq 200$ km at large time is consistent with this estimate of the localization length.

\section{Conclusion}

To conclude this work gives a new insight in the dynamics of
power oscillations, which are key to the stability of electric
grids. These power oscillations are traditionally studied
numerically with the central concept of inter-area or intra-area modes. Here we propose a new theoretical approach.
We show first that the classical swing equations, which
describe power oscillations, can be transformed exactly into
a Schrödinger-like equation. In a second step we treat the
complex distribution of susceptances and mass-inertia of a
grid as a stochastic disorder. The resulting stochastic
quantum models are then treated by mean field theories or
through the scaling theory of Anderson localization. This
gives a deep insight in the properties of power grids. As a
result three propagation regimes for power oscillations,
depending on the frequency, are expected for large grids.
They are identified for the first time on a model of the
European transmission grid. The present theoretical
approach can be developed further to investigate the effects of the massive integration of renewable energy source on power oscillations. To control their negative impacts will be indeed essential to achieve a secure "second electrical revolution".

\begin{acknowledgments}
We are grateful to Didier Georges and Vincent Rossetto for their pertinent feedback. We also thank the CNRS for funding the PhD thesis of P. Guichard through the 80 PRIME program.
\end{acknowledgments}

{\appendices

\section{The Hamiltonian formalism}
\label{appendix_hamiltonian}

The swing equation \eqref{LF_dyn} is a system of second order differential equations. It is transformed into the Schrödinger equation \eqref{schrodinger} of first order by extending the state-space. In a traditional Laplacian description of the grid, the unknowns are the angular phases $\theta$ at the nodes of the grid. In the Hamiltonian description introduced here the unknowns are stored in a state $\psi:=\left[\begin{array}{c}
         \psi^\mathrm{N} \\
         \hline
          \psi^\mathrm{L}\end{array}\right]$. The unknowns defined on the nodes (respectively, lines) and stored in $\psi^\mathrm{N}$ (respectively, $\psi^\mathrm{L}$) are proportional to the angular velocities (respectively, to the electrical currents). 
The state $\psi$ evolves according to the Schrödinger equation

\begin{equation}
\begin{split}
    i\frac{\partial}{\partial t}  \underbrace{\left[\begin{array}{c}
         i\mathbb{M}^{\frac{1}{2}}\dot{\theta} \\
         \hline
          \mathbb{H}_\mathrm{LN}\theta
    \end{array}\right]}_{\psi} = \underbrace{\left[\begin{array}{c|c}
         -i\mathbb{M}^{-1}\mathbb{D} &  \mathbb{M}^{-\frac{1}{2}}\mathbb{H}_\mathrm{LN}^\dag \\
         \hline
        \mathbb{H}_\mathrm{LN}\mathbb{M}^{-\frac{1}{2}} & 0
    \end{array}\right]}_{\mathbb{H}}& \underbrace{\left[\begin{array}{c}
         i\mathbb{M}^{\frac{1}{2}}\dot{\theta} \\
         \hline
          \mathbb{H}_\mathrm{LN}\theta
    \end{array}\right]}_{\psi} \\&- \underbrace{\left[\begin{array}{c}\mathbb{M}^{-\frac{1}{2}}P\\
         \hline 0
    \end{array}\right]}_{\mathbb{P}},
    \end{split}
    \label{h_def}
\end{equation}

where $\mathbb{M}^{\frac{1}{2}} := \text{Diag}(\sqrt{M_1}, \dots, \sqrt{M_{N_\mathrm{nodes}}})$ and $\mathbb{H}_\mathrm{LN}$ is a matrix of dimension $N_\mathrm{lines}\times N_\mathrm{nodes}$ that has only null coefficients except for

\begin{equation}
    \forall l=1\dots N_\mathrm{lines},\;
    \begin{cases}
       &\left[\mathbb{H}_\mathrm{LN}\right]_{l, i_l} = \sqrt{B_l},\\
        &\left[\mathbb{H}_\mathrm{LN}\right]_{l, j_l} = -\sqrt{B_l}.
    \end{cases}
\end{equation}

The eigenvalues and eigenvectors of $\mathbb{H}$ can directly be deduced from the matrices $\mathbb{L}_\mathrm{M}$ and $\Tilde{\mathbb{L}}_\mathrm{M}$, related to the square of $\mathbb{H}$ by
\begin{align}
    \mathbb{L}_\mathrm{M} &:= \mathcal{P}_\mathrm{N} \mathbb{H}^2 \mathcal{P}_\mathrm{N}, \label{laplacian_mass} \\
    \Tilde{\mathbb{L}}_\mathrm{M} &:=
    \mathcal{P}_\mathrm{L} \mathbb{H}^2 \mathcal{P}_\mathrm{L}, \label{laplacian_mass_2} 
\end{align}
where $\mathcal{P}_\mathrm{N}$ (respectively, $\mathcal{P}_\mathrm{L}$) is the projector on all the node (respectively, line) states.
Note that $\mathbb{L}_\mathrm{M}$ is related to the laplacian matrix $\mathbb{L}$ from equation \eqref{LF_dyn} by $\mathbb{L}_\mathrm{M} = \mathbb{M}^{-\frac{1}{2}} \mathbb{L}\mathbb{M}^{-\frac{1}{2}}$.
Now, let $\left[\begin{array}{c}
        X^\mathrm{N} \\
            \hline
            X^\mathrm{L}
    \end{array}\right]$ be an eigenvector of $\mathbb{H}$ related to the eigenvalue $E$, decomposed on the node and line states. Then,

\begin{equation}
   \left[\begin{array}{c|c}
            -i\mathbb{M}^{-1}\mathbb{D} & \mathbb{M}^{-\frac{1}{2}}\mathbb{H}_\mathrm{LN}^\dagger \\
            \hline
        \mathbb{H}_\mathrm{LN}\mathbb{M}^{-\frac{1}{2}}  & (0)
    \end{array}\right] \left[\begin{array}{c}
        X^\mathrm{N} \\
            \hline
            X^\mathrm{L}
    \end{array}\right] = E \left[\begin{array}{c}
        X^\mathrm{N}\\
            \hline
            X^\mathrm{L}
    \end{array}\right],
\end{equation}
which gives

\begin{align}
& \mathbb{L}_\mathrm{M}X^\mathrm{N}= E(E+i\mathbb{M}^{-1}\mathbb{D})X^\mathrm{N},\\
     & \Tilde{\mathbb{L}}_\mathrm{M} X^\mathrm{L} = E(E+i\mathbb{M}^{-1}\mathbb{D})X^\mathrm{L}.
\end{align}

This means that $X^\mathrm{N}$ and $X^\mathrm{L}$ are respectively eigenvectors of the matrices $\mathbb{L}_\mathrm{M}$ and $\Tilde{\mathbb{L}}_\mathrm{M}$,
corresponding to the same eigenvalue $E(E+i\mathbb{M}^{-1}\mathbb{D})$. In particular, $\mathbb{L}_\mathrm{M}$ and $\Tilde{\mathbb{L}}_\mathrm{M}$ have the same spectrum which is closely related to the spectrum of $\mathbb{H}$. When there is no damping, the eigenvalues of $\mathbb{H}$ are square roots of the eigenvalues of $\mathbb{L}_\mathrm{M}$ and the vector $\left[\begin{array}{c}
        -X^\mathrm{N} \\
            \hline
            X^\mathrm{L}
    \end{array}\right]$ is an eigenvector of $\mathbb{H}$ related to the eigenvalue $-E$.
However, the shapes of the matrices $\mathbb{L}_\mathrm{M}$ and $\Tilde{\mathbb{L}}_\mathrm{M}$ as well as the dimensions of their kernels are different. The kernel of $\mathbb{L}_\mathrm{M}$ is of dimension $1$ and spanned by the vector $\mathbb{M}^{-\frac{1}{2}} \begin{bmatrix}1 \\ \vdots \\ 1
    \end{bmatrix}$, while the kernel of $\Tilde{\mathbb{L}}_\mathrm{M}$ has a dimension equal to $N_\mathrm{lines}-N_\mathrm{nodes}+1$ and is composed on states localized on the loops of the network \cite{guichard}. 




Let us now consider that there is no damping in the system and that two site energies $\epsilon^\mathrm{N}$ and $\epsilon^\mathrm{L}$ are added on the node and line states, respectively. Up to a shifting of the scales of energies, this is equivalent to the addition of a site energy $\epsilon$ on the node states and $-\epsilon$ on the line states, and the resulting Hamiltonian $\mathbb{H}_\mathrm{tot}$ can be written 

\begin{equation}
    \mathbb{H}_\mathrm{tot} = \mathbb{H} + \epsilon (\mathcal{P}_\mathrm{N} - \mathcal{P}_\mathrm{L}).
\end{equation}

In this case, the square of $\mathbb{H}_\mathrm{tot}$ is

\begin{equation}
    \mathbb{H}^2_\mathrm{tot} = \mathbb{H}^2 + \epsilon^2 \mathbb{I},
    \label{H_self}
\end{equation}
where $\mathbb{I}$ is the identity matrix, hence $\mathbb{H}_\mathrm{tot}$ can still be written in function of the laplacian matrices $\mathbb{L}_\mathrm{M}$ and $\mathbb{L}_\mathrm{M}$, using formulas \eqref{laplacian_mass} and \eqref{laplacian_mass_2}.

An other operator defined with respect to the Hamiltonian $\mathbb{H}$ at the energy $z$ is the operator $G(z)$, given by
\begin{equation}
    G(z) := \left[z\mathbb{I} - \mathbb{H}\right]^{-1}.
\end{equation}
 It can be used for approximating the density of states $n(E)$ by

\begin{equation}
    \pi n(E) = - \lim_{\epsilon \longrightarrow 0}\text{Tr}(G(z = E + i\epsilon)),
\end{equation}
where $\text{Tr}(\cdot)$ is the trace operator.
The density of states $n(E)$ gives the breakdown of the eigenvalues of $\mathbb{H}$. When projected on a particular vector $X$, it is by definition the local density of states $n_X(E)$

\begin{equation}
      \pi n_X(E) = - \lim_{\epsilon \longrightarrow 0}X^\dagger G(z = E + i\epsilon)X.
\end{equation}
When $X$ is an elementary basis vector, i.e. when all the components of $X$ are zero, except for one particular component equal to one, then $X^\dagger G(z)X$ is the local Green function of an element of the lattice (node or line) and will be denoted in lower-case letters $g(z)$.
The Green function can be projected by the orthogonal projectors $\mathcal{P}$ and $\mathcal{Q}$ on subsets of states by the two following formulas

\begin{align}
    &\mathcal{P}G(z) \mathcal{P} = \frac{\mathcal{P}}{z - \mathcal{P}\mathbb{H}\mathcal{P} - \mathcal{P}\mathbb{H}\mathcal{Q}\frac{1}{z  -\mathcal{Q}\mathbb{H}\mathcal{Q}}\mathcal{Q}\mathbb{H}\mathcal{P}} ,\label{PGP}\\
    & \mathcal{P}G(z) \mathcal{Q} = - \mathcal{P}G(z) \mathcal{P} \mathbb{H}\mathcal{Q}\frac{1}{z -\mathcal{Q}\mathbb{H}\mathcal{Q}} \mathcal{Q}.
    \label{PGQ}
\end{align}





\section{Effect of the cumulation of the disorder of type ($\mathcal{N}$) and ($\mathcal{L}$)}

\begin{figure*}[ht!]
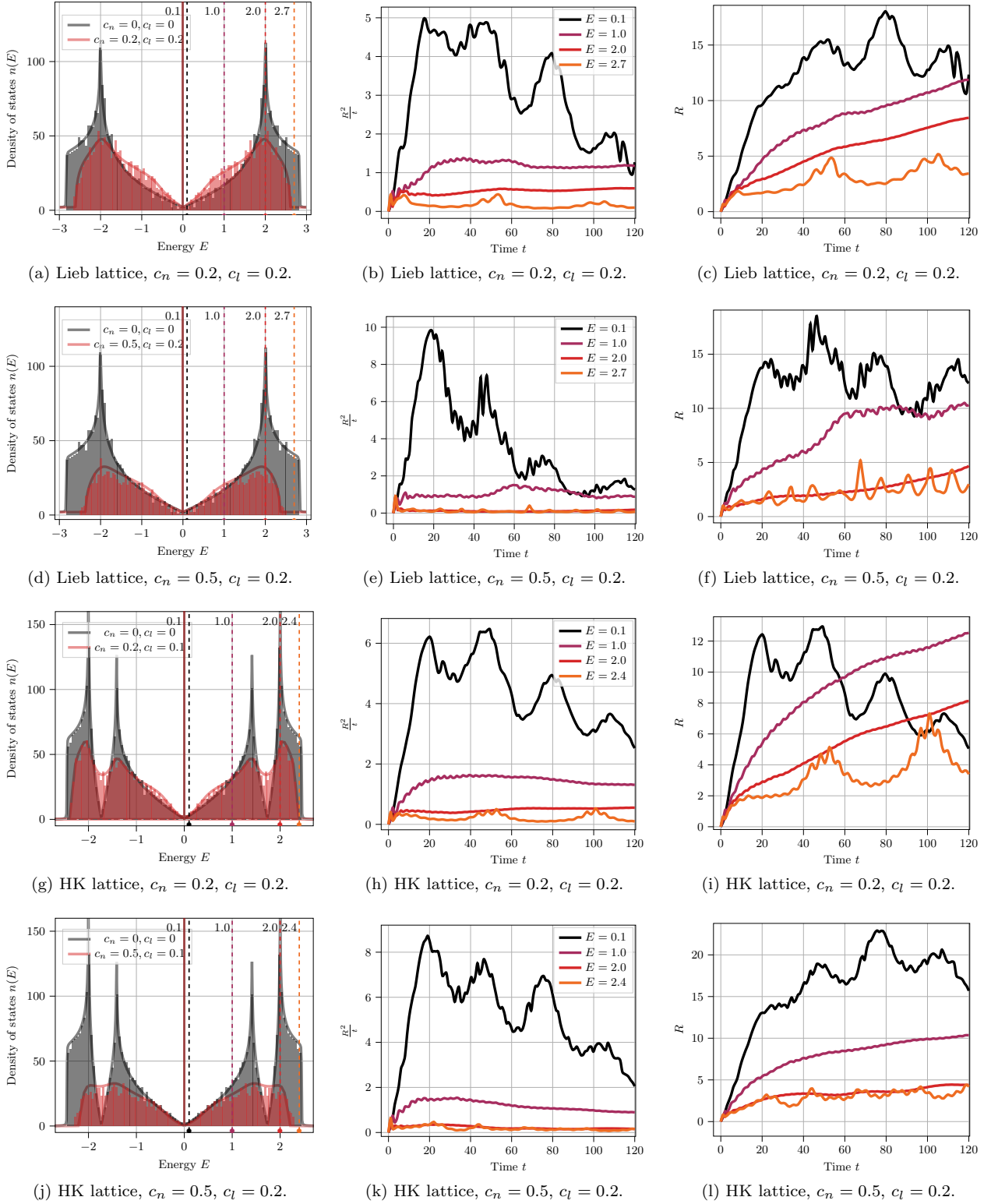

    \centering

  \subfloat[Lieb lattice, $c_n=0.2$, $c_l=0.2$.]{\begin{adjustbox}{clip,trim=0cm 0.cm 0.cm 0.cm,max width=0.3\linewidth}
\input{lieb_dos_mixed_0.2}
\end{adjustbox}}\quad
    \subfloat[Lieb lattice, $c_n=0.2$, $c_l=0.2$.]{\resizebox{0.3\linewidth}{!}{\input{w_mixed_noise_0.2}}}\quad
        \subfloat[Lieb lattice, $c_n=0.2$, $c_l=0.2$.]{\resizebox{0.3\linewidth}{!}{\input{w_mixed_noise_0.2_R}}}\\
                \subfloat[Lieb lattice, $c_n=0.5$, $c_l=0.2$.]{\begin{adjustbox}{clip,trim=0cm 0.cm 0.cm 0.cm,max width=0.3\linewidth}
\input{lieb_dos_mixed_0.5}
\end{adjustbox}}
\quad
            \subfloat[Lieb lattice, $c_n=0.5$, $c_l=0.2$.]{\resizebox{0.3\linewidth}{!}{\input{w_mixed_noise_0.5}}}\quad
     \subfloat[Lieb lattice, $c_n=0.5$, $c_l=0.2$.]{\resizebox{0.3\linewidth}{!}{\input{w_mixed_noise_0.5_R}}}\\
     \subfloat[HK lattice, $c_n=0.2$, $c_l=0.2$.]{\begin{adjustbox}{clip,trim=0cm 0.cm 0.cm 0.cm,max width=0.3\linewidth}
        \input{hk_mixed_0.2.tex}
        \end{adjustbox}}\quad \subfloat[HK lattice, $c_n=0.2$, $c_l=0.2$.]{\resizebox{0.3\linewidth}{!}{\input{w_hexa_mixed_noise_0.2}}}\quad \subfloat[HK lattice, $c_n=0.2$, $c_l=0.2$.]{\resizebox{0.3\linewidth}{!}{\input{w_hexa_mixed_noise_0.2_R}}}\\
        \subfloat[HK lattice, $c_n=0.5$, $c_l=0.2$.]{\begin{adjustbox}{clip,trim=0cm 0.cm 0.cm 0.cm,max width=0.3\linewidth}
            \input{hk_mixed_0.5.tex}
            \end{adjustbox}}\quad\subfloat[HK lattice, $c_n=0.5$, $c_l=0.2$.]{\resizebox{0.3\linewidth}{!}{\input{w_hexa_mixed_noise_0.5}}}\quad
            \subfloat[HK lattice, $c_n=0.5$, $c_l=0.2$.]{\resizebox{0.3\linewidth}{!}{\input{w_hexa_mixed_noise_0.5_R}}}
        \caption{\textbf{Cumulation of the disorders of type ($\mathcal{N}$) of concentration $c_n$ and ($\mathcal{L}$) of concentration $c_l$.} 
        The panels of the two upper layers correspond to the Lieb lattice, while the panels of the third and fourth layers correspond to the Honeycomb-Kagome (HK) lattice. 
    The left panels give the exact density of states of the perfect system (gray histogram) and of the disordered system (red histogram). The continuous lines are the densities of states given in Shiba's theory.
    The central (respectively, right) panels represents $\frac{R^2}{t}$ (respectively, $R$) in function of $t$ for various angular frequencies $E$, on the disordered system.
      }
    \label{fig:propagation_mixed_noise}
 \end{figure*}
The effect of the cumulation of the disorders ($\mathcal{N}$) and ($\mathcal{L}$) can be studied in Shiba's theory, analogously to what is done in equation \eqref{shiba}.
It comes back to combining the two previous models for the disorders of type $(\mathcal{N})$ and $(\mathcal{L})$.
 Four coefficients $\alpha_{1n}$, $\alpha_{2n}$, $\alpha_{1l}$, $\alpha_{2l}$ are simultaneously involved in the hopping terms of $\mathbb{H}$.
Their weighted averages are $\alpha_n := \sqrt{c_n \alpha^2_{1n} + (1-c_n) \alpha^2_{2n}}$ and $\alpha_l := \sqrt{c_l \alpha^2_{1l} + (1-c_l) \alpha^2_{2l}}$, where $c_n$ and $c_l$ are the concentration of the disordered sites,
for each type (node or line).
Two self-energies $\sigma_n(z)\simeq \sigma_{nr}z + i\sigma_{ni}z^3$ and $\sigma_l(z)\simeq \sigma_{lr}z + i\sigma_{li}z^3$ and effective interactors $\Delta_n(z)$ and $\Delta_l(z)$ coexist on the effective lattice,
whose hopping terms are equal to $\alpha_n\alpha_l \sqrt{\frac{B}{M}}$. All these coefficients are related by a system of two mean field equations, each one being exactly equivalent to equation \eqref{shiba} for each type of sites.
A major difference in the calculations is the formula \eqref{g_lap}, which is no longer valid, because all the sites of the effective lattice have a self-energy.
Let us defined a subscript $k$, a generic notation for one of the two types of sites $n$ or $l$, and a second subscribt $\bar{k}$ representing the other type of sites. 
The Green function of the effective lattice on a site of type $k$ is given by
\begin{equation}
    \Tilde{g}_k(z) = \frac{z-\sigma_{\bar{k}}(z)}{(\alpha_n \alpha_l)^2} g_\mathrm{lap}(Z), 
    \label{g_lap_nl}
\end{equation}
where $Z :=\frac{(z-\sigma_n(z))(z - \sigma_l(z))}{(\alpha_n \alpha_l)^2}$.
The calculations leading to the computation of the real part of $\sigma(z)$ are not affected by this modification, hence the values of $\sigma_{nr}$ and $\sigma_{lr}$
are the same than these obtained when only one type of disorder is applied.
However, the imaginary part of $\sigma(z)$ is changed by the combination of disorders.
The value $\sigma_{ki}$ is renormalised by a factor equal to $\frac{1-\sigma_{\bar{k}r}}{(\alpha_{\bar{k}})^2}$. In other words 
\begin{equation}
    \sigma_{ki} = \Tilde{\sigma}_{ki}\frac{1-\sigma_{\bar{k}r}}{(\alpha_{\bar{k}})^2},
\end{equation}
where $\Tilde{\sigma}_{ki}$ is the value obtained when only one type of disorder is applied.
Because the self-energies and the hopping coefficients of the effective Hamiltonian are changed, the equation giving $z$ in function of $E$ (equation \eqref{autoco}) is now

\begin{equation}
    (z - \sigma_n(z))(z - \sigma_l(z)) = \frac{B}{M}(\alpha_n\alpha_l E)^2.
\end{equation}

As a consequence, the velocity is now given by

\begin{equation}
    v = \sqrt{\frac{B}{M}}\frac{\alpha_n \alpha_l}{\sqrt{(1-\sigma_{nr})(1-\sigma_{lr})}}  a v_0,
\end{equation}
meaning that it is renormalised by a factor equal to the product of the factors obtained for the disorders taken separately.
Replacing the values of the parameters involved here leads to

\begin{equation}
    v = \sqrt{\frac{B}{M}} \left[\frac{1-\gamma - c_l}{(1-c_n)(1-\gamma)}\right]^{\frac{1}{2}} a v_0
\end{equation}
The inverse of the scattering time is a linear combinations of the inverses of the scattering times $\tau_n$ and $\tau_l$ obtained for the two disorders taken separately.
More precisely, it is given by

\begin{equation}
    \frac{1}{\tau} = \frac{\alpha_l}{\sqrt{1 - \sigma_{lr}}}\frac{1}{\tau_n}  + \frac{\alpha_n}{\sqrt{1 - \sigma_{nr}}}\frac{1}{\tau_l}
\end{equation}

Replacing the values of the parameters involved here leads to

\begin{equation}
    \frac{1}{\tau} = \sqrt{\frac{B}{M}} \left[\frac{1-\gamma - c_l}{(1-c_n)(1-\gamma)}\right]^{\frac{1}{2}} \left[\frac{c_n}{1-c_n}
     +\frac{\gamma c_l}{(1-\gamma)^2}\right] \eta E^3.
\end{equation}

The value of the velocity and of the scattering time are consistent with the expressions obtained for the two disorders taken separately, when $c_n$ or $c_l$ is set to zero.
These results also show that a disorder of type ($\mathcal{N}$) of magnitude $c_n$ has the same effect, in terms of scattering, than a disorder of type ($\mathcal{L}$), with a concentration of

\begin{equation}
    c_l = \frac{(1-\gamma)^2}{\gamma} \frac{c_n}{1-c_n}
\end{equation}

For example, setting $\frac{1}{4}$ of the masses to zero in the Lieb lattice is equivalent, in terms of scattering, to removing $\frac{1}{6}$ of the power lines.
In the Honeycomb-Kagome, the same inertia disorder is equivalent to the removal of $\frac{1}{9}$ of the lines only.

Numerical results for the cumulation of the two types of disorders are presented in figure \ref{fig:propagation_mixed_noise}. It shows in particular that the three expected regimes of propagation, ballistic, diffusive and localised, exist in the toy models at different frequencies. It shows also that the mean field theory gives an accurate approximation of the exact density of states, even in presence of two types of disorder.

    %

 \section{Perturbative results}

\label{appendix_fermi}

For small perturbations of the inertia or susceptance parameters, the results obtained with Shiba's theory introduced in equation \eqref{shiba} are now compared with the results provided by a perturbative analysis  known in the condensed matter community as "Fermi's golden rule" \cite{dirac1927quantum} \cite{fermi1950nuclear}.
This theory gives the scattering time for low values of disorder, when a perturbation $\mathbb{H}'$ is added to the initial Hamiltonian $\mathbb{H}$, as

\begin{equation}
    \frac{1}{\tau} = 2\pi <\vert Y^\dagger \mathbb{H}' X\vert^2>_Y n(\sqrt{\frac{B}{M}}E),
    \label{scattering}
\end{equation}

where $n(\cdot)$ is the density of states of $\mathbb{H}$, $X$ and $Y$ are eigenvectors of $\mathbb{H}$ related to the energy $\sqrt{\frac{B}{M}} E$, and $<\vert Y^\dagger \mathbb{H}' X\vert^2>_Y$ is the average of $\vert Y^\dagger \mathbb{H}' X\vert^2$ over all eigenvectors $Y$ with energy $\sqrt{\frac{B}{M}} E$.

 \begin{figure*}[ht!]
    \centering
  \subfloat{\resizebox{0.9\linewidth}{!}{\input{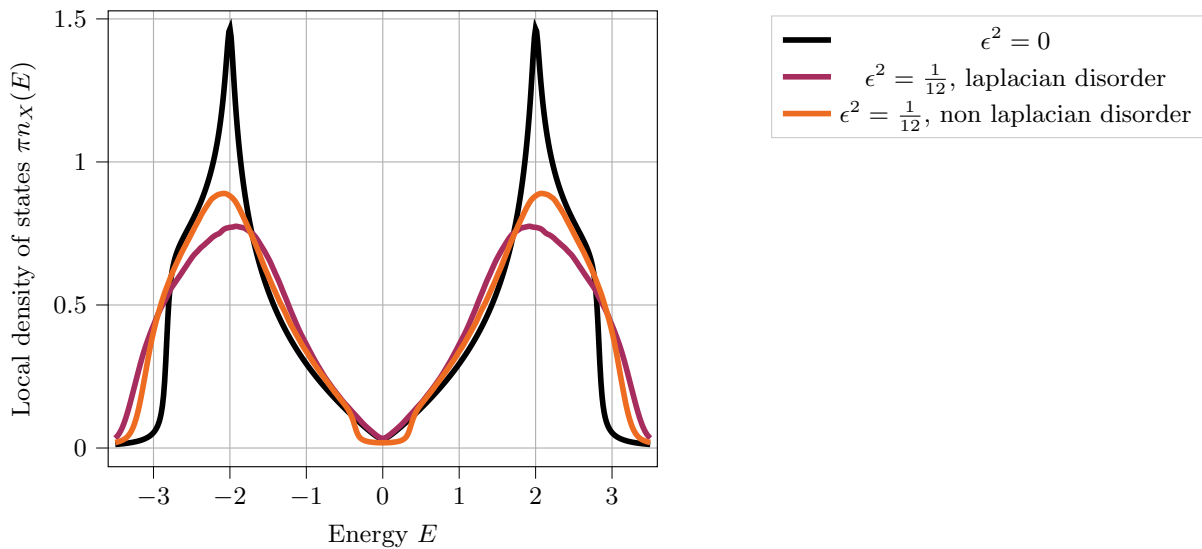}}}
        \caption{\textbf{Comparison between the laplacian disorder and the non laplacian disorder of the same magnitude $\epsilon^2=\frac{1}{12}$, on the Lieb lattice.}
        The local densities of states $\pi n_X(E)$ on a node state $X$ are plotted.
         The laplacian disorder consists in tiny modification of the susceptance parameters.
          The non Laplacian disorder is the addition of non correlated coefficients on the hopping terms.}
    \label{fig:dos_fermi}
 \end{figure*}

For small values of disorders with a Laplacian structure (i.e with a Hamiltonian in the separable form of equation \eqref{H_alpha}), we have checked that the results of Shiba's theory are the same than the results computed with Fermi's golden rule. In particular, the scattering time is inversely proportional to the cube of the energy. This is characteristic of disorders preserving the fact that the square of the Hamiltonian is related to a laplacian matrix, in other words that the Hamiltonian verifies formulas \eqref{laplacian_mass} and \eqref{laplacian_mass_2}. 
We show now the consequences of the removal of this property.
We consider that a specific disorder is added to our system having initially constant inertia $M$ and susceptance $B$ parameters. A coefficient $\delta_{i, l}$ is added to the coefficient $\sqrt{\frac{B}{M}}$ of the Hamiltonian linking the node $i$ to the line $l$. The normalised standard deviation of this disorder is called $\epsilon$ and is given by

\begin{equation}
    \epsilon^2 := \frac{M}{ B N_\mathrm{lines}} \sum_{n=1}^{N_\mathrm{lines}} \delta_{i, l}^2.
\end{equation}
This disorder is specifically chosen such that $\delta_{i, l}$ has no correlation with the coefficient linking the same line $l$ to its other node $j$, in other words $\delta_{i, l} \neq \delta_{j, l}$.
 Because of this, the disordered Hamiltonian is not related to a laplacian matrix, thus to a power system. Fermi's golden rule, introduced in equation \eqref{scattering} gives the following value for the inverse of the scattering time

\begin{equation}
    \frac{1}{\tau} = 4\eta \epsilon^2 \sqrt{\frac{B}{M}} E.
    \label{tau_non_laplac}
\end{equation}

This result indicates that when there is no laplacian correlation in the Hamiltonian, i.e when this Hamiltonian does not model a power system,
the inverse of the scattering time is linear in the energy, instead of being cubic.
 It emphasizes the specific and particular good protection of the low energy states of power systems relatively to disorder, in comparison with other systems.

Formula \eqref{tau_non_laplac} is valid in the range of low energies where the density of states remains unchanged by the disorder.
Therefore, is is valid only outside of the gap induced by the non laplacian disorder, as shown in figure \ref{fig:dos_fermi}.
We approximate the value of this gap by using the  Courant-Fisher-Weyl pinciple.
It gives the smallest eigenvalue of $\mathcal{P}_\mathrm{N} \mathbb{H}^2 \mathcal{P}_\mathrm{N}$ as $\lambda_1 = \underset{X \in \mathbb{C}^n}{\text{min}}\frac{X^\dagger \mathcal{P}_\mathrm{N} \mathbb{H}^2 \mathcal{P}_\mathrm{N} X}{X^\dagger X}$. We approximate this quantity by the Rayleigh quotient on the initial ground state $X_0$, such that

\begin{equation}
    \lambda_1 \simeq  \frac{X_1^\dagger \mathcal{P}_\mathrm{N} \mathbb{H}^2 \mathcal{P}_\mathrm{N} X_1}{X_1^\dagger X_1} = 4\epsilon^2,
\end{equation}
for the Lieb lattice. Taking the square root yields an approximation of the smallest eigenvalue of $\mathbb{H}$, hence of the gap of the density of states, of $2 \epsilon$.
Other non laplacian disorders, namely the addition of random site-energies to the Hamiltonian, are not considered here.

\section{Additional figures}

In this appendix are presented additional figures supplementing our work. 

Figure \ref{fig:lieb_shiba_lignes} shows the densities of states $n_X(E)$ on the disordered states (nodes or lines, depending on the type of disorder) computed with Shiba's theory. It proves to be remarkably similar to the exact values.

Figure \ref{fig:lieb_lattice_40_0.2} represents the shape of the signal in the grid, for the toy models, at the time $t=20~\text{s}$ and $t = 100~\text{s}$. The shape of the signal are less scattered in the ballistic regime of frequencies. It shows also that localised signals exist at high frequency.

Figure \ref{fig:panta_eigenvectors} represents eigenvectors of the Pantagruel model, for various frequencies. In particular, the presence of localized states is confirmed. For the same model, figures \ref{fig:panta_isar} and \ref{fig:panta_crey} extend the study on diffusivity presented in figure \ref{fig:panta_fessenheim}, starting from other points of the European grid. 

 \begin{figure*}[ht!]
    \centering
    \subfloat[Lieb lattice, disorder of type ($\mathcal{N}$), $X$ is a node state.]{\resizebox{0.4\linewidth}{!}{\input{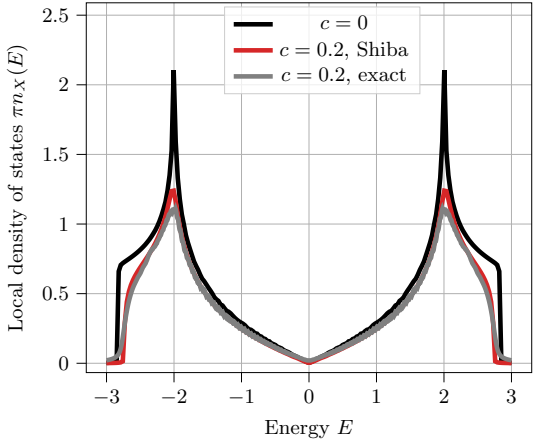}}}\quad
\subfloat[HK lattice, disorder of type ($\mathcal{N}$), $X$ is a node state.]{\resizebox{0.4\linewidth}{!}{\input{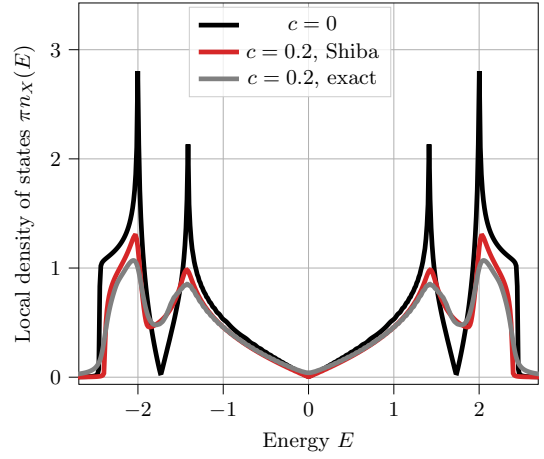}}}\\
    \subfloat[Lieb lattice, disorder of type ($\mathcal{L}$), $X$ is a line state.]{\begin{adjustbox}{clip,trim=0cm 0.cm 0.cm 0.cm,max width=0.4\linewidth}
\input{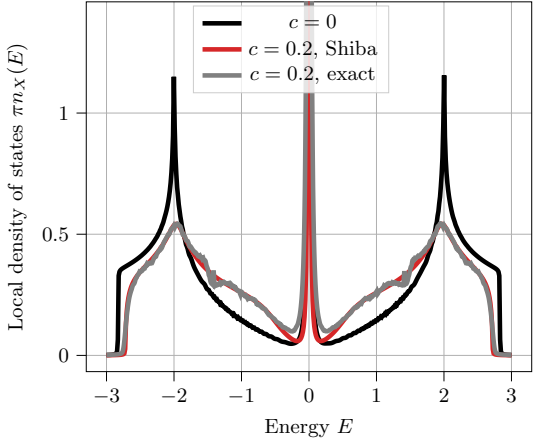}
\end{adjustbox}}\quad
  \subfloat[HK lattice, disorder of type ($\mathcal{L}$), $X$ is a line state.]{\begin{adjustbox}{clip,trim=0cm 0.cm 0.cm 0.cm,max width=0.4\linewidth}
\input{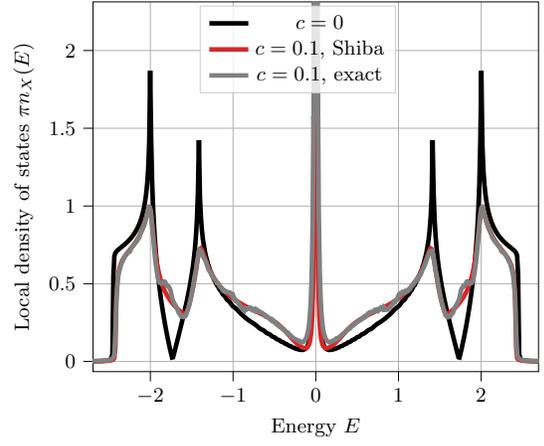}
\end{adjustbox}}\\
        \caption{\textbf{Densities of states in Shiba's theory.}
        The upper (respectively, bottom) panels compare the local density of states $n_X(E)$ computed in Shiba's theory with the exact value, for a disorder of type ($\mathcal{N}$) (respectively, type ($\mathcal{L}$)) of magnitude $c$.
        The upper left and bottom left panels are computed on the Lieb lattice while the upper right and bottom right panels involve the Honeycomb-Kagome (HK) lattice.}
      \label{fig:lieb_shiba_lignes}
 \end{figure*}

 \begin{figure*}[ht!]
    \centering
        \subfloat[Lieb lattice, $E=0.1$, $t=20~\text{s}$.]{\begin{adjustbox}{clip,trim=1.35cm 5.2cm 1cm 2cm,max width=0.25\linewidth}
        \includegraphics[width=0.7\linewidth]{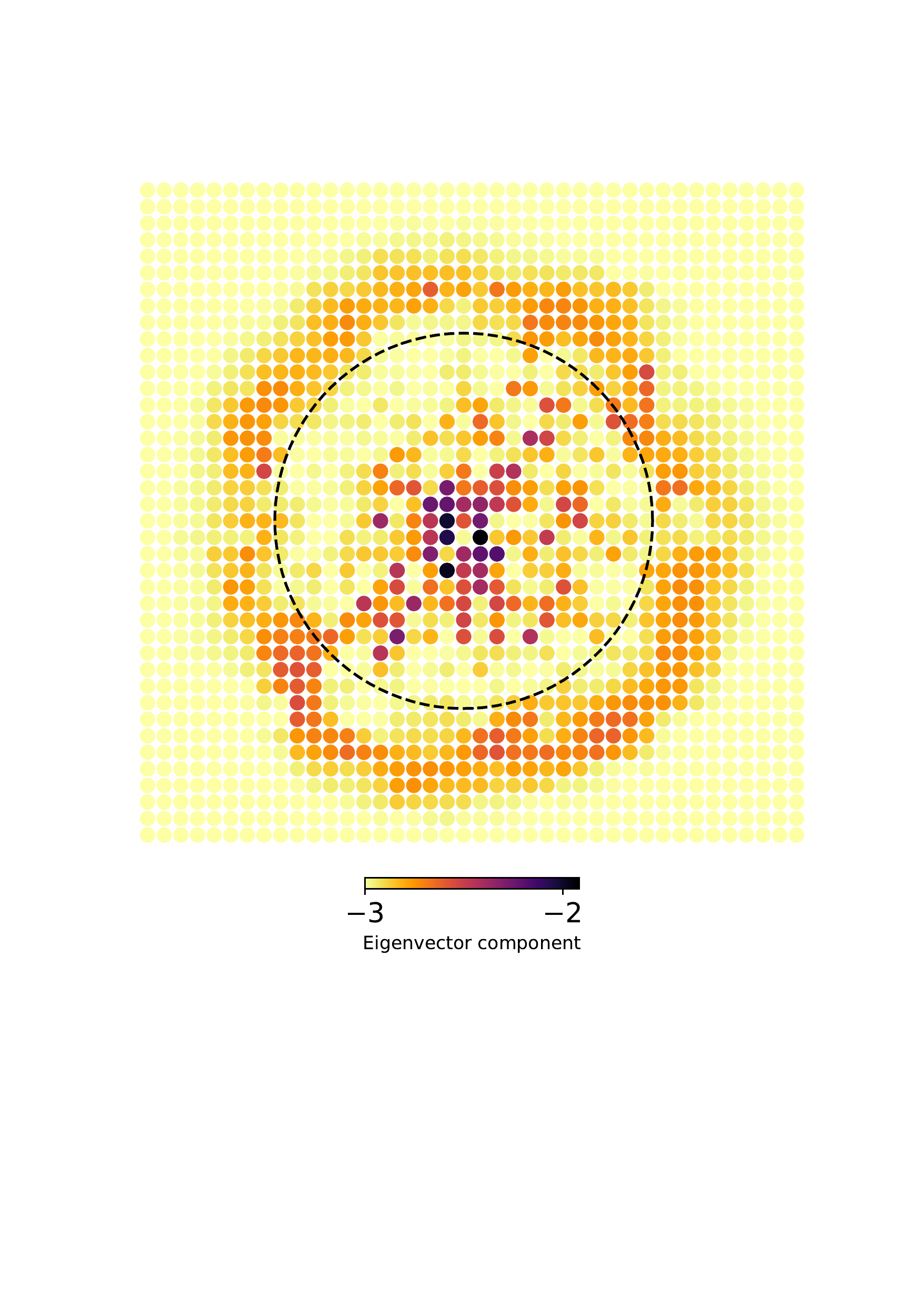}
    \end{adjustbox}}\quad\quad
    \subfloat[Lieb lattice, $E=1.0$, $t=20~\text{s}$.]{\begin{adjustbox}{clip,trim=1.35cm 5.2cm 1cm 2cm,max width=0.25\linewidth}
        \includegraphics[width=0.7\linewidth]{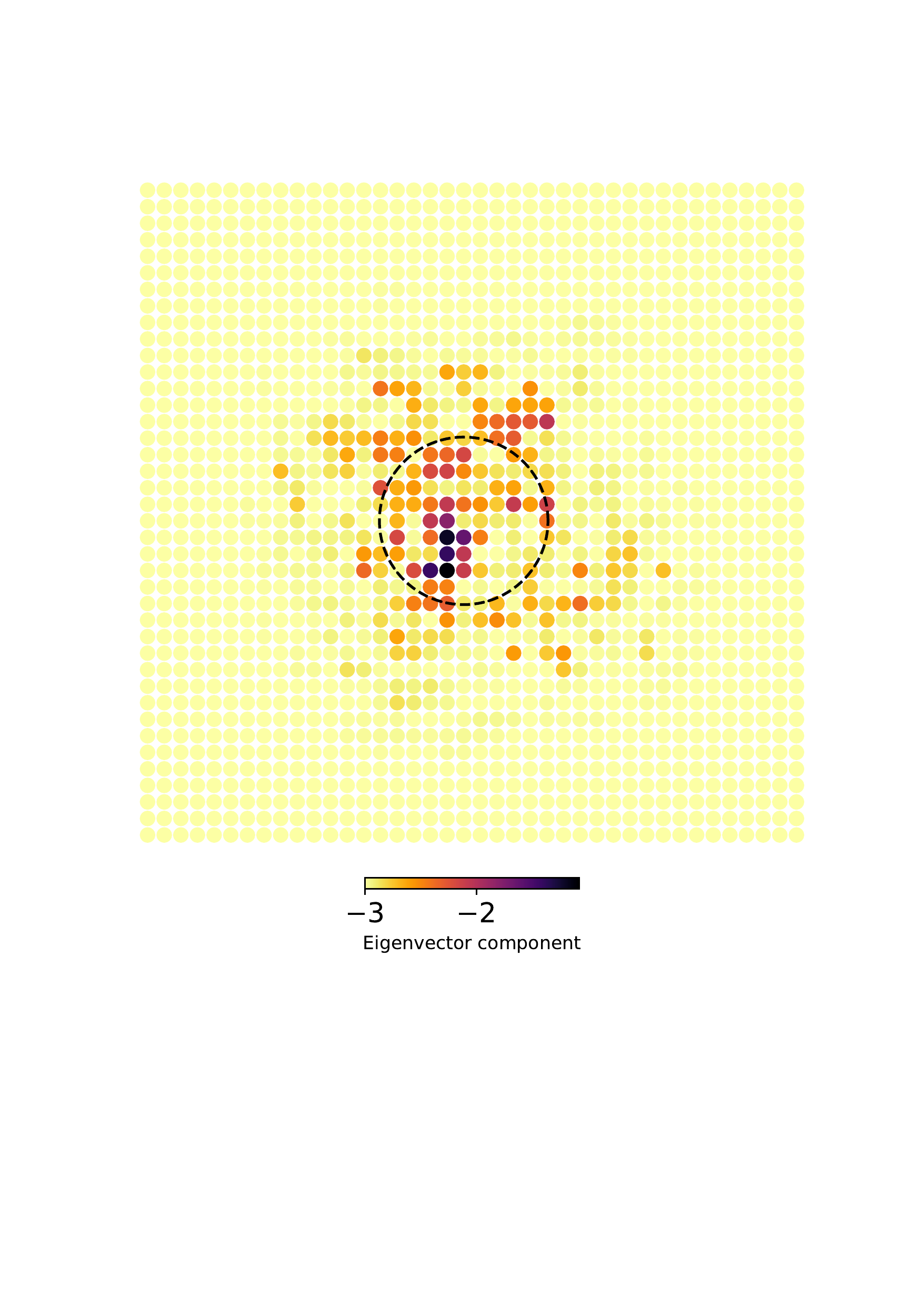} 
    \end{adjustbox}}\quad\quad
    \subfloat[Lieb lattice, $E=2.7$, $t=20~\text{s}$.]{\begin{adjustbox}{clip,trim=1.35cm 5.2cm 1cm 2cm,max width=0.25\linewidth}
        \includegraphics[width=0.7\linewidth]{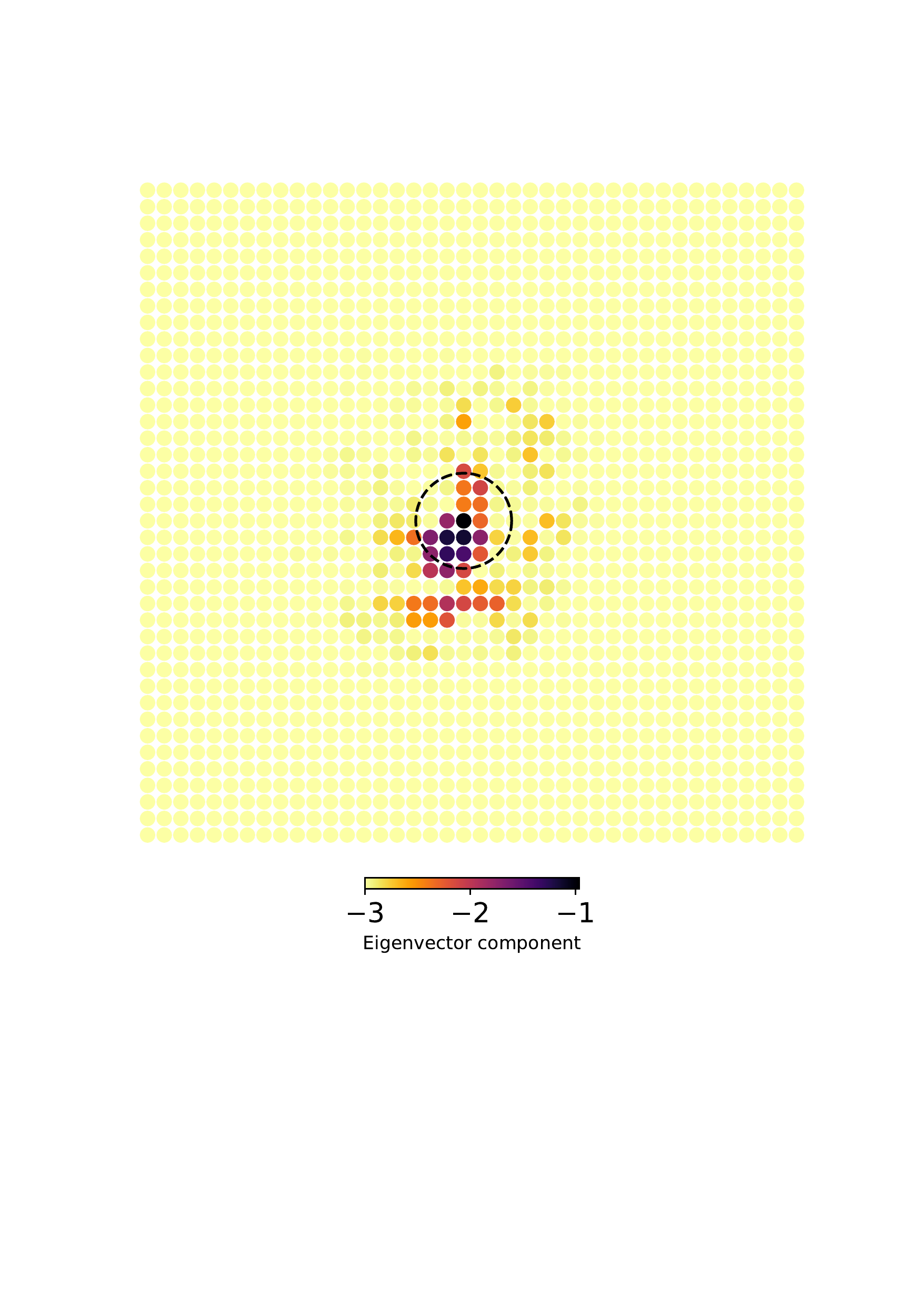}
    \end{adjustbox}}
    \\
     \subfloat[Lieb lattice, $E=0.1$, $t=100~\text{s}$.]{\begin{adjustbox}{clip,trim=1.35cm 5.2cm 1cm 2cm,max width=0.25\linewidth}
        \includegraphics[width=0.7\linewidth]{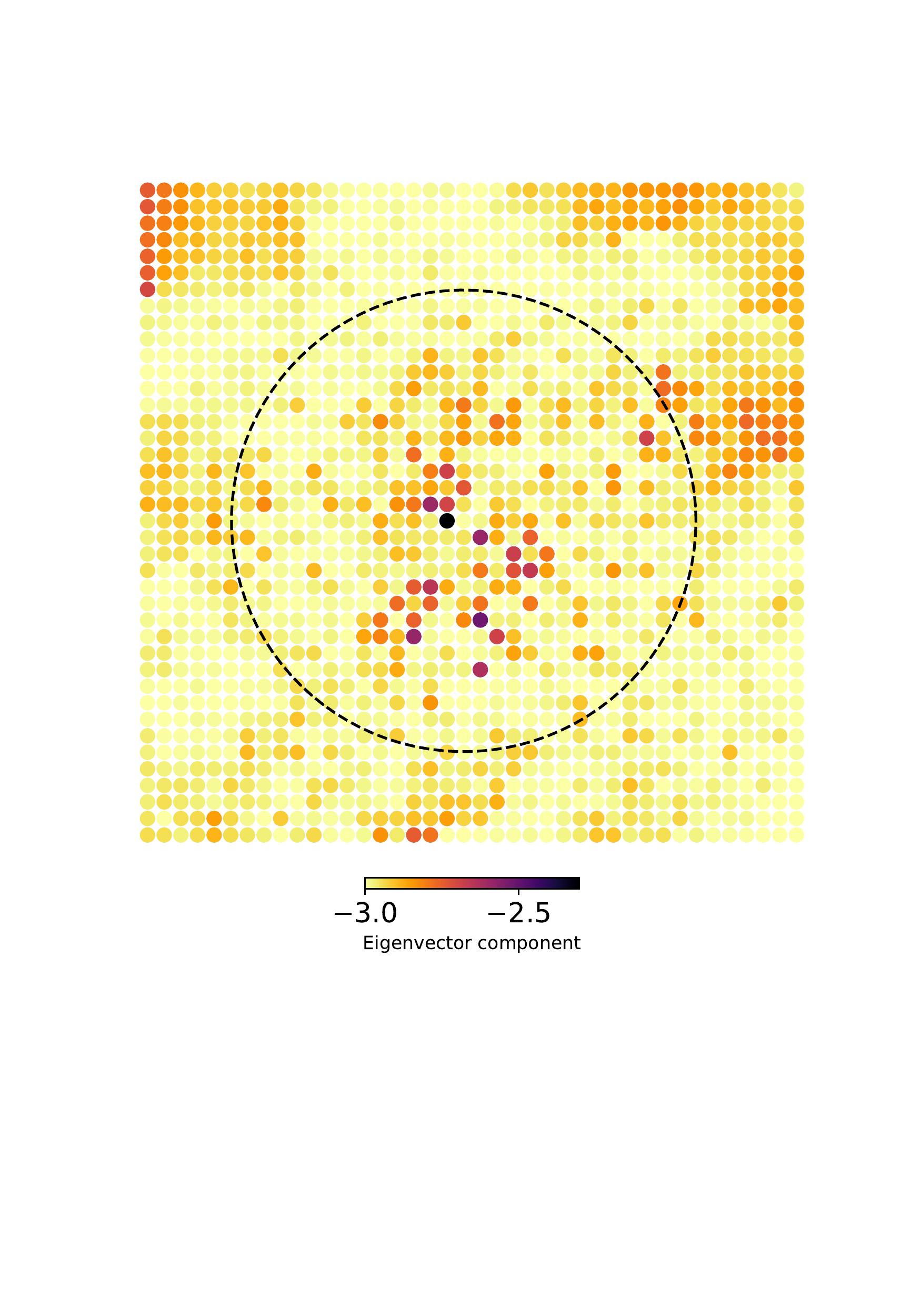}
    \end{adjustbox}}\quad\quad
     \subfloat[Lieb lattice, $E=0.6$, $t=100~\text{s}$.]{\begin{adjustbox}{clip,trim=1.35cm 5.2cm 1cm 2cm,max width=0.25\linewidth}
        \includegraphics[width=0.7\linewidth]{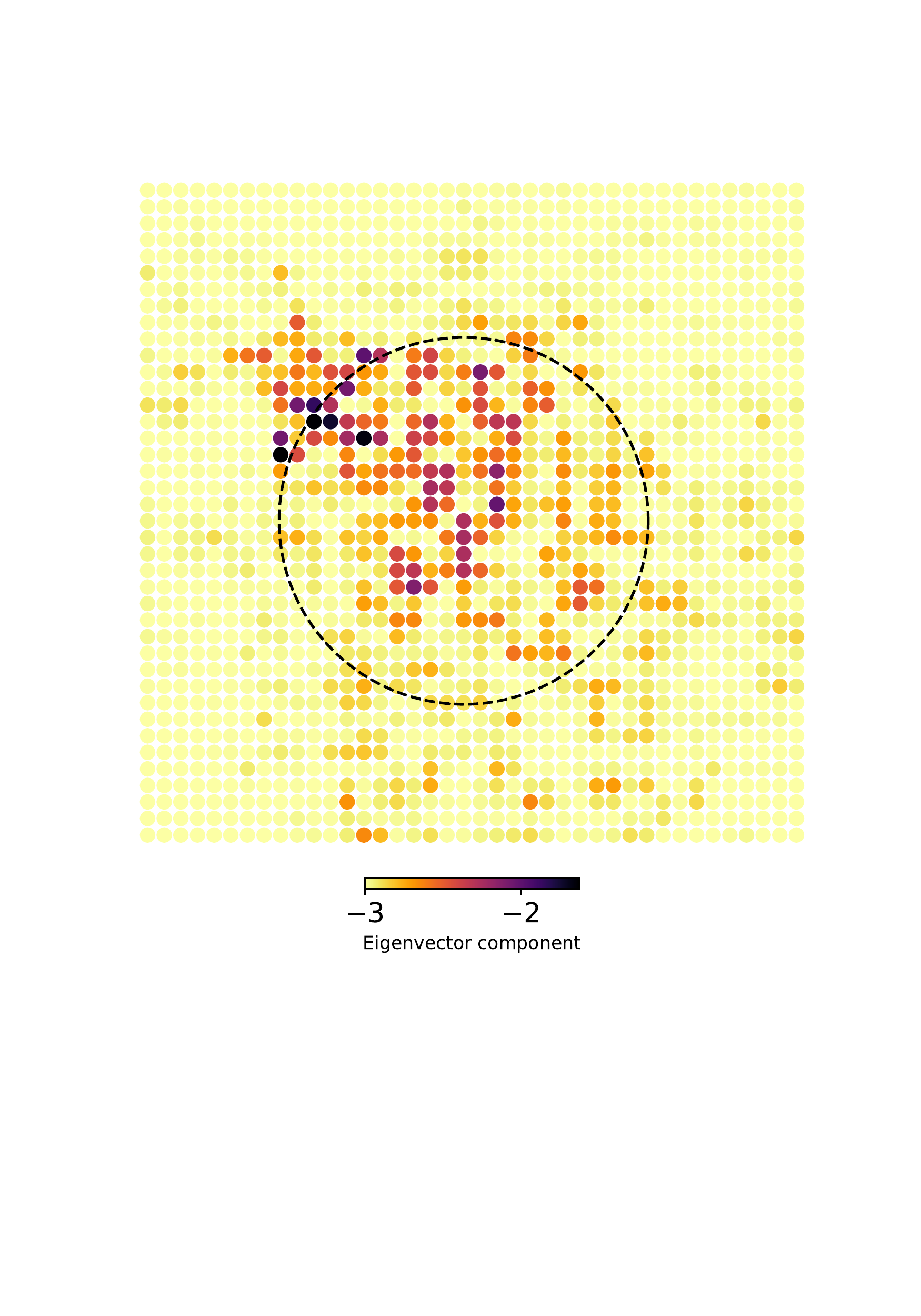} 
    \end{adjustbox}}\quad\quad
    \subfloat[Lieb lattice, $E=2.7$, $t=100~\text{s}$.]{\begin{adjustbox}{clip,trim=1.35cm 5.2cm 1cm 2cm,max width=0.25\linewidth}
        \includegraphics[width=0.7\linewidth]{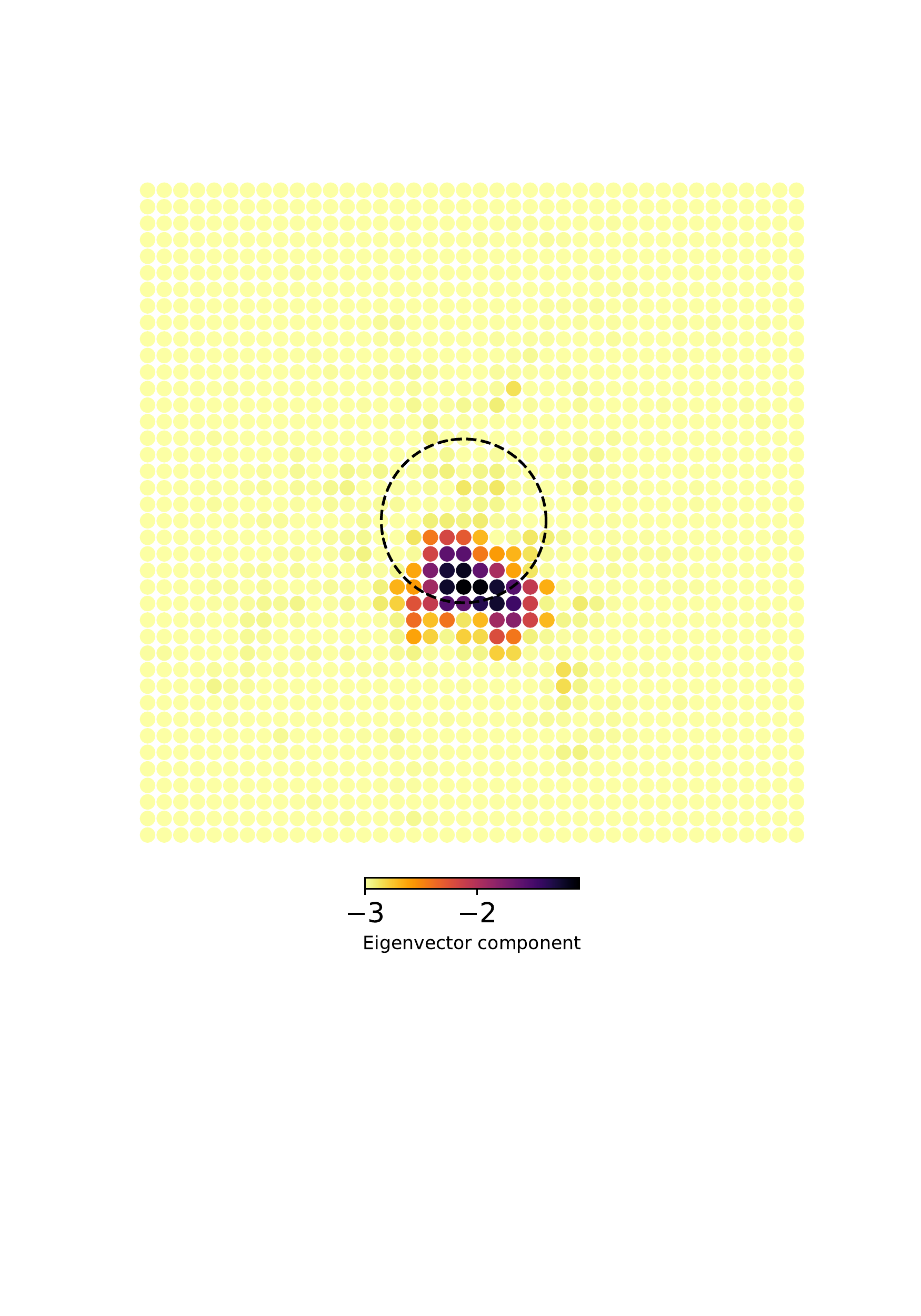}
    \end{adjustbox}}\\
     \subfloat[HK lattice, $E=0.1$, $t=20~\text{s}$.]{\begin{adjustbox}{clip,trim=1.4cm 2.7cm 1.2cm 1.cm,max width=0.3\linewidth}
        \includegraphics[width=0.7\linewidth]{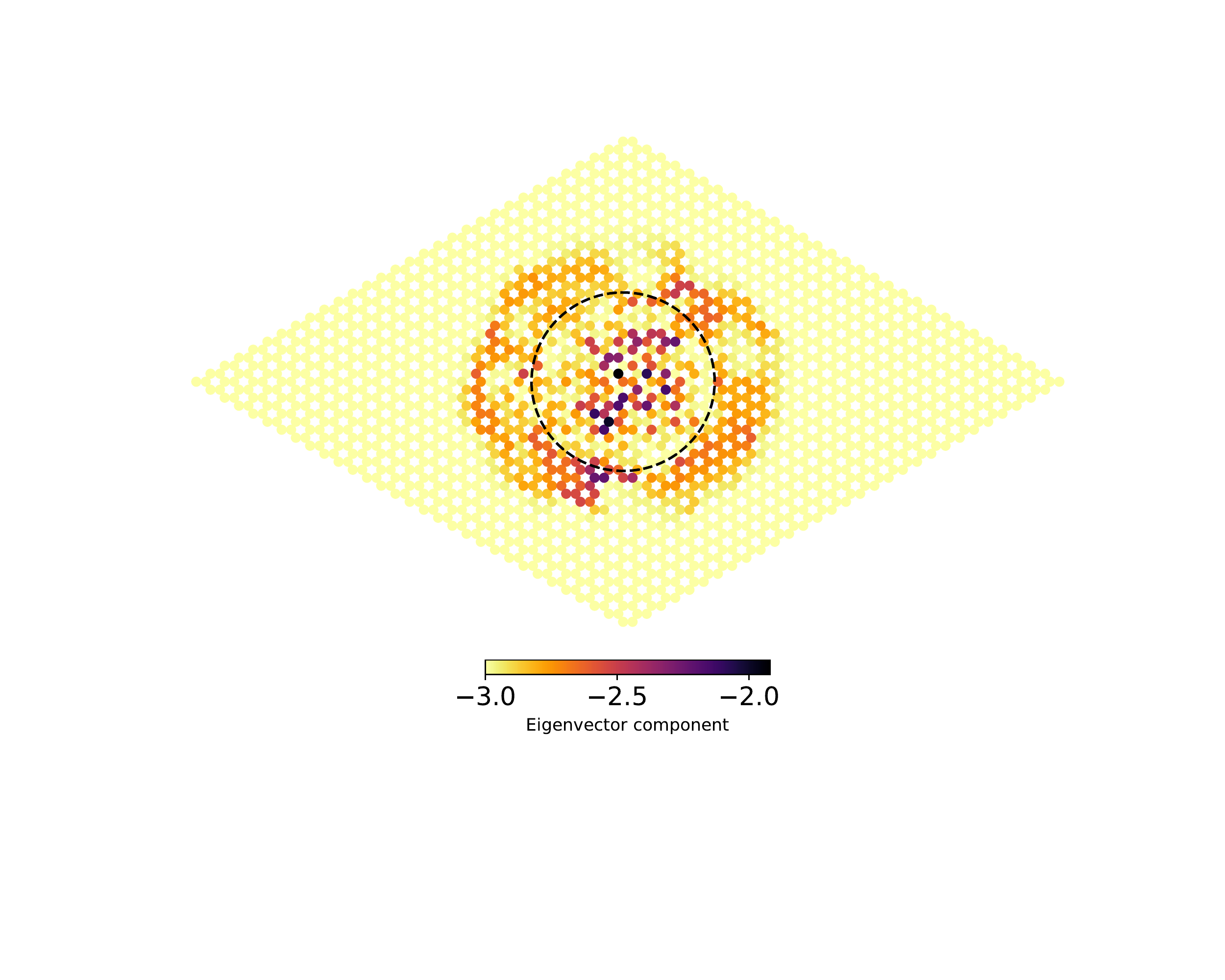}
    \end{adjustbox}}
     \subfloat[HK lattice, $E=0.6$, $t=20~\text{s}$.]{\begin{adjustbox}{clip,trim=1.4cm 2.7cm 1.2cm 1.cm,max width=0.3\linewidth}
        \includegraphics[width=0.7\linewidth]{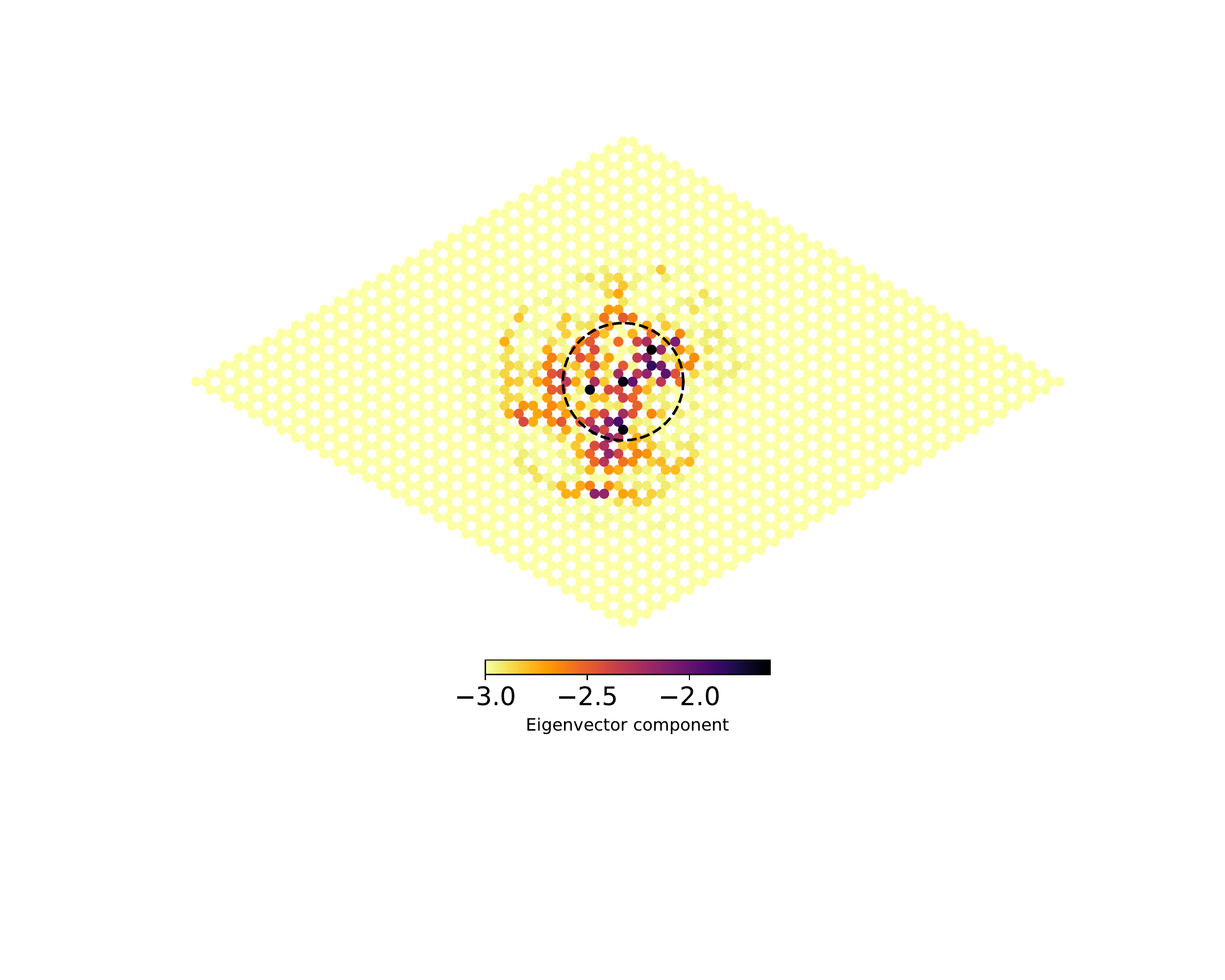}
    \end{adjustbox}}
     \subfloat[HK lattice, $E=2.4$, $t=20~\text{s}$.]{\begin{adjustbox}{clip,trim=1.4cm 2.7cm 1.2cm 1.cm,max width=0.3\linewidth}
        \includegraphics[width=0.7\linewidth]{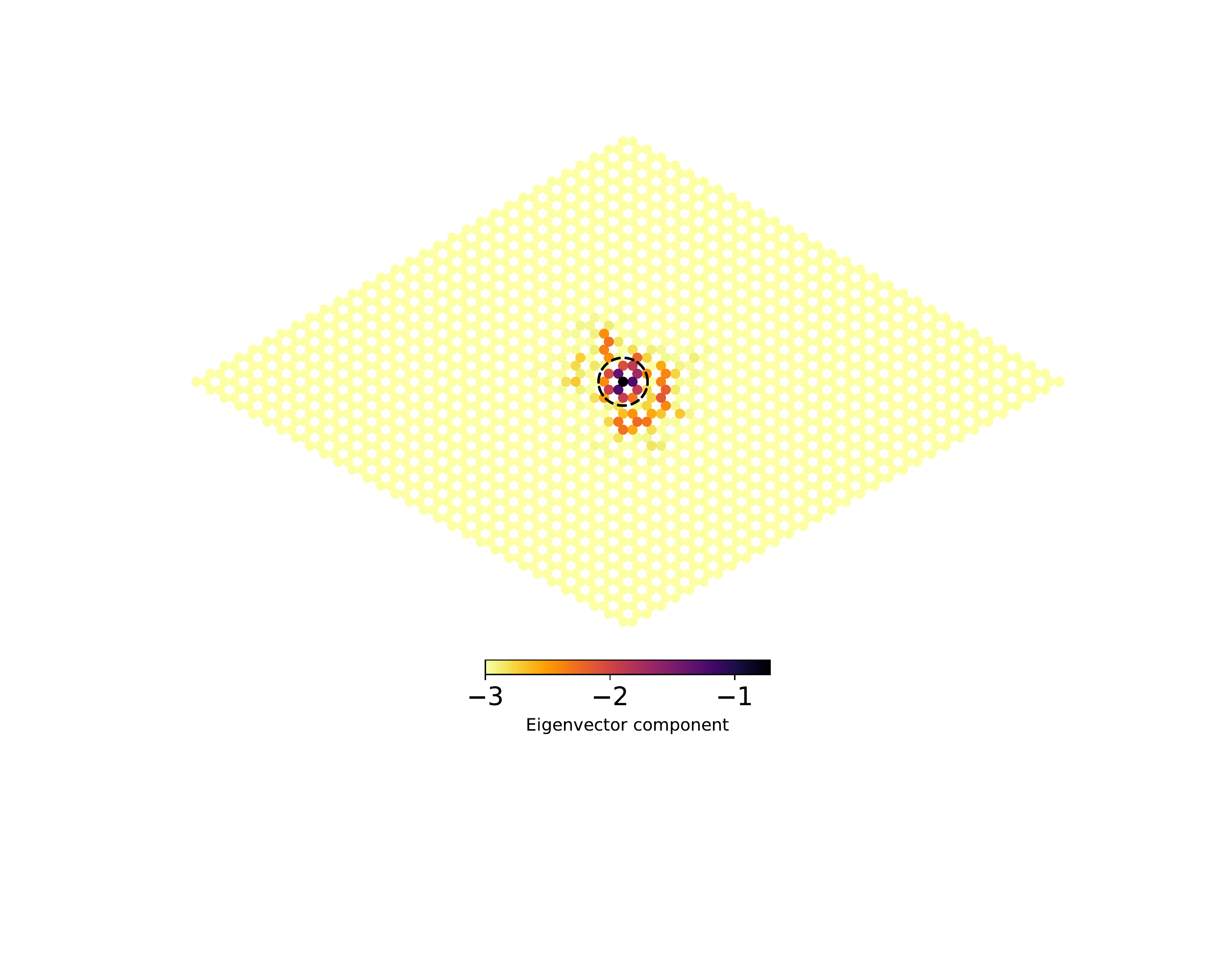}
    \end{adjustbox}}\\
     \subfloat[HK lattice, $E=0.1$, $t=100~\text{s}$.]{\begin{adjustbox}{clip,trim=1.4cm 2.7cm 1.2cm 1.cm,max width=0.3\linewidth}
        \includegraphics[width=0.7\linewidth]{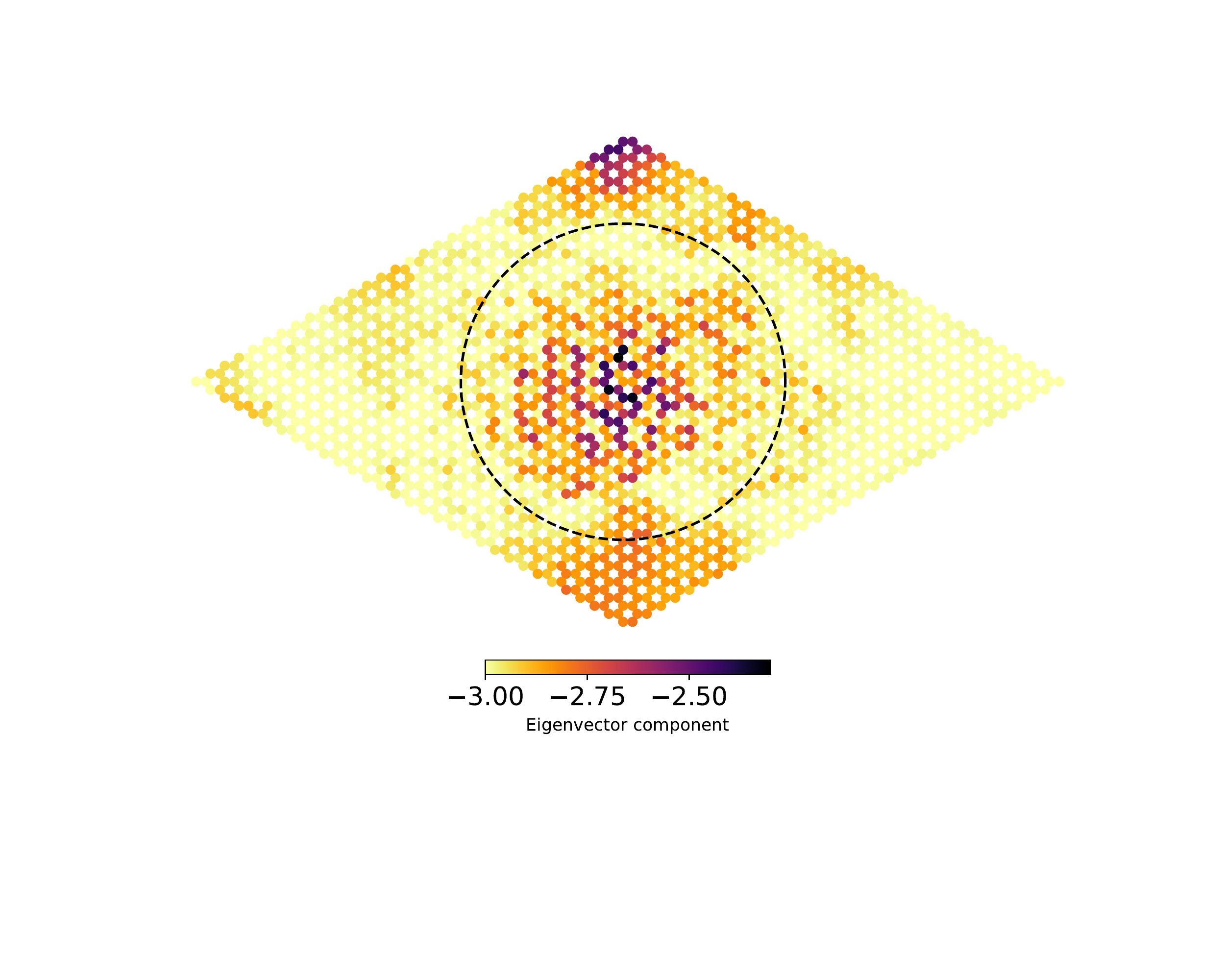}
    \end{adjustbox}}
     \subfloat[HK lattice, $E=0.6$, $t=100~\text{s}$.]{\begin{adjustbox}{clip,trim=1.4cm 2.7cm 1.2cm 1.cm,max width=0.3\linewidth}
        \includegraphics[width=0.7\linewidth]{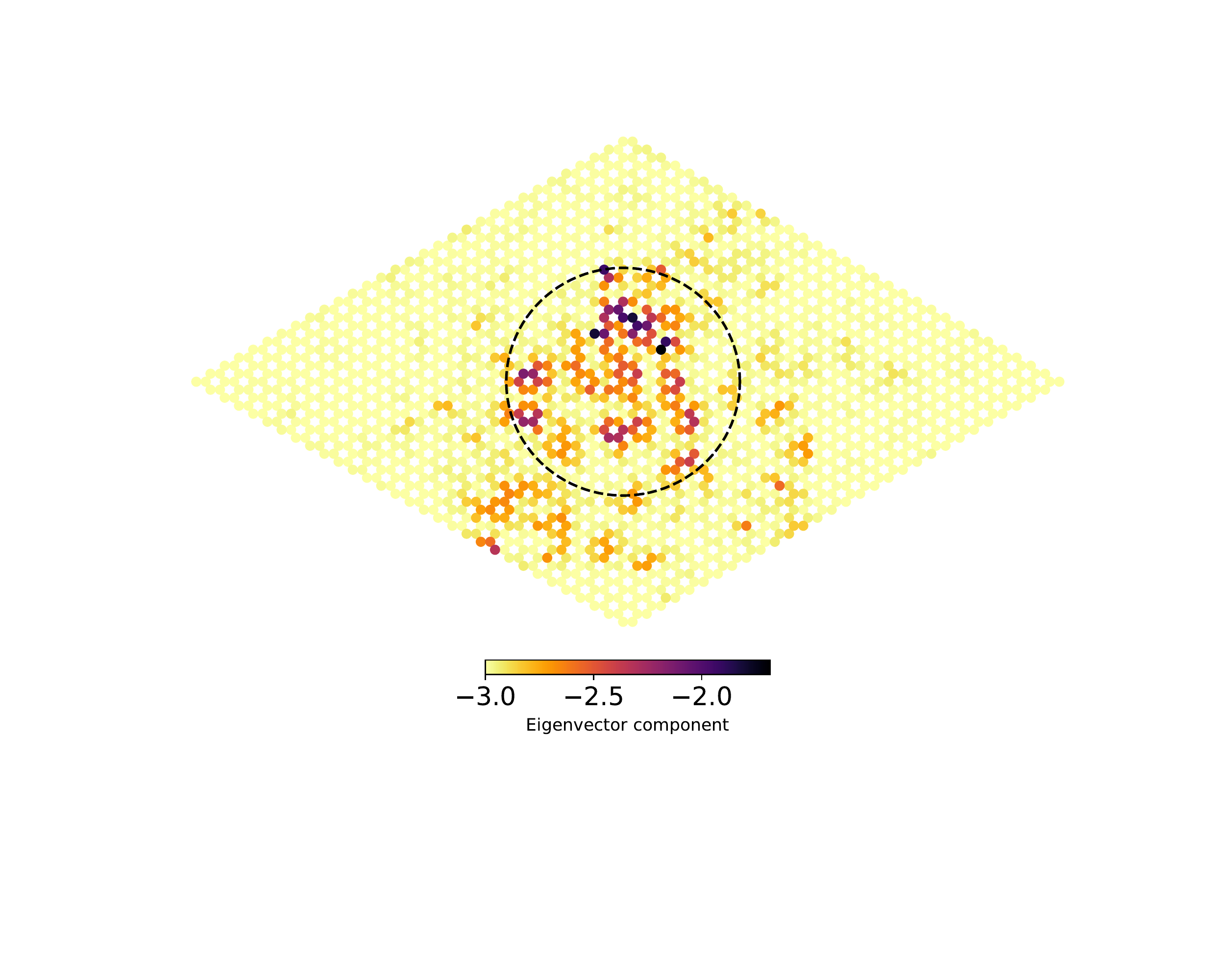}
    \end{adjustbox}}
     \subfloat[HK lattice, $E=2.4$, $t=100~\text{s}$.]{\begin{adjustbox}{clip,trim=1.4cm 2.7cm 1.2cm 1.cm,max width=0.3\linewidth}
        \includegraphics[width=0.7\linewidth]{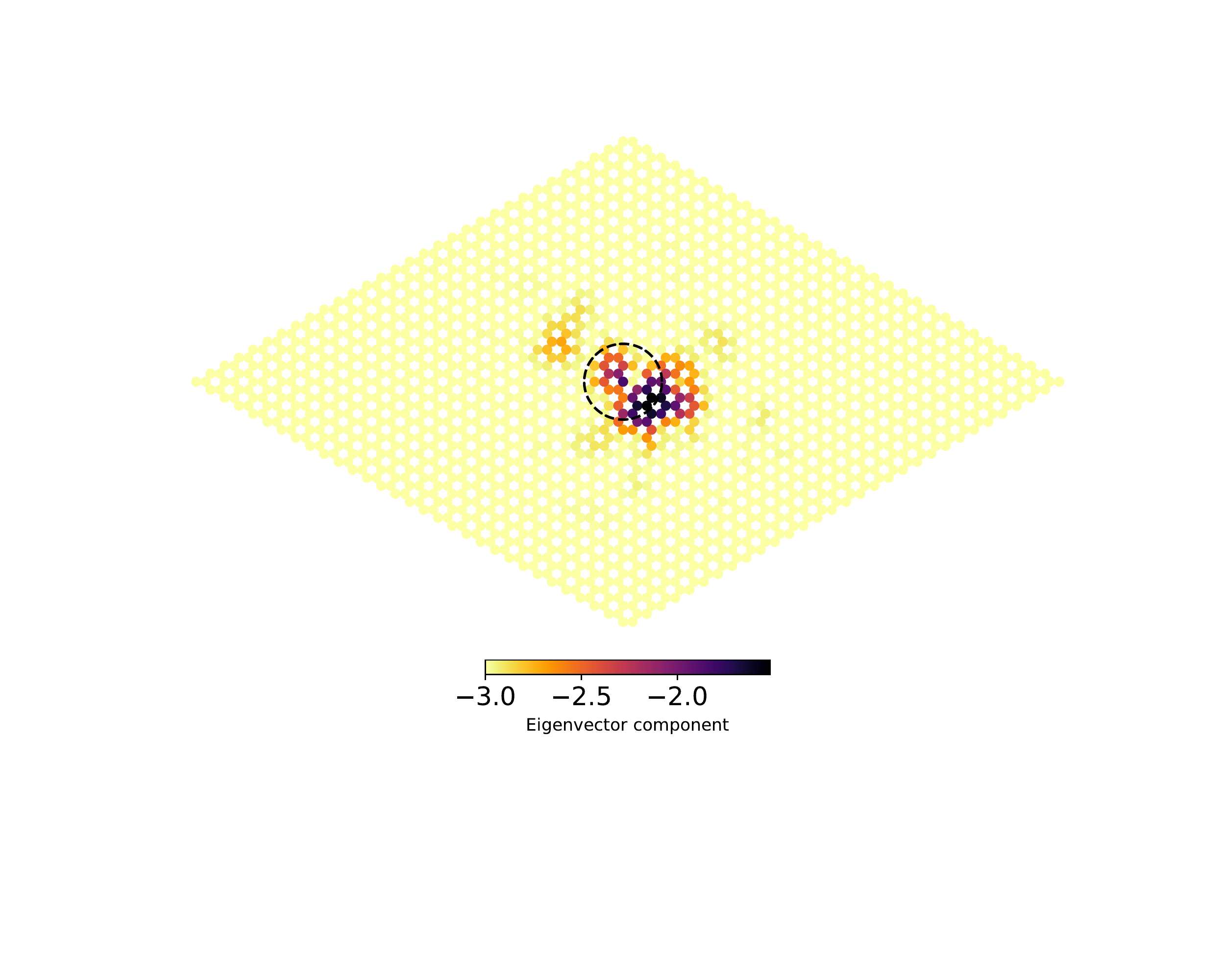}
    \end{adjustbox}}
    \caption{\textbf{Colormaps of the electrical state of the grid, for the injection of a sinusoidal signal at the central node of the toy models with a disorder of type ($\mathcal{L}$).} 
    The panels of the two upper layers correspond to the Lieb lattice, with $c=0.2$. The panels of the third and fourth layers correspond to the Honeycomb-Kagome (HK) lattice, with $c=0.1$. 
On each panel is represented as a colormap the value of $\log(\frac{\vert \psi_n\vert^2 }{\vert\vert \psi\vert \vert^2})$ for every node $n$ of the lattice.
   }
    \label{fig:lieb_lattice_40_0.2}
 \end{figure*}

  \begin{figure*}[ht!]
    \centering
    
    \subfloat[$E=2$.]{\begin{adjustbox}{clip,trim=1.4cm 3.8cm 1.2cm 1.3cm,max width=0.5\linewidth}
        \includegraphics[width=0.7\linewidth]{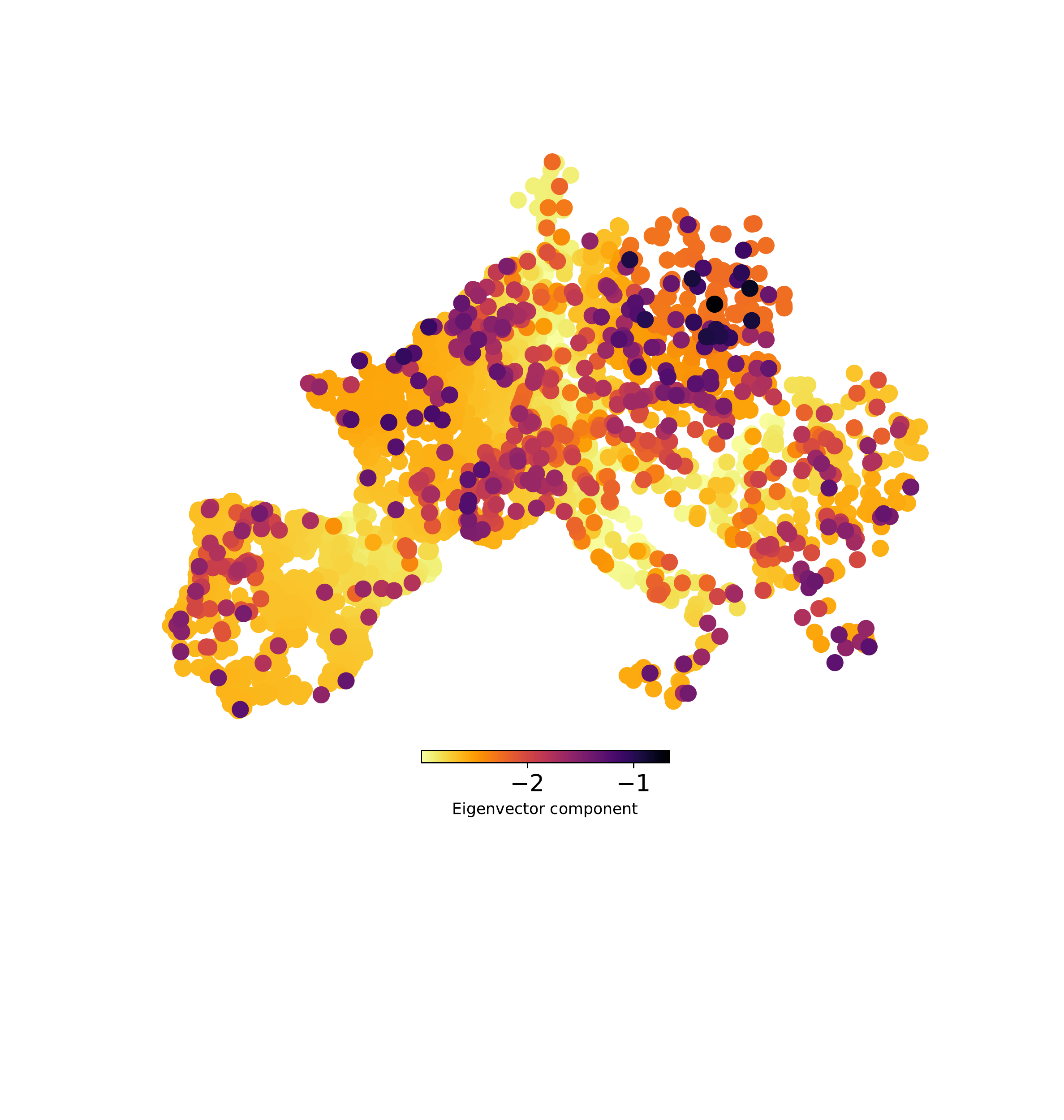}
    \end{adjustbox}}
    \subfloat[$E=10$.]{\begin{adjustbox}{clip,trim=1.4cm 3.8cm 1.2cm 1.3cm,max width=0.5\linewidth}
        \includegraphics[width=0.7\linewidth]{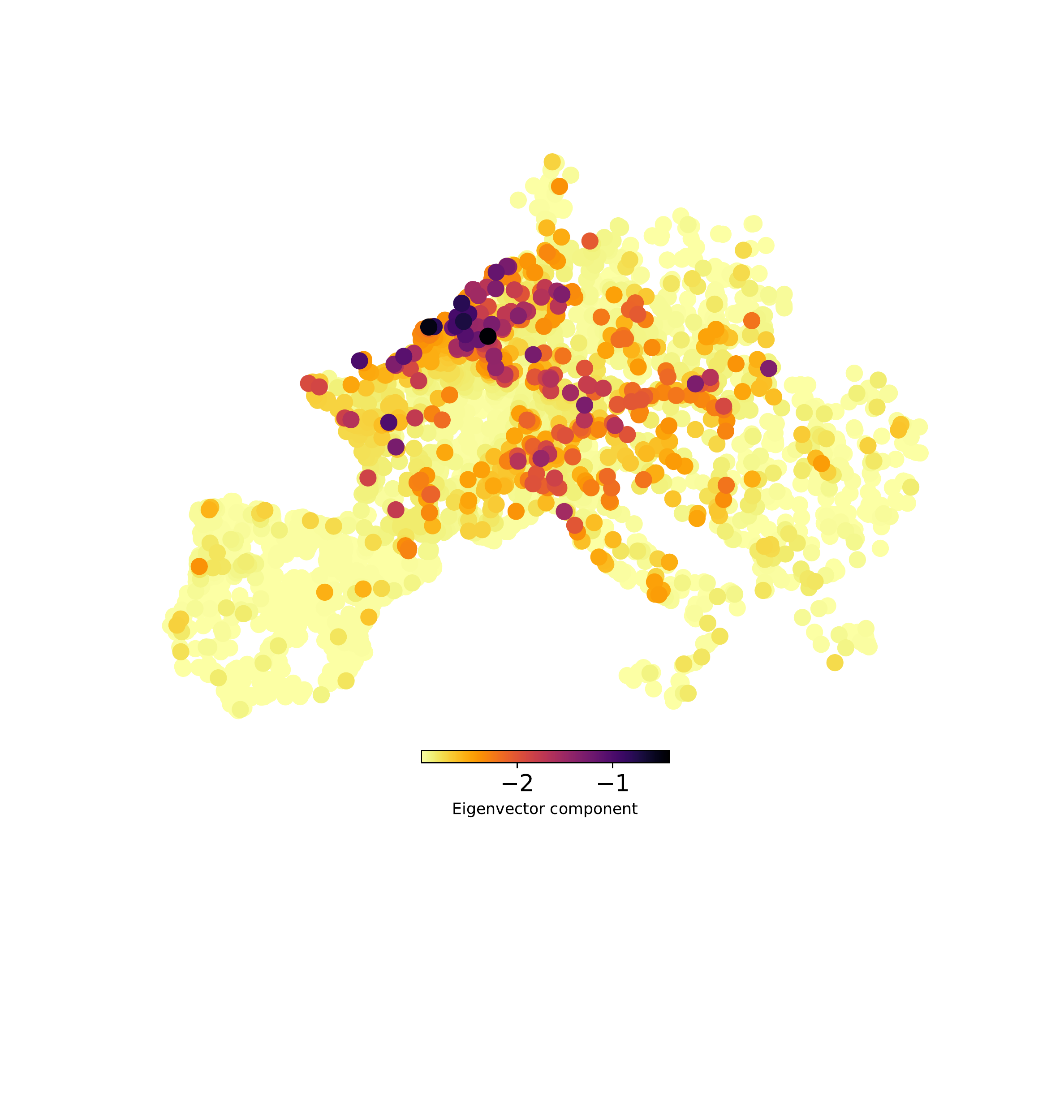}
    \end{adjustbox}}\\
    \subfloat[$E=20$.]{\begin{adjustbox}{clip,trim=1.4cm 3.8cm 1.2cm 1.3cm,max width=0.5\linewidth}
        \includegraphics[width=0.7\linewidth]{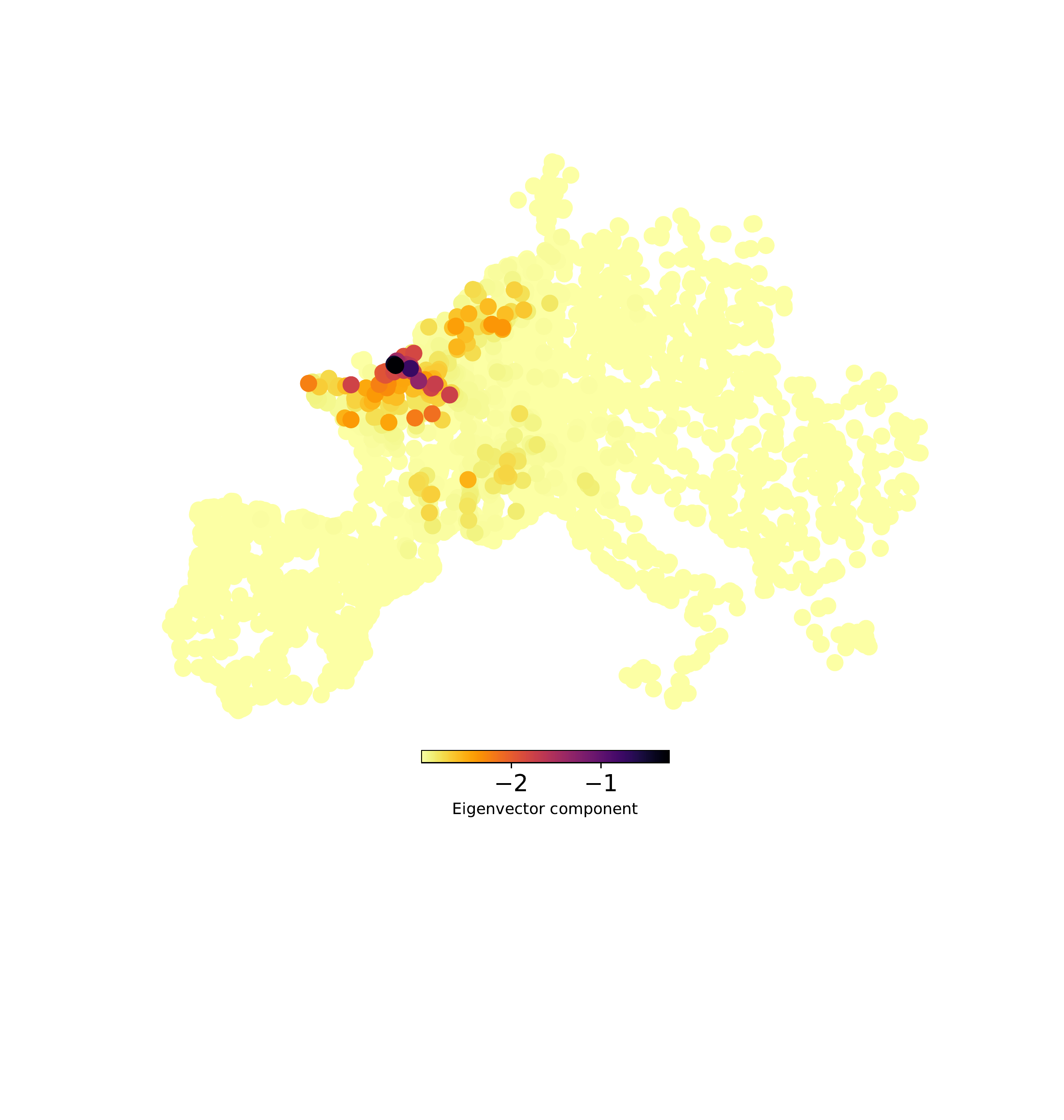}
    \end{adjustbox}}
    \subfloat[$E=40$.]{\begin{adjustbox}{clip,trim=1.4cm 3.8cm 1.2cm 1.3cm,max width=0.5\linewidth}
        \includegraphics[width=0.7\linewidth]{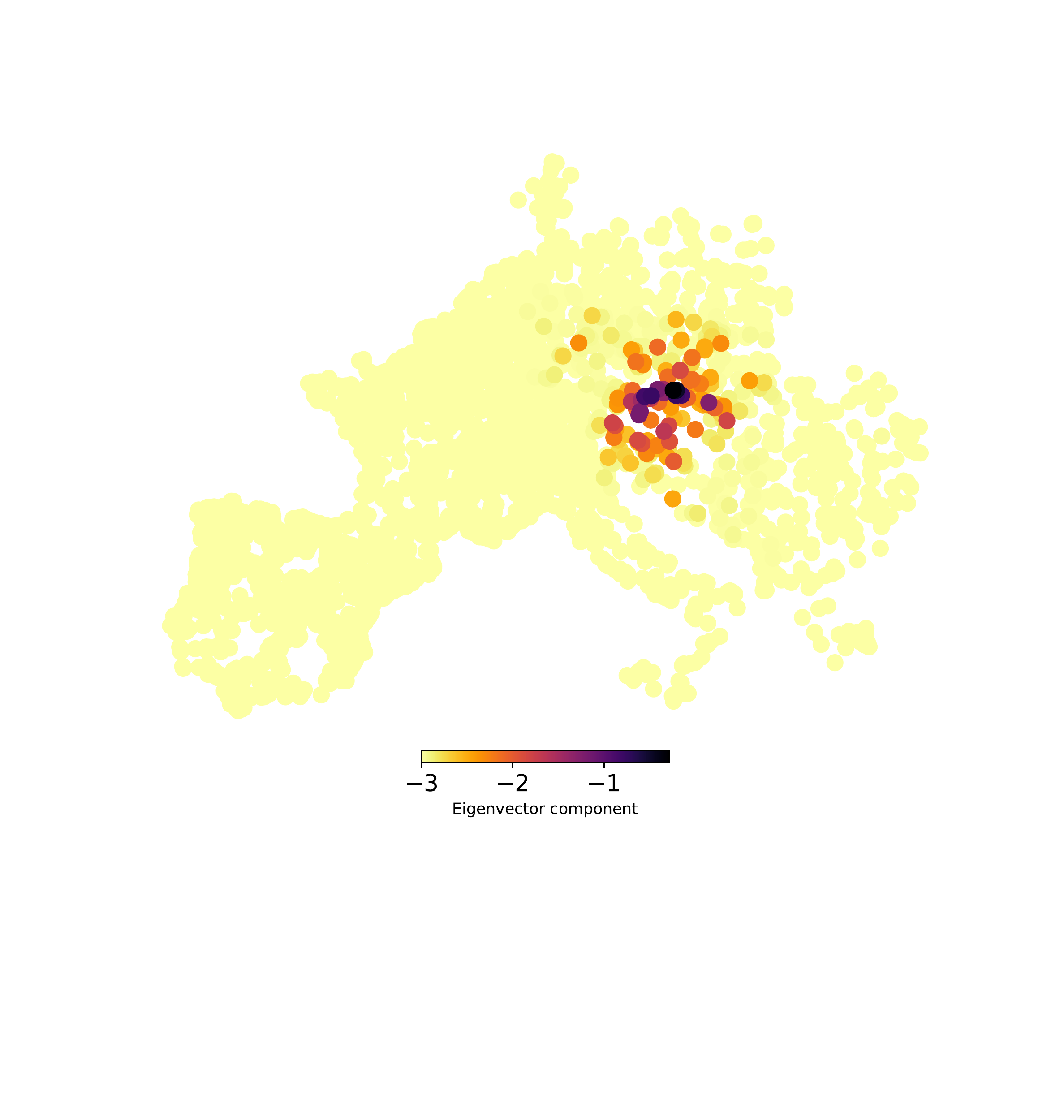}
    \end{adjustbox}}
        \caption{\textbf{Four eigenvectors of the Pantagruel model of the European grid.}
         In color scale are plotted the node components of $\log(\vert X\vert )$, where $X $ is an eigenvector associated with the eigenvalues $E=2$, $E=10$, $E=20$ $E=40$ (from left to right and from top to bottom).
          The first eigenvalues correspond respectively to the ballistic and diffusive regimes. The two last ones are localised states.}
    \label{fig:panta_eigenvectors}
 \end{figure*}

\begin{figure*}[ht!]
    \centering
    \subfloat{\resizebox{0.45\linewidth}{!}{\input{w_2_5_20_isar}}}\quad
    \subfloat{\resizebox{0.45\linewidth}{!}{\input{w_2_5_10_20_isar_R}}}\\
          \subfloat[$E=2$, $t=1~\text{s}$.]{\begin{adjustbox}{clip,trim=1.4cm 3.8cm 1.2cm 1.3cm,max width=0.3\linewidth}
        \includegraphics[width=0.7\linewidth]{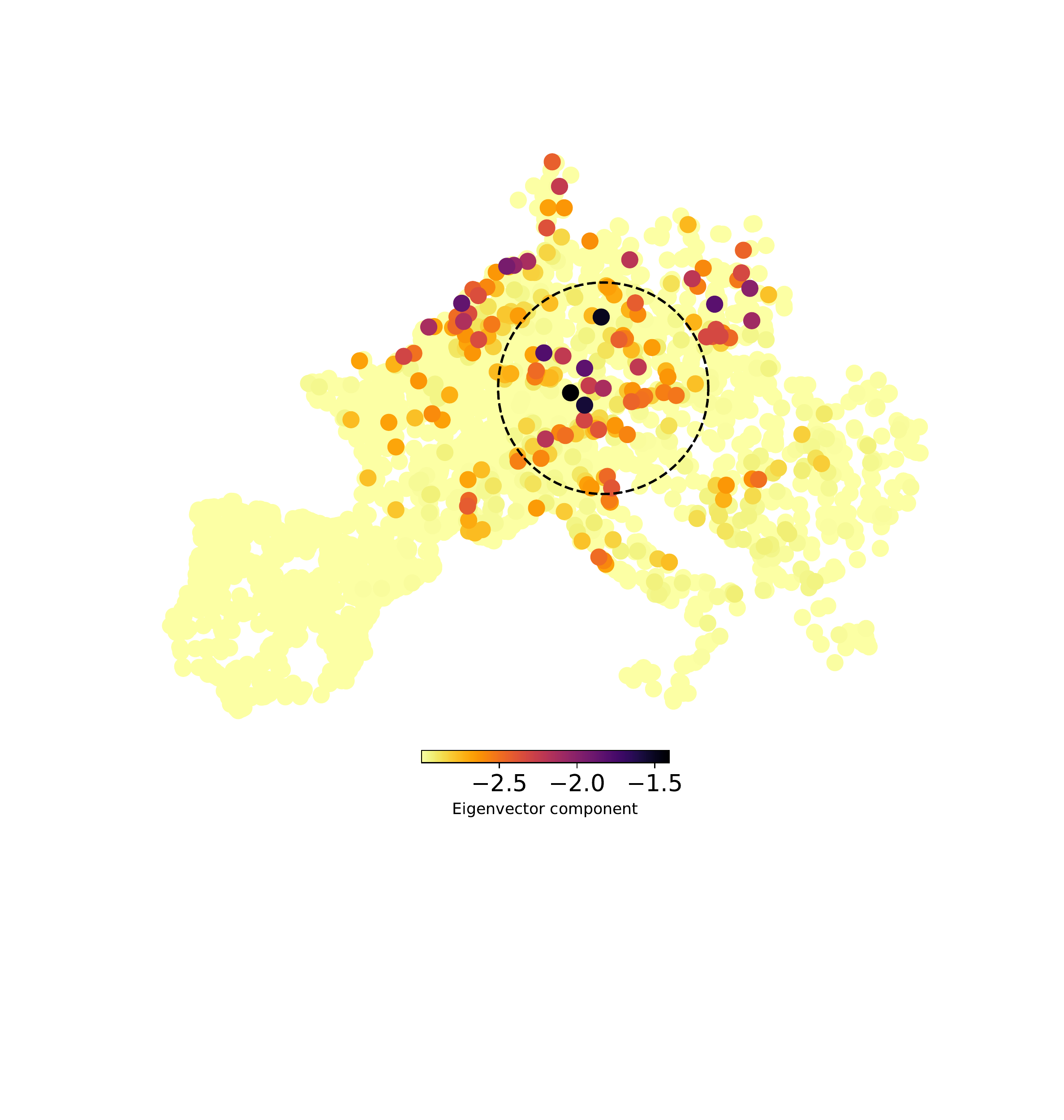}
    \end{adjustbox}}
    \subfloat[$E=10$, $t=1~\text{s}$.]{\begin{adjustbox}{clip,trim=1.4cm 3.8cm 1.2cm 1.3cm,max width=0.3\linewidth}
        \includegraphics[width=0.7\linewidth]{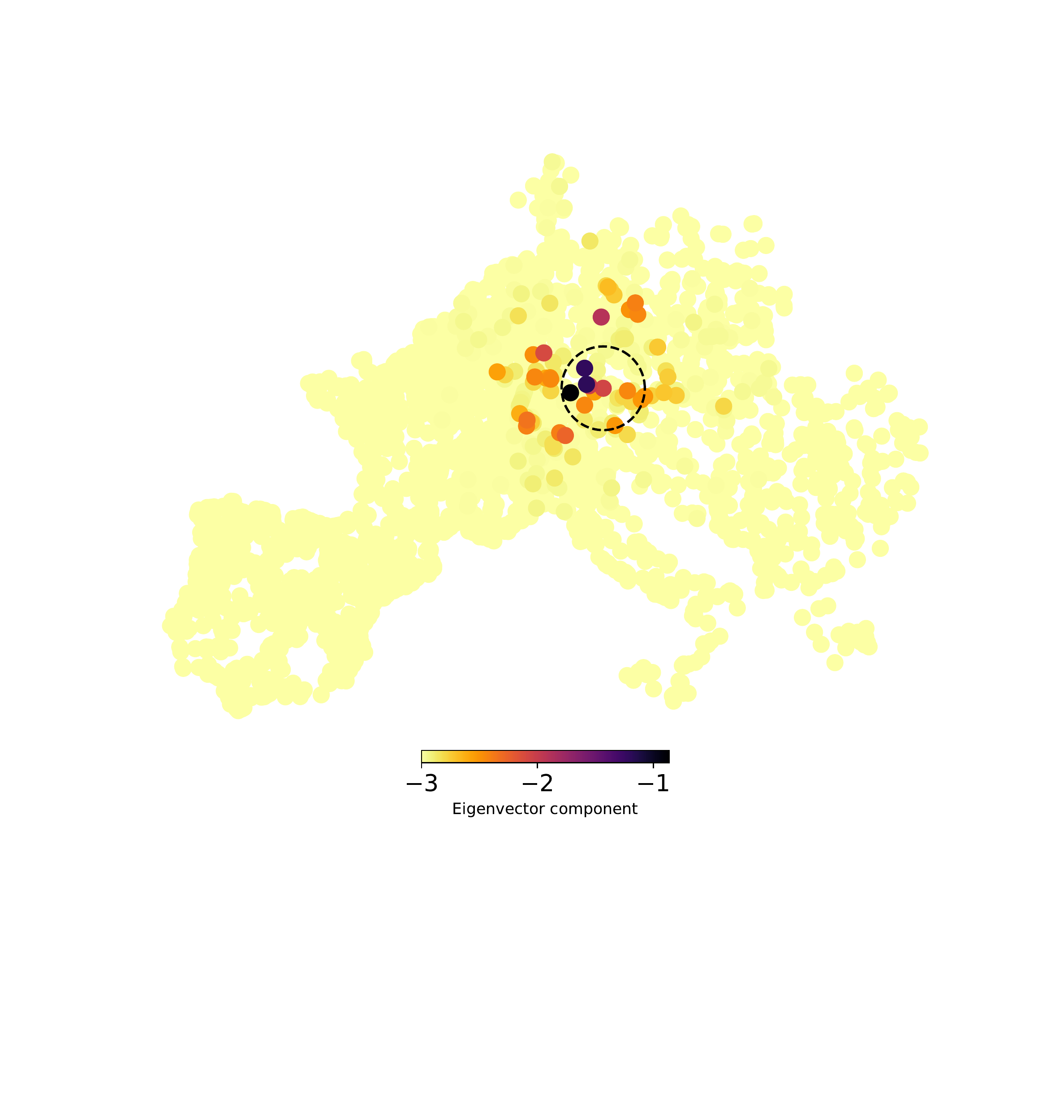}
    \end{adjustbox}}
    \subfloat[$E=20$, $t=1~\text{s}$.]{\begin{adjustbox}{clip,trim=1.4cm 3.8cm 1.2cm 1.3cm,max width=0.3\linewidth}
        \includegraphics[width=0.7\linewidth]{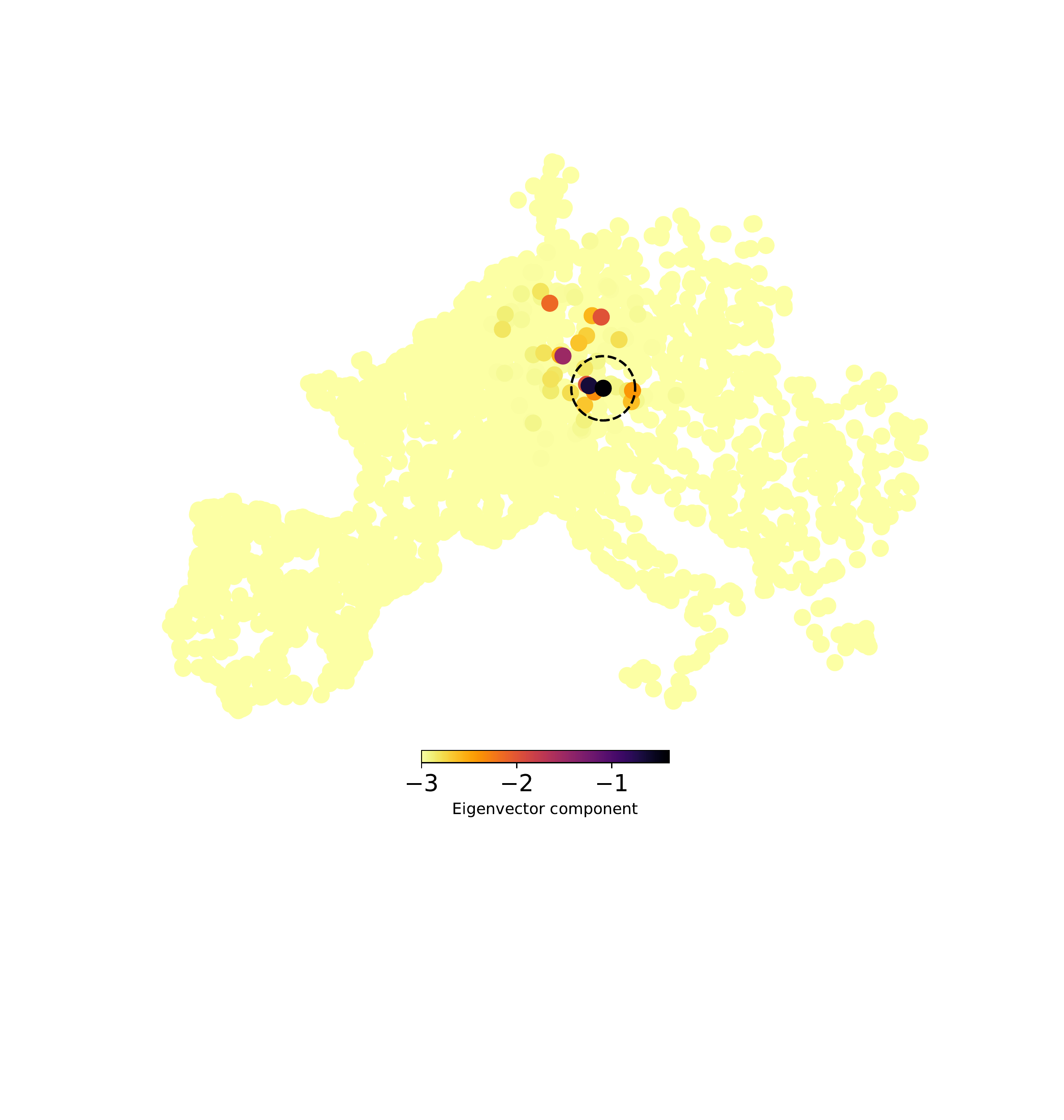}
    \end{adjustbox}}
    \\
    \subfloat[$E=2$, $t=5~\text{s}$.]{\begin{adjustbox}{clip,trim=1.4cm 3.8cm 1.2cm 1.3cm,max width=0.3\linewidth}
        \includegraphics[width=0.7\linewidth]{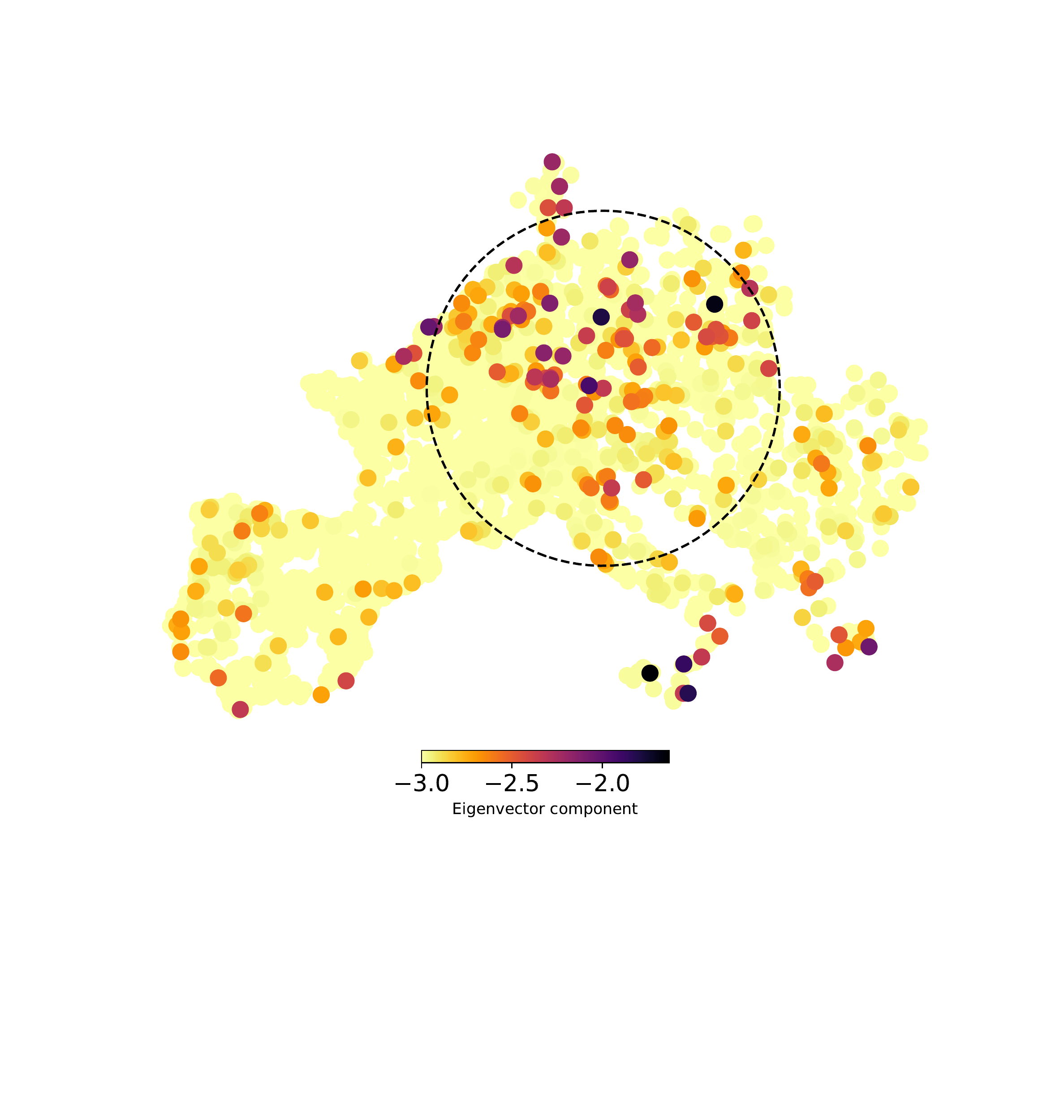}
    \end{adjustbox}}
    \subfloat[$E=10$, $t=5~\text{s}$.]{\begin{adjustbox}{clip,trim=1.4cm 3.8cm 1.2cm 1.3cm,max width=0.3\linewidth}
        \includegraphics[width=0.7\linewidth]{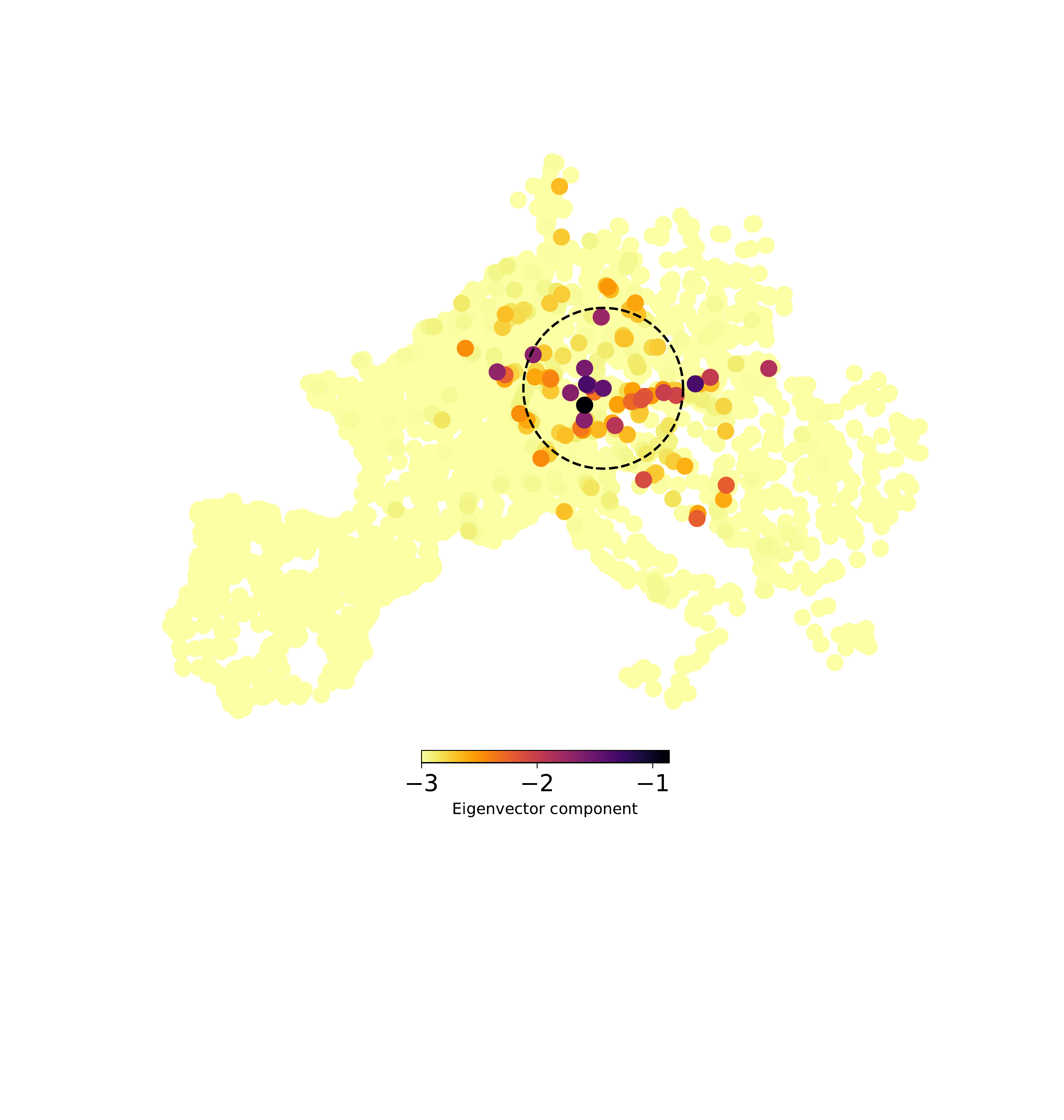}
    \end{adjustbox}}
    \subfloat[$E=20$, $t=5~\text{s}$.]{\begin{adjustbox}{clip,trim=1.4cm 3.8cm 1.2cm 1.3cm,max width=0.3\linewidth}
        \includegraphics[width=0.7\linewidth]{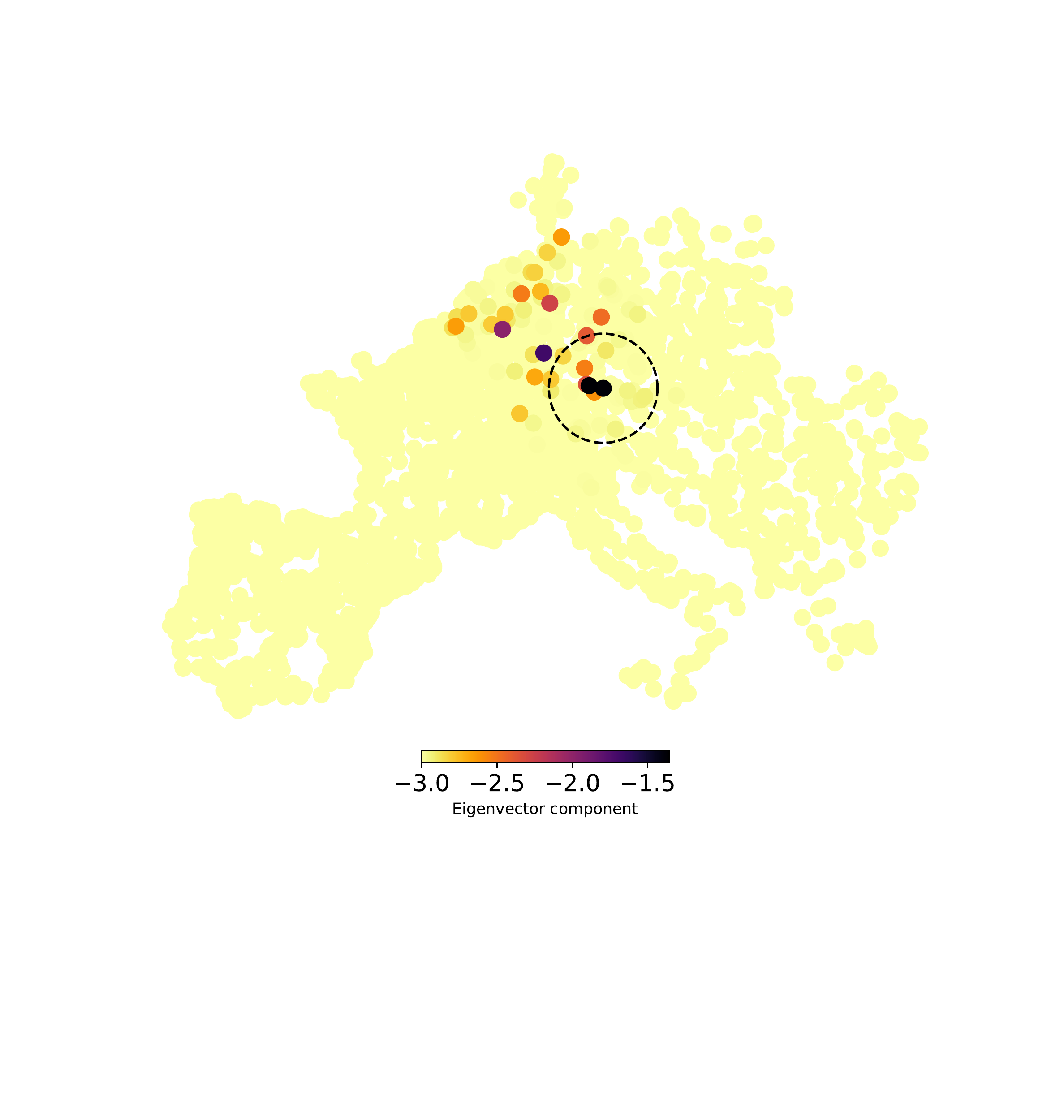}
    \end{adjustbox}}
        \caption{\textbf{Propagation of signals in the Pantagruel model of the European transmission grid, from the Isar node.} The upper left (respectively, right) panel represents $\frac{R^2}{t}$ (respectively, $R$) in function of $t$ for various angular frequencies $E$.
    The central and bottom panels show as a colormap the value of $\log(\frac{\vert \psi_n\vert^2}{\vert\vert \psi\vert \vert^2})$ for every node $n$ of the network. The central panels represent the signals at $t=1s$ for $E=2$, $E=10$ and $E=20$ (from left to right). The bottom panels represent the same quantities at $t=5s$.}
    \label{fig:panta_isar}
 \end{figure*}
 
  \begin{figure*}[ht!]
    \centering
    \subfloat{\resizebox{0.45\linewidth}{!}{\input{w_2_10_20_creys}}}\quad
    \subfloat{\resizebox{0.45\linewidth}{!}{\input{w_2_10_20_creys_R}}}\\
      \subfloat[$E=2$, $t=1~\text{s}$.]{\begin{adjustbox}{clip,trim=1.4cm 3.8cm 1.2cm 1.3cm,max width=0.3\linewidth}
        \includegraphics[width=0.7\linewidth]{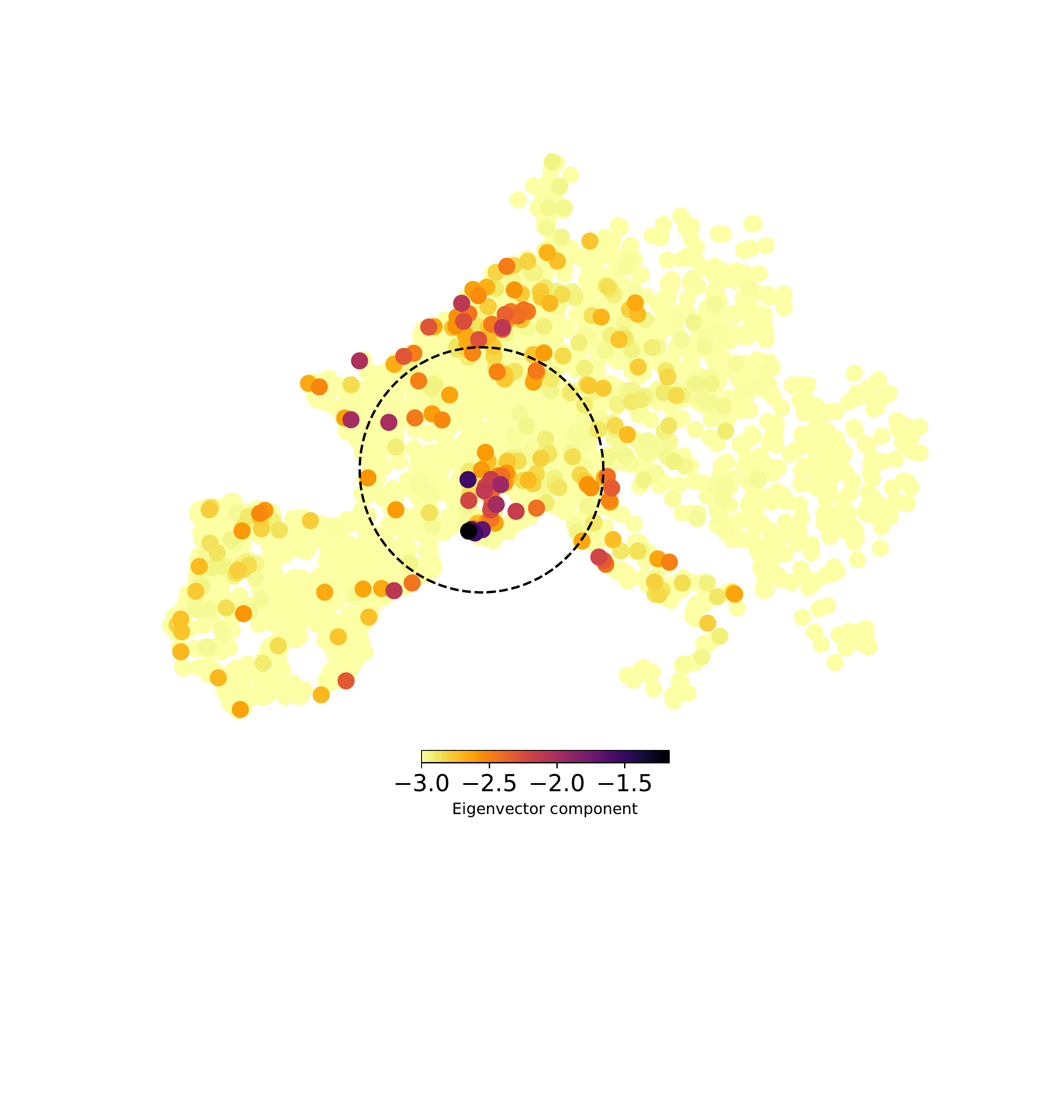}
    \end{adjustbox}}
    \subfloat[$E=10$, $t=1~\text{s}$.]{\begin{adjustbox}{clip,trim=1.4cm 3.8cm 1.2cm 1.3cm,max width=0.3\linewidth}
        \includegraphics[width=0.7\linewidth]{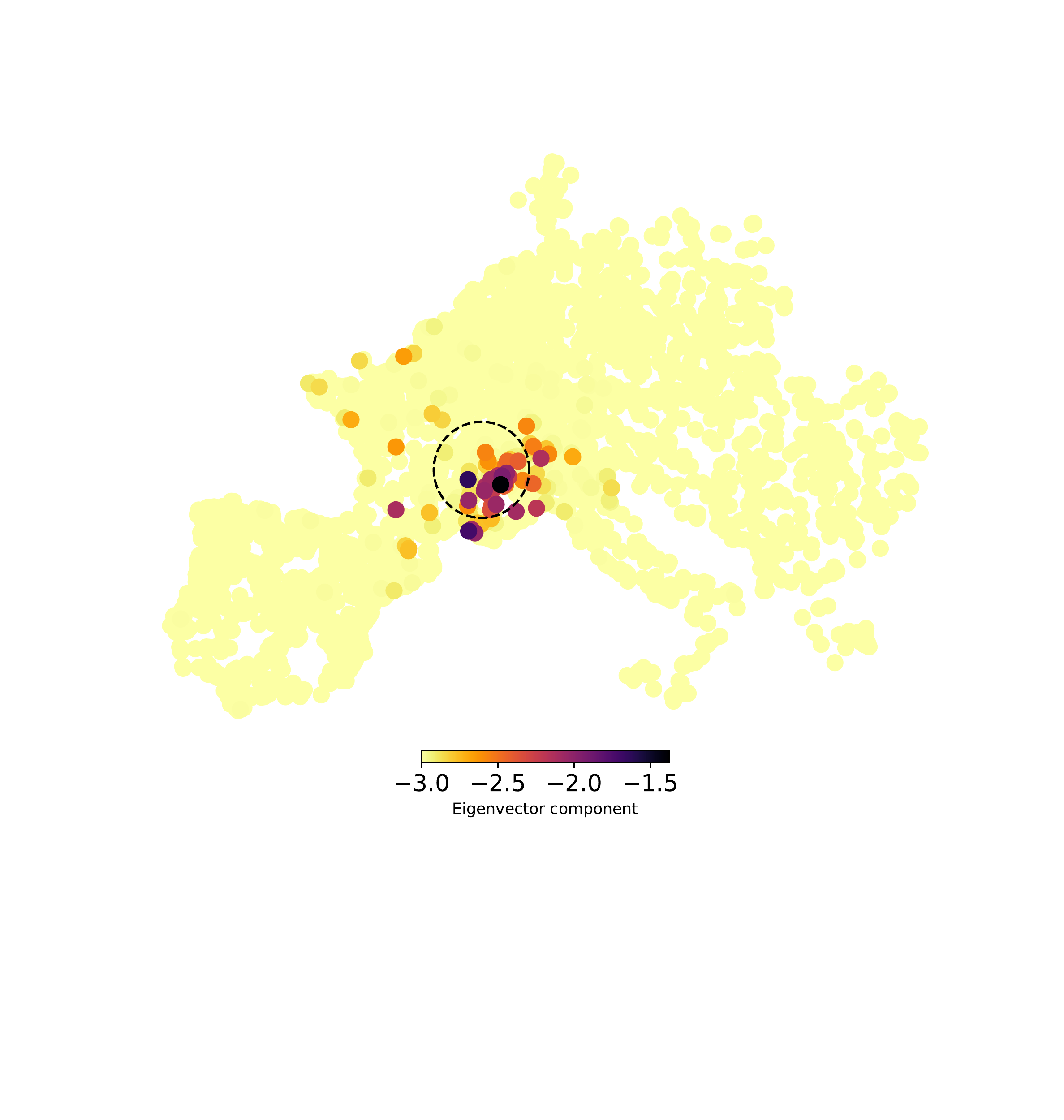}
    \end{adjustbox}}
    \subfloat[$E=20$, $t=1~\text{s}$.]{\begin{adjustbox}{clip,trim=1.4cm 3.8cm 1.2cm 1.3cm,max width=0.3\linewidth}
        \includegraphics[width=0.7\linewidth]{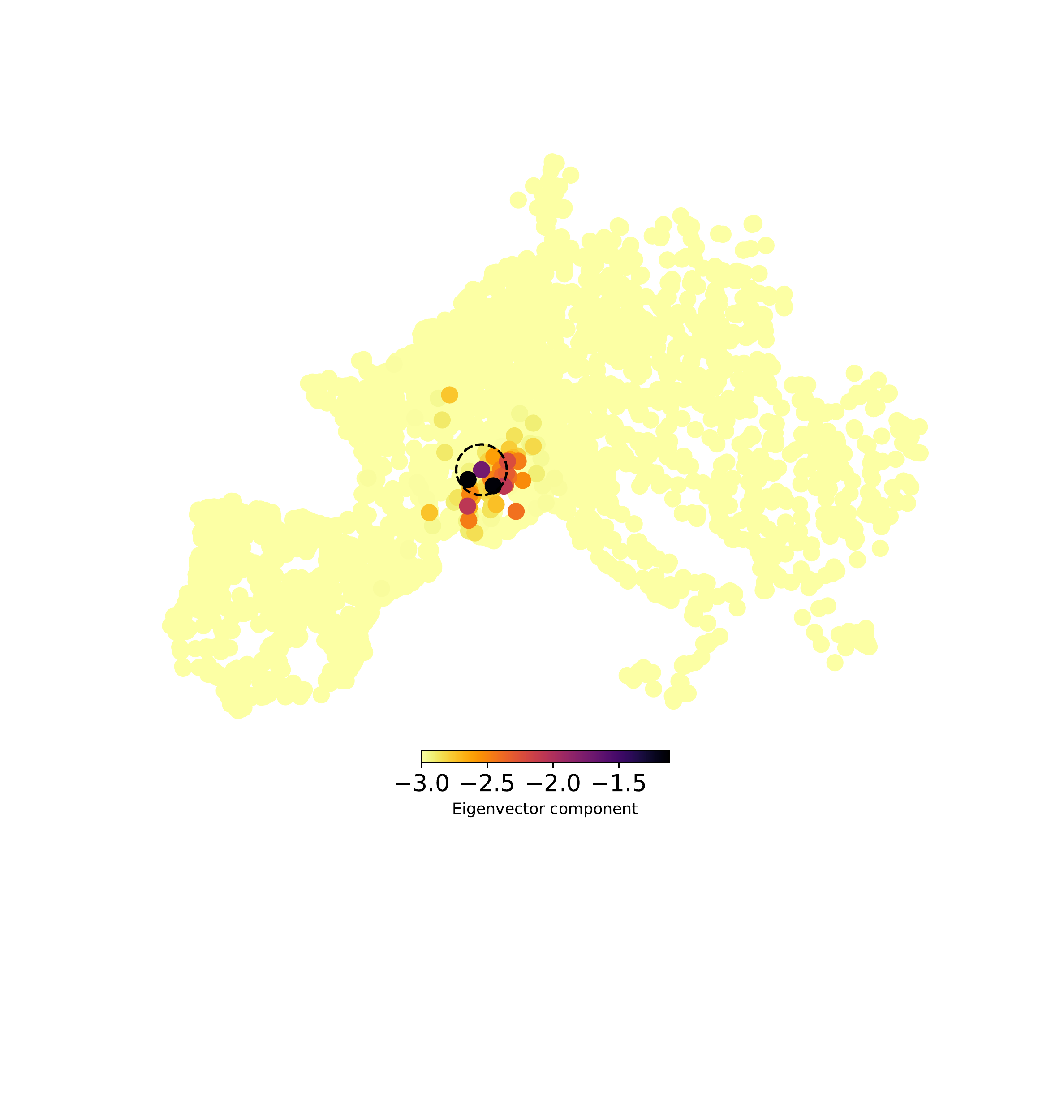}
    \end{adjustbox}}
    \\
    \subfloat[$E=2$, $t=5~\text{s}$.]{\begin{adjustbox}{clip,trim=1.4cm 3.8cm 1.2cm 1.3cm,max width=0.3\linewidth}
        \includegraphics[width=0.7\linewidth]{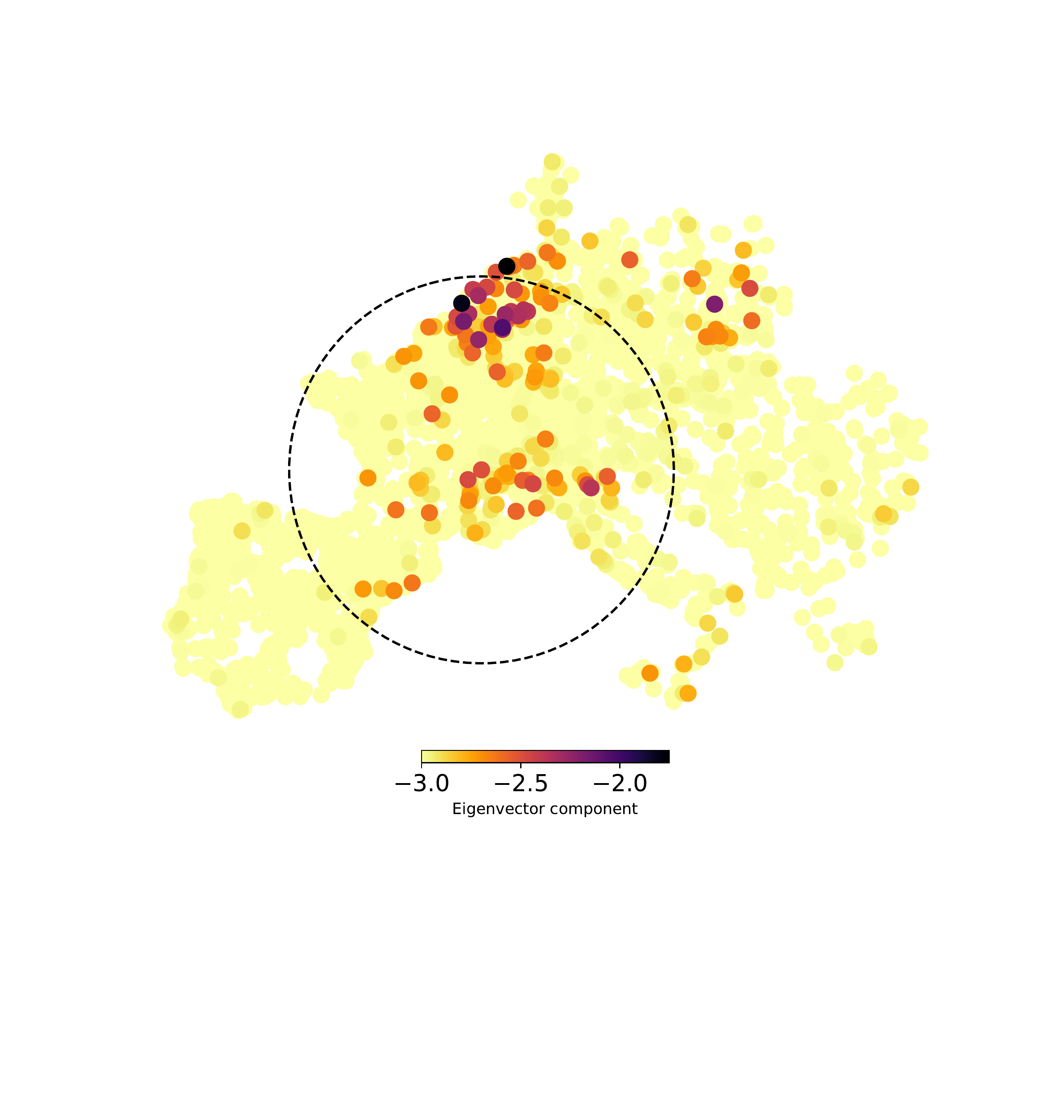}
    \end{adjustbox}}
    \subfloat[$E=10$, $t=5~\text{s}$.]{\begin{adjustbox}{clip,trim=1.4cm 3.8cm 1.2cm 1.3cm,max width=0.3\linewidth}
        \includegraphics[width=0.7\linewidth]{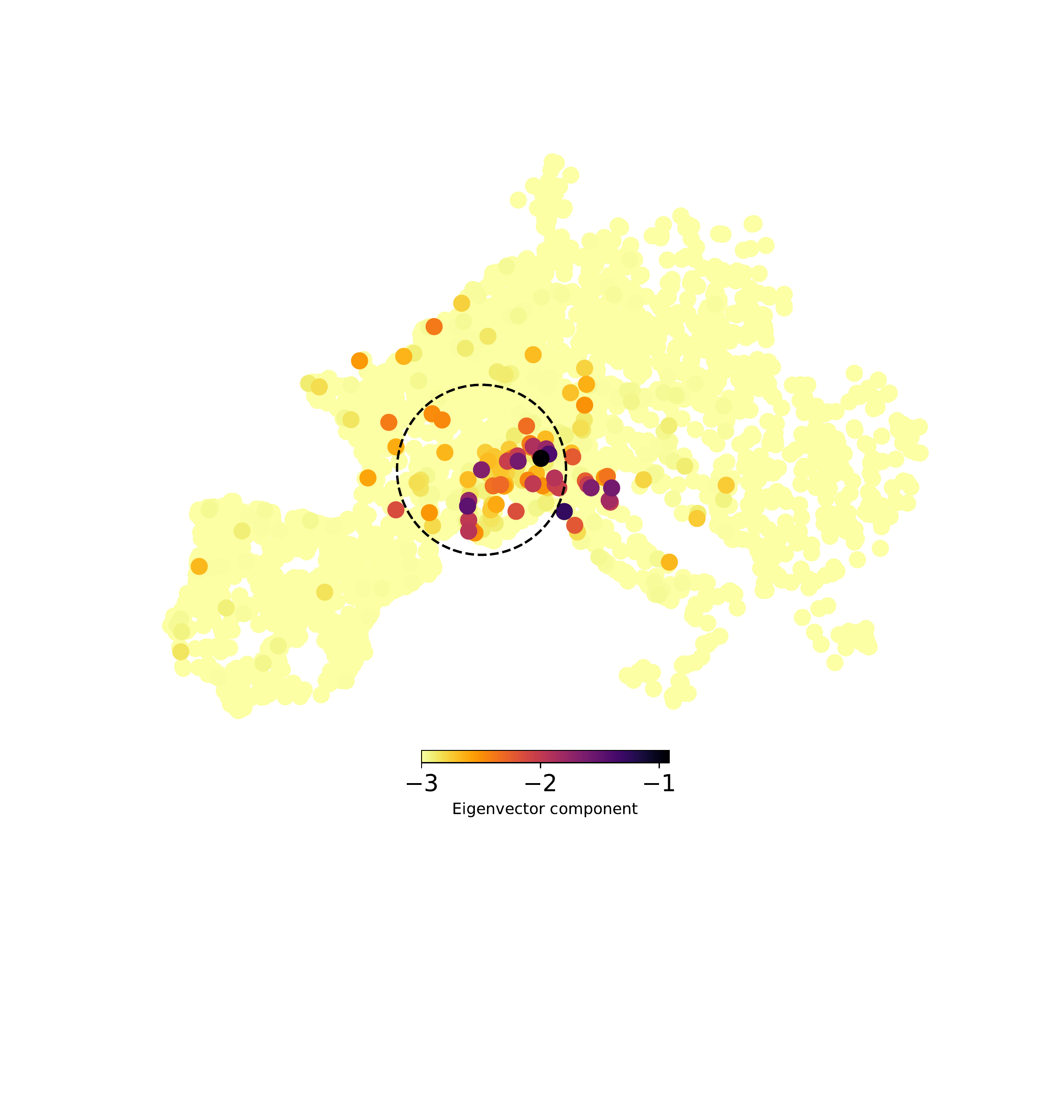}
    \end{adjustbox}}
    \subfloat[$E=20$, $t=5~\text{s}$.]{\begin{adjustbox}{clip,trim=1.4cm 3.8cm 1.2cm 1.3cm,max width=0.3\linewidth}
        \includegraphics[width=0.7\linewidth]{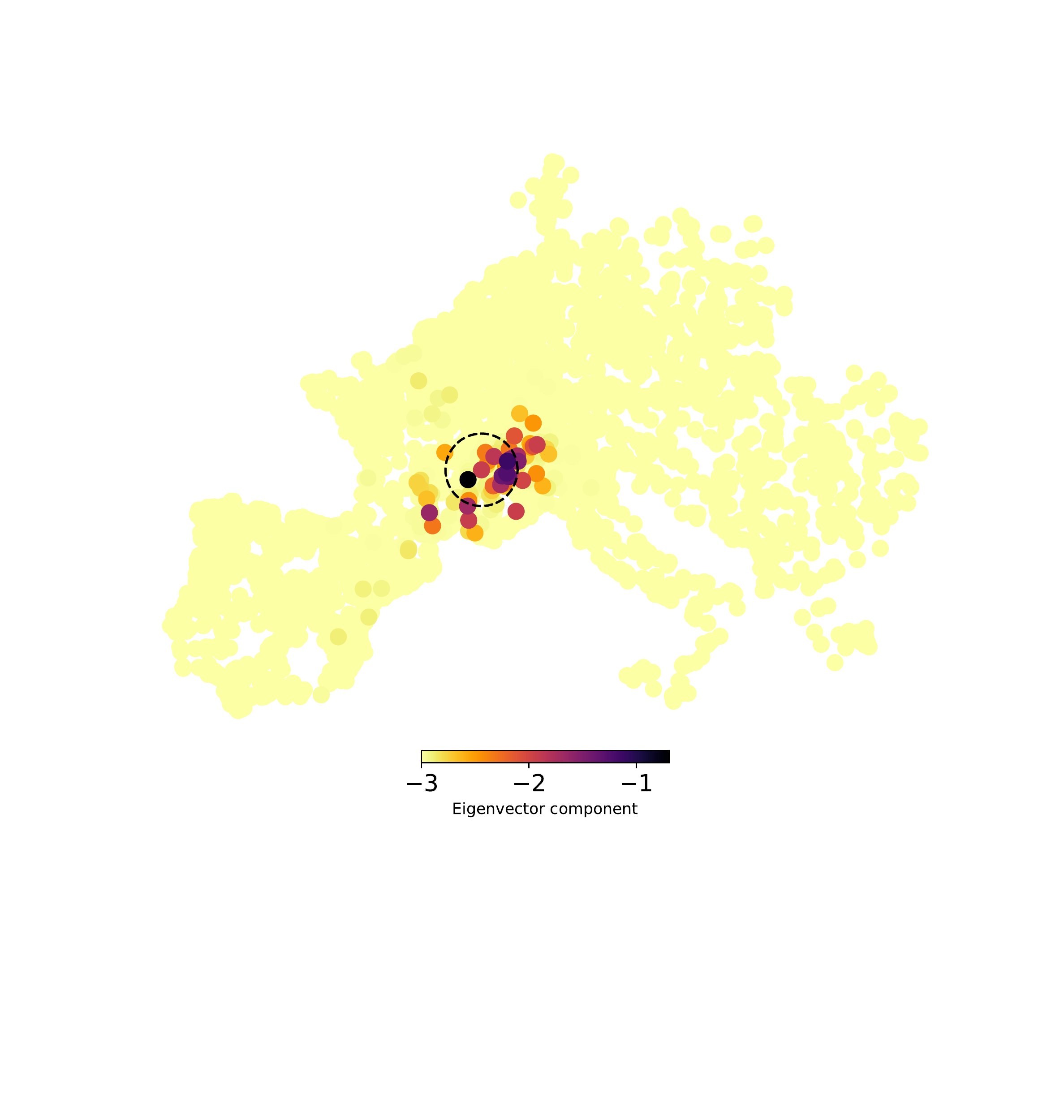}
    \end{adjustbox}}
      \caption{\textbf{Propagation of signals in the Pantagruel model of the European transmission grid, from the Creys-Malville node.} The upper left (respectively, right) panel represents $\frac{R^2}{t}$ (respectively, $R$) in function of $t$ for various angular frequencies $E$.
    The central and bottom panels show as a colormap the value of $\log(\frac{\vert \psi_n\vert^2}{\vert\vert \psi\vert \vert^2})$ for every node $n$ of the network. The central panels represent the signals at $t=1s$ for $E=2$, $E=10$ and $E=20$ (from left to right). The bottom panels represent the same quantities at $t=5s$.}
    \label{fig:panta_crey}
 \end{figure*}

\end{document}